\documentclass[aps,prx,superscriptaddress,twocolumn,showpacs,longbibliography]{revtex4-1}
\usepackage{amsmath, amsthm, amssymb}
\usepackage{bbold}
\usepackage{graphicx}
\usepackage{color}
\usepackage{times, verbatim}

\usepackage{pstricks}
\usepackage{pst-all}
\usepackage{bbold}
\newcommand{\bra}[1]{\left\langle #1 \right|}
\newcommand{\ket}[1]{\left| #1 \right\rangle}

\newcommand{\Top}{t_\text{op}}
\usepackage[english]{babel}
\usepackage{natbib}
\usepackage{hyperref}
\hypersetup{
        colorlinks=true,
}

\newcommand{\di}{\mathrm{d}}
\renewcommand{\vr}{{\mathbf{r}}}

\begin{document}

\title{How quickly can anyons be braided?
\\ Or: How I learned to stop worrying about diabatic errors and love the anyon}

\author{Christina Knapp}
\affiliation{Physics Department, University of California,  Santa Barbara, California 93106, USA}

\author{Michael Zaletel}
\affiliation{Station Q, Microsoft Research, Santa Barbara, California 93106-6105, USA}

\author{Dong E. Liu}
\affiliation{Station Q, Microsoft Research, Santa Barbara, California 93106-6105, USA}

\author{Meng Cheng}
\affiliation{Station Q, Microsoft Research, Santa Barbara, California 93106-6105, USA}

\author{Parsa Bonderson}
\affiliation{Station Q, Microsoft Research, Santa Barbara, California 93106-6105, USA}

\author{Chetan Nayak}
\affiliation{Station Q, Microsoft Research, Santa Barbara, California 93106-6105, USA}
\affiliation{Physics Department, University of California,  Santa Barbara, California 93106, USA}

\begin{abstract}
Topological phases of matter are a potential platform for the storage and processing of
quantum information with intrinsic error rates that decrease exponentially with inverse temperature and with the
 length scales of the system, such as the distance between quasiparticles. However, it is less well-understood how error rates depend on the speed with which
non-Abelian quasiparticles are braided. In general, diabatic corrections to the holonomy or Berry's matrix
vanish at least inversely with the length of time for the braid, with faster decay occurring
as the time-dependence is made smoother. We show that such corrections will not
affect quantum information encoded in topological degrees of freedom, unless they involve the creation of topologically nontrivial quasiparticles.
Moreover, we show how measurements that detect unintentionally created quasiparticles
can be used to control this source of error.
\end{abstract}

\maketitle

\tableofcontents

\section{Introduction}

Topological phases of matter can protect quantum information indefinitely at zero temperature, so long as all quasiparticles in the system are kept infinitely far apart and all processes are performed infinitely-slowly~\cite{Kitaev97,Nayak08}. If the temperature is not zero and quasiparticles are a finite distance $L$ apart, then errors will occur with a rate $\Gamma \sim \text{max}(e^{-\beta\Delta},e^{-L/\xi})$, where $\beta$ is the inverse temperature, $\Delta$ is the energy gap to topologically nontrivial
quasiparticles, and  $\xi \sim 1/\Delta$ is the correlation length~\cite{Kitaev97,Bonderson13}. The exponential suppression of thermal and finite-size errors makes topological phases a promising avenue for quantum computing, provided
that it is possible to control errors caused by moving quasiparticles in a finite duration of time. These ``diabatic errors''
are the subject of this paper.

For a system in a topological phase, the energy gap to topologically nontrivial quasiparticles determines a natural time scale, $1/\Delta$. In order to avoid unintentionally exciting quasiparticles,
all operations should be performed in a time $t_\text{op}$ that is much larger
than this time scale. On the other hand, the topological degeneracy of non-Abelian anyons
is not exact, except when all length scales are infinite, as there will generically be a small energy splitting
$\delta E \sim {E_0}e^{-L/\xi}$ between all nearly-degenerate states~\cite{Bonderson09}. (Here $E_0$ is an energy scale related to the
kinetic energy of quasiparticles, i.e. an ``attempt frequency'' for quantum tunneling events.)
Rotations between states in this nearly-degenerate state space will only occur so long as braiding is {\it fast} compared to $1/\delta E$.
Attempting to drag charged anyons through a disordered environment  presents a similar upper limit on the braiding time~\cite{Khemani2015}.
Therefore, we narrow our focus to the regime $1/\Delta \ll t_\text{op} \ll  1/\delta E$ and ask the question: within this range of time scales, how does the error rate decrease as $t_\text{op}$ is increased?

The unitary transformations effected by braiding non-Abelian quasiparticles in a gapped topological phase
can be understood as a manifestation of the non-Abelian generalization~\cite{Wilczek84}
of Berry's geometric phase~\cite{Berry84}.
More specifically, in the adiabatic limit, the unitary time evolution can be split into contributions from the dynamical phase, the Berry's matrix, and the instantaneous energy eigenbasis transformation. The dynamical phase is $t_{\text{op}}$-dependent. The combination of the Berry's matrix and the instantaneous energy eigenbasis transformation is known as the \emph{holonomy} and
is $t_\text{op}$-independent.
Consequently, corrections to the braiding transformations due to the finite
completion time for a braiding operation can be viewed as a special case of diabatic corrections to the holonomy.
In considering such corrections, it is important to keep in mind that, away from the adiabatic limit~\cite{Born28},
the time evolution of states does not cleanly separate into a $t_\text{op}$-independent holonomy
and a $t_\text{op}$-dependent dynamical phase.
In other words, for diabatic evolution, what one considers to be the dynamical phase is somewhat arbitrary. For the purpose of comparing with the adiabatic limit, it will be most convenient for us to call the quantity
$-\int_{0}^{t_\text{op}} dt E(t)$ the ``dynamical phase,'' where $E(t)$ is the instantaneous ground-state energy of the time-dependent Hamiltonian, even when we are not working in the adiabatic limit. Factoring this dynamical phase out of the (diabatic) time evolution operator, the remainder will generally depend strongly on the details of the Hamiltonian and will no longer simply be equivalent to the holonomy (which it equals in the adiabatic limit). The deviation of the remainder from its adiabatic limit is precisely what we wish to analyze for braiding transformations of topological quasiparticles.

Generically, diabatic corrections to the transition amplitude from a ground state to an excited state
vanish as $\mathcal{O}(1/t_\text{op})$ as $t_\text{op}$ is taken to infinity~\cite{Born28}. However,
 the scaling of diabatic corrections is sensitive to the precise time-dependence of the parameters in the Hamiltonian. In particular, the corrections are $\mathcal{O}(1/t^{k+1}_\text{op})$ when the time-dependence is $C^k$ smooth~\cite{Lidar09,Wiebe12,Lidar15}, and are exponentially suppressed when the time-dependence is analytic~\cite{Garrido62,Joye91a, Joye91b, Nenciu93, Joye98}. (Infinitely smooth $C^\infty$ time-dependence may result in stretched exponential decay of corrections.)  As transitions out of the ground state subspace may affect the topological degrees of freedom, diabatic corrections to braiding do not appear to exhibit the nice topological protection, i.e. exponential suppression of errors, that thermal and finite-size corrections exhibit.  Moreover, they seem to depend on details to a worrisome extent, though one may question whether this dependence is stable against noise in these parameters, as may arise from coupling to a bath.

On the other hand, quantum information encoded in a topological state space is expected to be corrupted
only by the uncontrolled motion of quasiparticles. This is the reason for the temperature
and length dependence of error rates: the density of thermally-excited quasiparticles, which decohere the topological states by diffusing through the system, scales as $e^{-\beta\Delta}$; the amplitude for virtual quasiparticles to be transferred between two quasiparticles separated by a distance $L$ scales as $e^{-L/\xi}$, which generically splits degeneracies of their topological states. Hence, one would expect that diabatic corrections to the holonomy would only
affect the overall phase of a state, rather than the quantum information encoded in it, unless quasiparticles
are created or braided in an unintended manner. In other words, it seems possible for diabatic
corrections to be large, but only entering as overall phases when there is no uncontrolled quasiparticle motion,
allowing the encoded quantum information to remain topologically-protected.

This is, indeed, the case. Diabatic errors are due to the uncontrolled creation
or motion of quasiparticles; other diabatic corrections to the holonomy do not affect the topologically-encoded quantum information.
Since these quasiparticles are created by the diabatic variation of specific terms in the Hamiltonian,
they can only occur in specific places, i.e. in the vicinity of the quasiparticles' motion paths. These errors can, therefore, be diagnosed by corresponding measurements
and corrected. Such protocols apply to diabatic errors, but they cannot correct all errors, such as
those due to tunneling or thermally-excited quasiparticles, which must be minimized by increasing quasiparticle separations and lowering the temperature, or by engineering a shorter correlation length and larger energy gap. If all of these different sources of errors were significant, it would require a full-blown error-correcting code to contend with them.  In this paper, we focus on corrections which are not exponentially suppressible and we leave implicit errors due to non-zero correlation length and finite gaps.

Previous studies have considered the effects of diabatic evolution on particular topological systems.  Refs.~\onlinecite{Cheng11, Karzig13, Scheurer13} have investigated the stability of Majorana zero modes (MZMs)~\cite{Read00,Volovik99,Kitaev01} outside the adiabatic limit and other papers have suggested methods of reducing the diabatic error for MZMs~\cite{Karzig14,Karzig15} and for Kitaev surface codes and color codes~\cite{Cesare14}.  In this paper,  we consider diabatic error for braiding more broadly.  We present results on the magnitude, origin, and correction of diabatic errors for general anyonic braiding.  We further apply our results to the braiding of MZMs.~\footnote{MZMs are actually extrinsic defects, not anyons, but their topological properties closely resemble those of the non-Abelian Ising anyons, so it is clear that our analysis should apply equally to both cases. In particular, MZMs have the same non-Abelian fusion rules and a projective version of the braiding transformations of Ising anyons. More generally, our analysis and discussion can be applied to extrinsic defects, such as parafermionic zero modes, for which braiding is projective and described by $G$-crossed theory~\cite{Barkeshli14}, but does not necessarily correspond to projective braiding of an anyon theory.} (See Ref.~\onlinecite{Alicea12a} for an excellent review on MZMs and proposed physical realizations.) In particular, we concentrate on MZMs in topological superconducting nanowires~\cite{Kitaev01, Lutchyn10, Oreg10, Alicea12a}, both for concreteness and also because experimentally such systems have been successfully realized and signatures of MZMs have been observed~\cite{Mourik12,Das12,Deng12,Finck12,Rokhinson12,Churchill13,Nadj-Perge14,Deng2014,Higginbotham15}. The braiding transformations of MZMs in such systems are implemented in a quasi-one-dimensional geometry by slow variations of the couplings in a nanowire $T$-junction~\cite{Alicea11, SauPRB2011, Heck12, Hyart13}. We will critically analyze the practical aspects of our theory applied to the braiding and measurement schemes introduced in Refs.~\onlinecite{Hassler11, Heck12, Hyart13}.

This paper is structured as follows. After briefly reviewing previous literature on quasi-adiabatic evolution of two-level systems in Section~\ref{sec:LZ}, we investigate the effect of dissipative coupling to a thermal bath in Section~\ref{sec:dissipation}. In Section~\ref{sec:CS-theory}, we consider the motion of one anyon around a second anyon fixed at the origin within a Chern-Simons effective field theory with fixed anyon number. We show that diabatic corrections to the holonomy do not affect the braiding phase unless diabatic variation of the Hamiltonian parameters causes the moving anyon to have a non-vanishing amplitude of following trajectories that wind a different number of times than intended around the stationary anyon.  In~Section~\ref{sec:MZM-LZ}, we compute the diabatic corrections to the braiding transformation of MZMs.  We show that these corrections are of the form of generic diabatic corrections: the transition amplitude vanishes as $1/t^2_{\text{op}}$.  In Section~\ref{sec:Measurement=Correction}, we show that these errors can be diagnosed by measurements and corrected by a repeat-until success protocol.  We generalize this error-correction protocol to generic non-Abelian anyon braiding in Section~\ref{sec:general-braiding}.  In Section~\ref{sec:Hyart-system}, we apply our results to the proposal of Ref.~\onlinecite{Hyart13} and introduce a variation of the qubit therein to facilitate measurements.  We critically assess the feasibility of such a correction scheme with current technology in Section~\ref{sec:feasibility}.  Finally, we address the question posed in the title of this paper in Section~\ref{sec:discussion}.

\section{Quasi-Adiabatic Evolution of Two-Level Systems}
\label{sec:general-theory}

\subsection{Landau-Zener Effect and the Dependence on Turn-On/Off}
\label{sec:LZ}

Diabatic corrections to the adiabatic limit asymptotically decrease with the operation time $\Top$ with a functional form which depends on the smoothness of the time dependence in the Hamiltonian.  In particular,
if the time dependence of the Hamiltonian is analytic (within a strip around the real axis), diabatic corrections decay exponentially in the inverse of the rate at which the Hamiltonian evolves.
A classic example was provided by  Landau~\cite{Landau32} and Zener~\cite{Zener32},
who considered a two-level system with the following time-dependent Hamiltonian:
\begin{equation}
H_{\text{LZ}}(t) = ct \sigma_z -\lambda \sigma_x.
\label{eqn:L-Z-model}
\end{equation}
We will assume $c>0$ in the following.
The state of the system takes the form
\begin{equation}
\left| \psi(t) \right\rangle = a(t)\, \left| \uparrow \right\rangle + b(t)\,\left| \downarrow \right\rangle.
\end{equation}
We consider a time evolution starting from $t=-\infty$ and ending at $t$ large, given by
\begin{equation}
	\begin{pmatrix}
		a(t)\\
		b(t)
	\end{pmatrix}
=
\left[
	\begin{matrix}
		S_1 & S_2\\
		-S_2^* & S_1^*
	\end{matrix}
\right]
	\begin{pmatrix}
		a(-\infty)\\
		b(-\infty)
	\end{pmatrix}.
	\label{}
\end{equation}
Then, as we review in Appendix~\ref{sec:Landau-Zener}, the matrix elements are found to be (dropping subleading contributions)
\begin{eqnarray}
S_1 &=& e^{-\frac{\pi}{2} \Lambda  } \\
S_2 &=& - 2\sqrt{\frac{\pi}{\Lambda}}  \frac{e^{-\frac{\pi}{4}\Lambda}}{ \Gamma(-i \frac{\Lambda}{2})}e^{i \frac{\pi }{4} -i\Phi(t) },
	\label{}
\end{eqnarray}
where we have defined $\Lambda=\frac{\lambda^2}{c}$ and $\Phi(t)={ct^2}+\Lambda \ln |2ct|$. In the above we take the $t\rightarrow\infty$ limit, but keep the time dependence in the oscillatory phase $\Phi(t)$ as it does not have a well-defined limit (this does not affect the diabatic transition probability).

When the system is initially in the ground state, i.e. $a(-\infty)=1$ and $b(-\infty)=0$, the final state's probability for a transition into the excited state is given by
\begin{equation}
P_{\text{G} \rightarrow \text{E} } = |a(t\rightarrow\infty)|^2 = |S_1|^2 = e^{-\pi\frac{{\lambda^2}}{c}}.
\end{equation}
If the goal is to remain in the ground state, then this is an error, but it is
an error that is exponentially small in $\Lambda$, the inverse of the
speed with which the system is moved through the avoided crossing.

A few comments are in order. In the model in Eq.~(\ref{eqn:L-Z-model}), the spectral gap goes to infinity at
large times. One might worry that the exponential protection in the Landau-Zener model is an artifact of
an infinite asymptotic gap. Since we will generally be interested in Hamiltonians which
have a spectral gap that is approximately constant, it is important to see that such protection applies to
such Hamiltonians as well. To this end, consider the family of Hamiltonians
\begin{equation}
H_{\theta}(t) = {E_0} \cos(\theta(t))\, \sigma_z + {E_0}\sin(\theta(t))\, \sigma_x
\label{eqn:theta-Hamiltonian}
\end{equation}
The Hamiltonian $H_{\text{LZ}}(t)/\sqrt{{c^2}{t^2} + {\lambda^2}}$ is of this form, with
$\cos(\theta(t)) = ct/\sqrt{{c^2}{t^2} + {\lambda^2}}$, $\sin(\theta(t))=-\lambda/\sqrt{{c^2}{t^2}+\lambda^2}$, and $E_0=1$. A change of variables
to $\tilde{t}(t)$ with $d\tilde{t}/dt = \sqrt{{c^2}{t^2} + {\lambda^2}}$ applied
to Schr\"odinger's equation brings the Hamiltonian $H_{\text{LZ}}(t)$ to the form ${H_\theta}(\tilde{t})$. If the function
$\tilde{t}(t)$ is bounded by a polynomial, then the protection will remain exponential in the new time
variable, in terms of which the Hamiltonian has a constant gap. Since $\tilde{t} \sim \lambda t$ for small $t$
and $\tilde{t} \sim \pm \frac{1}{2}ct^2$ for large $t$, this is satisfied.

Although the speed with which the Hamiltonian evolves, as measured by $|\dot{H}|/|H|$, is roughly $c/\lambda$
near the avoided crossing, the total time of the adiabatic evolution is infinite.
This was the price that we paid in order to evolve the system in a completely analytic manner.
If the time dependence changes more sharply, so that the total operation time is finite, then the exponential protection will disappear.
To see an example of this, we modify the Hamiltonian of Eq.~(\ref{eqn:L-Z-model})
to one in which the time dependence occurs over a finite interval. There are several ways to do this;
we focus on one that will have relevance to later sections of the paper.
We consider a time dependent Hamiltonian of the form
\begin{equation}
H(t) =  h(t) \sigma_z - \lambda\sigma_x
\label{eqn:piece-wise-model}
\end{equation}
with
\begin{equation}
h(t) = \left\{
\begin{array}{rrrl}
-c{t_\text{op}} & & \text{ for } & t \leq -{t_\text{op}} \\
ct_{\phantom{\text{op}}} & & \text{ for } & -t_\text{op} \leq  t \leq  t_\text{op} \\
c{t_\text{op}} & & \text{ for } &  t_\text{op} \leq t
\end{array}
\right .
\end{equation}
In the adiabatic limit, this Hamiltonian rotates the state of the system between non-orthogonal initial and final states.
In the long-time regime, where $\sqrt{c}\Top\gg 1$ and $c\Top\gg \lambda$, we find that the time evolution operator acquires a correction to its diagonal components (see Appendix~\ref{sec:Landau-Zener} for a derivation):
\begin{equation}
S_1=e^{-\frac{\pi}{2} \Lambda } -\sqrt{\frac{\pi}{c}}\frac{e^{-\frac{\pi}{4}\Lambda }}{\Gamma(-i \frac{\Lambda}{2}) {t_\text{op}}} e^{-i \frac{\pi}{4} +i\Phi(t) } .
	\label{eqn:S1-LZ-sudden}
\end{equation}
The transition probability is given by
\begin{equation}
	P_{\text{G} \rightarrow \text{E}} = \frac{\lambda^2}{4c^2\Top^2} + \mathcal{O}\Big(\frac{e^{-\frac{\pi \lambda^2}{2c} }}{\sqrt{c}\Top}, e^{-\frac{\pi\lambda^2}{c}}\Big).
	\label{}
\end{equation}
Here we only worry about the corrections that do not decay exponentially with $\Lambda$.

The $\mathcal{O}(\Top^{-2})$ diabatic transition probability is characteristic of any continuous, but otherwise generic,  time dependence.  A set of more general results show that errors become smaller as the evolution becomes smoother~\cite{Lidar09,Wiebe12,Lidar15}.  If the first $k$ derivatives of the Hamiltonian exist and are continuous, then the diabatic corrections to the transition probability vanish as ${\cal O}({t^{-2 k-2}_\text{op}})$.  Our primary interest will be diabatic corrections to the holonomy, the scaling of which we will return to at the end of Section~\ref{sec:CS-theory}. Previous studies were done in the context of adiabatic quantum computing and thus did not address diabatic corrections to the holonomy.

\subsection{Effects of Dissipation due to Coupling to a Bath}
\label{sec:dissipation}

Although this dependence on the differentiability of the Hamiltonian is mathematically correct, one may worry about its relevance to experimental solid state systems, for which noise and dissipation are unavoidable.
At the turn-on and turn-off of the time dependence, when the time derivatives of the Hamiltonian are small, but perhaps not quite zero (hence, requiring a discontinuity in the next higher derivative), noise could wash out some of the sensitivity to the precise values of these derivatives.
Hence, it is interesting to study the effect of coupling to a dissipative bath,
which is effectively like randomly adding discontinuities to the time dependence of the system Hamiltonian.

In anticipation of our eventual application to MZMs, we consider the product of \emph{two} two-level systems, which we can think of as spins with the corresponding Pauli operators $\vec{\sigma}$ and $\vec{\tau}$.
The two-level systems are coupled to a bath through bath operators $B_j$ as described by the Hamiltonian
\begin{equation}
H = \sum_{j=1}^3 \left[  - \Delta_j(t)(1+B_j) \sigma_j \otimes \tau_z + H_{Bj} \right ].
\label{eqn:spin-hamiltonian-with-diss}
\end{equation}
The system has an exact two-fold degeneracy labeled by ${\tau_z=\pm 1}$, which we think of as distinct ``sectors.''
The bosonic bath, which is a proxy for all of the environmental degrees of freedom other than the two spins, is modeled by a collection of oscillators through the terms
\begin{align}
B_{j} &=\sum_{\alpha}\tilde{\lambda}_{j\alpha}(a_{j\alpha}^{\dagger}+a_{j\alpha})
\\
H_{Bj} &= \sum_{\alpha}\omega_{j\alpha}a_{j\alpha}^{\dagger}a_{j\alpha}.
\end{align}
The bath couplings $\tilde{\lambda}$ are chosen to model a  zero-temperature Ohmic bath.
Each spin component $\sigma_j$ couples to a different subset of the oscillators $a_{j\alpha}$.
The crucial features of this Hamiltonian, which are not generic to all two-level systems, are that $\sigma_j$ is only coupled to the bath when ${\Delta_j}(t)\neq 0$ and that the bath is uncorrelated for different $\sigma_{j}$.
The first feature was chosen for reasons that will become clear in Section~\ref{sec:MZM-LZ}, when we discuss the braiding of MZMs, the choice of uncorrelated noise will be explained in Section~\ref{sec:DiabaticErrorTopTransmon}.

\begin{figure}
\begin{centering}
\includegraphics[width=.85\columnwidth]{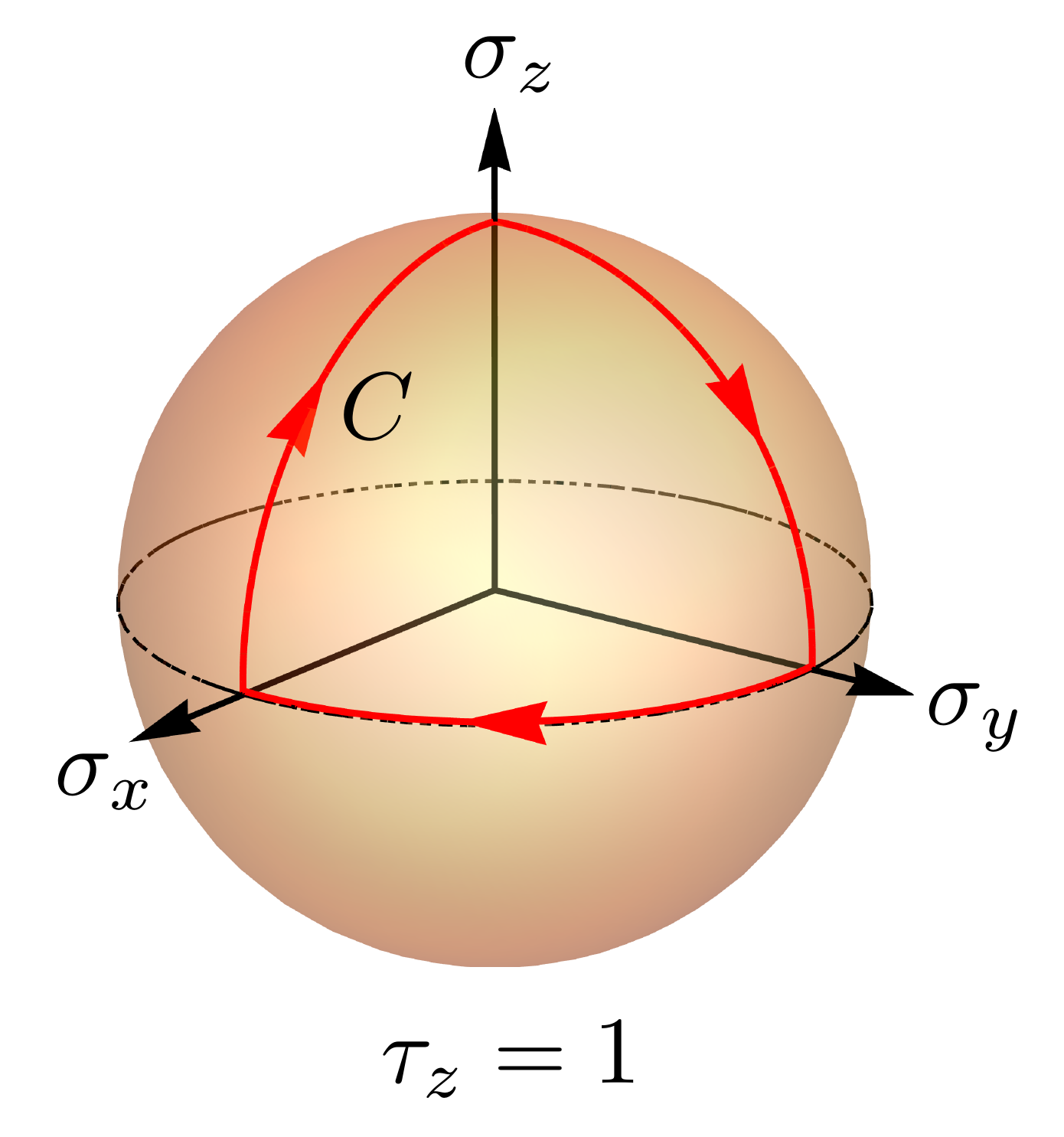}
\end{centering}
\caption{In the $\tau_z=1$ sector the instantaneous eigenstates of $H(t)$ trace out an octant of the Bloch sphere, shown above as the contour $C$.  At times $t=0, t_1, t_2, t_{\text{op}}$ only one of the $\Delta_i$ is non-zero.  At these times, $\sigma_i$ commutes with the Hamiltonian and the corresponding point on the contour is one of the corners or ``turning points." The holonomic phase at the end of the evolution is half the solid angle traced out by the contour $C$, $\frac{\Omega(C)}{2}=-\frac{\pi}{4}$.}
\label{fig:bloch-sphere}
\end{figure}

We choose the time dependence of the ${\Delta_j}(t)$ to consist of three steps through which the instantaneous eigenstates of $H$ circumscribe an octant of the Bloch sphere, as shown in Fig.~\ref{fig:bloch-sphere}.
Specifically, we interpolate linearly in time between $({\Delta_1}, {\Delta_2}, {\Delta_3}) = (0,\Delta,0)$ at time $t=0$ and $(\Delta,0,0)$ at $t = t_1$; between $( \Delta,0, 0)$ at $t=t_1$ and $(0,0, \Delta)$ at $t=t_2$; and finally between $(0,0,\Delta)$ at $t=t_2$ and $(0,  \Delta,0)$ at $t=t_{\text{op}}$. This evolution is similar to ``adiabatic gate teleportation,'' as discussed in Ref.~\onlinecite{Bacon09}.  In the $\tau_z = 1$ sector,  the ground state acquires the holonomic (geometric) phase $-\pi/4$.
In the $\tau_z = -1$ sector, the handedness is reversed, and the ground state acquires the holonomic phase $\pi/4$.
The dynamical phase, on the other hand, is identical for the two sectors, since they are related by an anti-unitary symmetry which takes $\sigma_{j} \to - \sigma_{j}$.
Thus, the dynamical phase can be canceled by comparing the $\tau_z = 1$ and $\tau_z = -1$ sectors, and the $\tau_z=-1$ sector picks a $\pi/2$ holonomic phase relative to the $\tau_z=1$ sector during the time evolution in the adiabatic limit.

\begin{figure}
\begin{centering}
\includegraphics[width=\columnwidth]{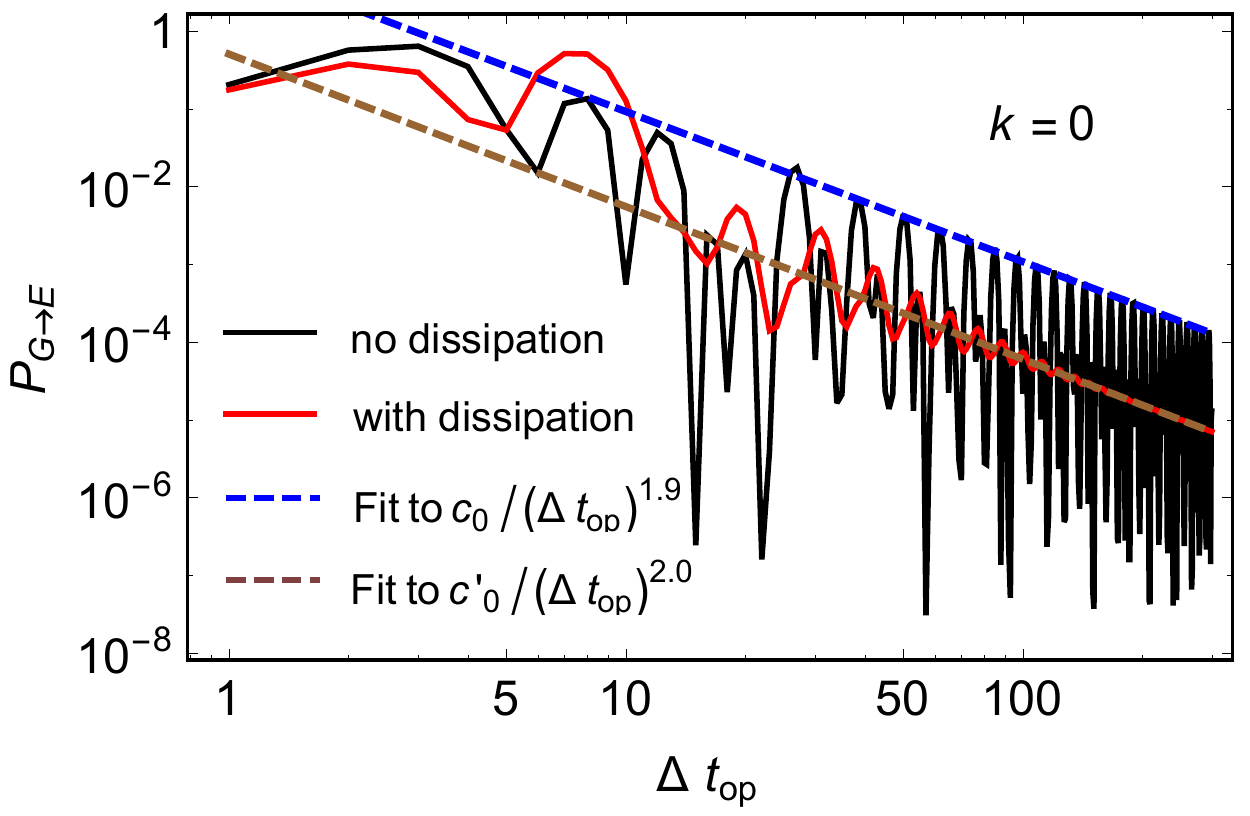}
\includegraphics[width=\columnwidth]{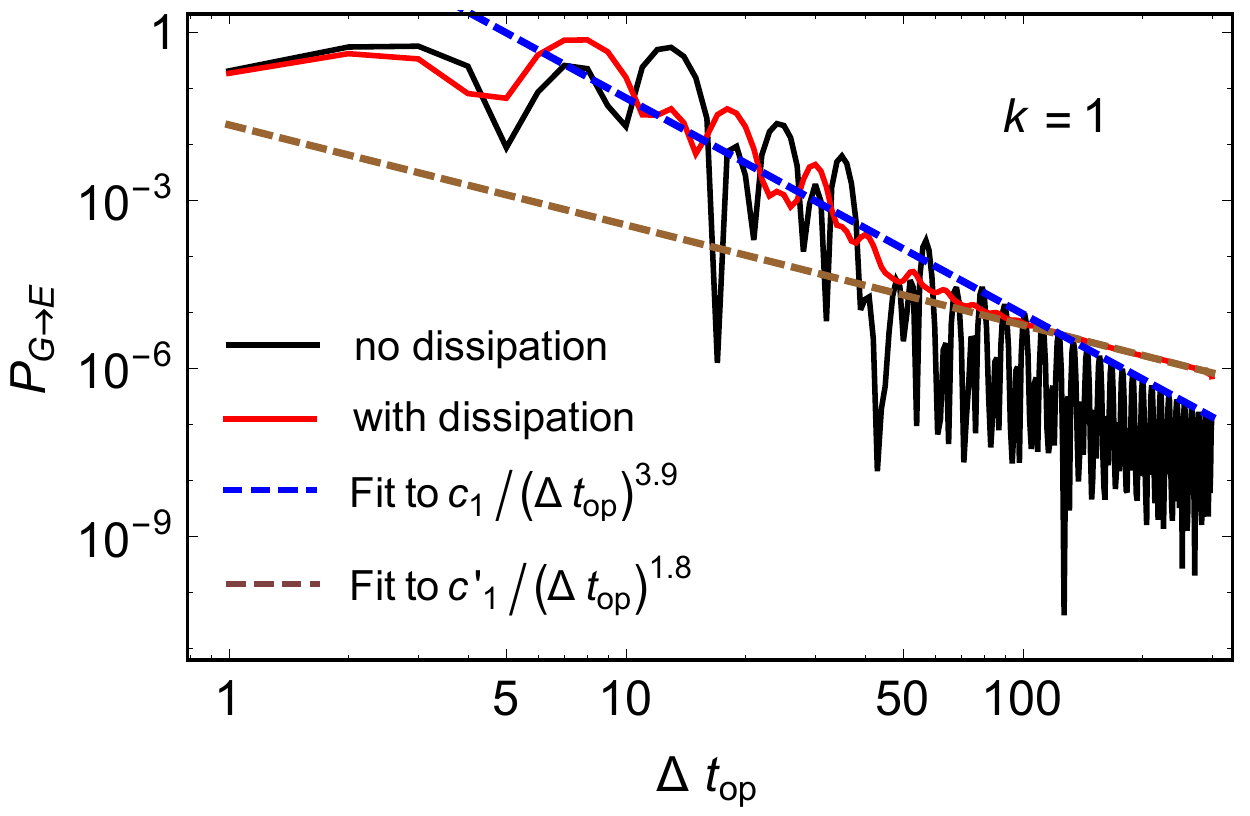}
\par\end{centering}
\caption{With dissipation (red solid line), transition probability $P_{\text{G} \rightarrow \text{E}}$ vs the gap multiplied by the
total evolution time $\Delta {\Top}$, due to diabatic effects for k=0 (top),
and k=1 (bottom). The long time tail is fitted to $c_0/(\Delta t_{\rm op})^x$ with $x\approx 2$ (brown dashed line).
We choose the cutoff $\omega_{c}=10 \Delta,$
Ohmic bath at low temperature $T=1/\beta=0.001 \Delta$, system-bath coupling $\lambda_{1}=\lambda_{2}=\lambda_{3}=0.01 \Delta$.
The black solid line shows the results without dissipation, and the envelope function for long time is
is fitted to  $c'_0/(\Delta t_{\rm op})^x$ with $x\approx 2k+2$ (blue dashed line).
}
\label{fig:transition-rate-dissipation}
\end{figure}

\begin{figure}
\begin{centering}
\includegraphics[width=\columnwidth]{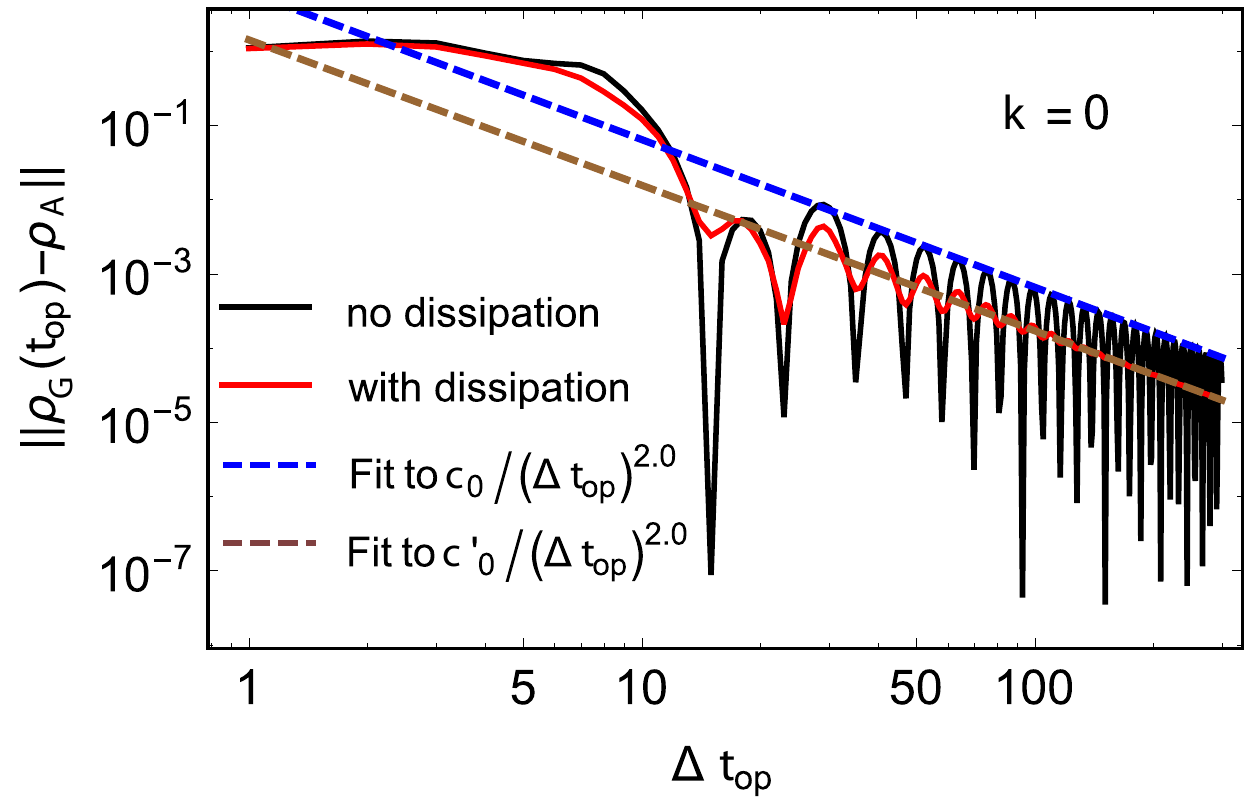}
\includegraphics[width=\columnwidth]{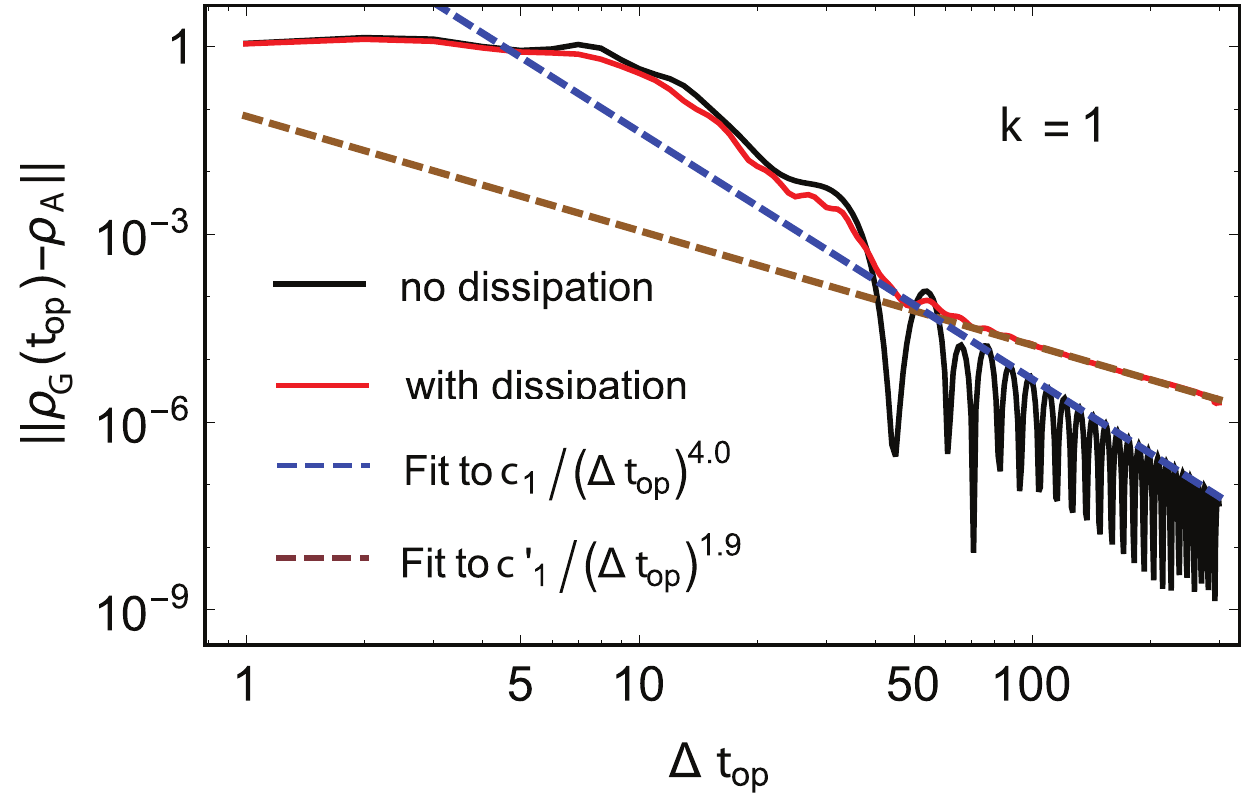}
\par \end{centering}
\caption{ The deviation of the density matrix after projection onto the ground state, $||\rho_{\text{G}}(t_{\text{op}})-\rho_A||$, vs the gap muliplied by the total evolution time, $\Delta t_{\text{op}}$.  The parameters are the same as in Fig.~\ref{fig:transition-rate-dissipation}.
}
\label{fig:Berry-correction-with dissipation}
\end{figure}

In order to quantitatively study the effects of the bath, we initialize the system in a certain superposition of $\ket{\sigma_y=+1; \tau_z=+1}$ and $\ket{\sigma_y=-1; \tau_z=-1}$
and numerically solve the master equation derived in Appendix~\ref{sec:master-equation}.
(The results for this initial state should be qualitatively representative of a general input.)
We first compute the probability of a transition out of the ground state manifold into an excited state
for the $\tau_z=1$ sector, $P_{\text{G} \rightarrow \text{E}}$; the $\tau_z=-1$ sector has similar behavior. In Fig.~\ref{fig:transition-rate-dissipation}, we plot $P_{\text{G} \rightarrow \text{E}}$ as a function of the total evolution time ${\Top}$, both with and without dissipation.
The upper panel shows it for a stepwise linear time dependence ($k=0$).
The lower panel shows it for a smoothed-out time dependence ($k=1$) in which the first derivatives exist and are continuous everywhere, i.e. they vanish at the beginning and end of each time step.
Details are given in Appendix~\ref{sec:dissipation-details}.
In the absence of dissipation, the envelope of the decay follows the expected scaling as $\Top^{-2}$ and $\Top^{-4}$ for $k = 0$ and $1$, respectively.
As may be seen in the plots, the dissipation suppresses oscillations in the transition probability.
For $k = 0$, the dissipative case has the same $\Top^{-2}$ falloff at long times.
For $k = 1$, however, dissipation has an important qualitative effect at long times: the excitation probability again goes as $\Top^{-2}$, rather the $\Top^{-4}$ behavior that occurs without dissipation.
This can be understood as follows: the suppressed excitation rate for the non-dissipative $k=1$ protocol relies on the smoothness of the time evolution of the system's Hamiltonian, i.e. the smoothness of the $\Delta_j(t)$.  With dissipation, this smoothness is washed out by the random discontinuities added by the bath. For shorter $\Top$, however, there remains a quantitative difference between the $k = 0$ and $1$ protocols, which suggests some level of engineering the time dependence of the system Hamiltonian remains beneficial.

If we measure the system and find that it remains in the two-fold degenerate ground state manifold, then a phase gate has been applied to this subspace, due to the sector-dependent holonomic phase of $\pm \frac{\pi}{4}$.
However, there may have been intermediate diabatic excitations which relaxed, causing the final state to deviate from the adiabatic result. This deviation is quantified by $||\rho_{\text{G}}(t_{\text{op}})-\rho_A||$,
where $\rho_A$ is the final density matrix obtained in the adiabatic limit,
\begin{equation}
 \rho_{\text{G}}(t_{\text{op}}) = \frac{\Pi_G \rho(t_{\text{op}}) \Pi_G}{\rm{Tr} \left( \Pi_G \rho(t_{\text{op}}) \right)}
\end{equation}
is the density matrix for finite $t_{\text{op}}$ projected into the ground state manifold,
where $\Pi_G$ is the projection operator into the ground state and $\rho(t_{\text{op}})$ is the density matrix before the projection measurement, and $||\ldots ||$ denotes the trace norm.
$||\rho_{\text{G}}(t_{\text{op}})-\rho_A||$ measures the deviation of the state from the ideal/adiabatic limit result.  As shown in Fig.~\ref{fig:Berry-correction-with dissipation}, ${||\rho_{\text{G}}(t_{\text{op}})-\rho_A||}$ exhibits behavior similar to that of $P_{\text{G} \rightarrow \text{E}}$. In particular, without dissipation, the long-time asymptotics exhibit $\Top^{-2k-2}$ scaling, while the inclusion of dissipation suppresses oscillations in $||\rho_{\text{G}}(t_{\text{op}})-\rho_A||$ and leads to a power-law decay $\Top^{-2}$ at long times. 

We believe the $\Top^{-2}$ is universal for diabatic transitions in the presence of disssipation. A heuristic explanation is to consider the rate equation for the occupation number of the excited level $N_E$ in the instantaneous basis. Phenomelogically, we postulate that the time evolution of $N_E$ is governed by the following rate equation:
\begin{equation}
	\frac{\di N_E}{\di t}=h(t) - \Gamma(t) N_E.
	\label{}
\end{equation}
Here $h(t)$ describes the generation of excitations due to the matrix element between the ground state $\ket{G}$ and excited state $\ket{E}$, and $\Gamma(t)$ characterizes the relaxation of the excitations. Importantly, in the model Eq. \eqref{eqn:spin-hamiltonian-with-diss}, the bath coupling is assumed to be synchronized with the time-dependent couplings of the Hamiltonian, whose time variation is responsible for diabatic transitions. Therefore, as a zeroth order approximation we can assume that $h(t)$ and $\Gamma(t)$ have the same time dependence. Furthermore, we have $h(t)\sim \mathcal{O}\left(|\langle E|\partial_t H|G\rangle|^2\right)$. We expect that if $\Top$ becomes longer, the speed at which the Hamiltonian changes on average should decrease as $\Top^{-1}$. To capture this dependence on $\Top$ we make a crude estimate of $h(t)$ to be $h(t)=\frac{\Delta}{\Top^2}f(t)$, where $f(t)$ is a dimensionless function whose range is $[0,1]$, and write $\Gamma(t)=\Gamma f(t)$. The rate equation can now be integrated with the initial condition $N_E(t=0)=0$, and the result is
\begin{equation}
	N_E(t)=\frac{\lambda}{\Gamma \Top^2}\left[ 1-e^{-F(t)} \right],
	\label{}
\end{equation}
where $F(t)=\int_0^t \di s f(s)$. It is not hard to see that $F(\Top)$ grows at least linearly with $\Top$, so asymptotically we find $N_E(\Top)=\mathcal{O}( \Top^{-2})$.

To summarize, diabatic corrections (to both the transition probability from the ground state to an excited state and to the phase acquired if the system remains
in the ground state) are non-universal and dependent on the detailed time dependence of the Hamiltonian in the absence of dissipation. In the presence of
dissipation, however, the scaling of diabatic corrections  appears to become universal in the limit of large operation time.

\section{Diabatic Corrections to Braiding Transformations of Anyons}
\label{sec:CS-theory}

In the previous section, we saw that diabatic corrections to the holonomy are only polynomially
suppressed in the time ${t_\text{op}}$ of the evolution and, for the system of Eq.~(\ref{eqn:spin-hamiltonian-with-diss}), can be as bad as $\mathcal{O}(t^{-2}_\text{op})$. This is especially worrisome if the holonomy in question determines the braiding transformations in a topological quantum computer. However, we argue in this section that diabatic corrections to the braiding transformations of anyons originate from the uncontrolled creation or motion of anyons.

We justify this claim by studying the diabatic time evolution for two theories with fixed anyon number, where one anyon braids around the other. We perform these calculations using Maxwell-Chern-Simons theory~\cite{Deser82}, which has a finite gap to gauge field excitations. In the first theory, the anyons are forced to move along a specific trajectory.
In this case, we find that the corrections to the braiding transformations are independent of the braiding time and are exponentially suppressed by the separation of anyons.
In the second theory, the anyons are transported via a pinning potential. In this case, the anyons have some amplitude to tunnel out of the potential trap and possibly wind around the other anyon a number of times that does not match that of the trap motion. The sum over such topologically distinct trajectories, i.e. with different winding numbers, destroys the quantization of the braiding transformation.

Consider an Abelian Maxwell-Chern-Simons theory for two anyons carrying charges $a$ and $b$, respectively.  Anyon $b$ sits at the origin for all time and anyon $a$ sits a distance $R$ away until time $t=0$, at which it circles the origin and then returns to its initial position.  We use $x=(t,\vr)$ to denote the space-time coordinates collectively.  The action is
\begin{equation}
{\mathcal S} = \int d^3 x\, \Big(\frac{k}{4\pi}\epsilon_{\mu \nu \lambda} a^\mu \partial^\nu a^\lambda-\frac{1}{4g^2} f_{\mu \nu}f^{\mu \nu} -j^\mu a_\mu \Big).
\label{eqn:CS-action}
\end{equation}
Between $t=0$ and $t=\Top$ the moving anyon has current (in the polar coordinate $(r,\theta)$):
\begin{eqnarray}
j^0_a(x)&=& \frac{a}{r}\delta(r-R) \delta\Big(\theta-\frac{2\pi t}{t_\text{op}}\Big)
\\ j^\theta_a(x) &=& a\frac{2\pi R}{r{t_\text{op}}} \delta(r-R) \delta\Big(\theta-\frac{2\pi t}{t_\text{op}}\Big)
\end{eqnarray}
and the stationary anyon has current
\begin{equation}
j^0_b(x)=b\delta^{(2)}(\vr).
\end{equation}
All other currents vanish.  For a pure Abelian Chern-Simons theory we would expect the braiding transformation to be the phase factor $e^{i  2\pi a b  /k}$.  Adding the Maxwell term gives the interaction a non-topological component, which is exponentially-decaying.  Hence, the braiding transformation is expected to have corrections that are exponentially-small in $R$.

Integrating out $a_\mu$ gives the effective action
\begin{equation}\label{Seff}
\begin{split}
	{\mathcal S}_\text{eff} &= \int d^3x d^3x'\, \Big[j^\mu(x)G_{\mu\nu}^{(1)}(x,x') j^\nu(x')
\\ &~~~ -\frac{g^2}{2}j_\alpha(x)G^{(2)}(x,x')j^\alpha(x')\Big].
\end{split}
\end{equation}
Here, the two propagators are given by
\begin{eqnarray}
 G^{(1)}_{\mu\lambda}(x, x') &=& \frac{\pi}{k}\langle x\big|\frac{\epsilon_{\mu\nu\lambda}\partial^\nu}{\partial^2(1+\frac{\partial^2}{g^4k^2/4\pi^2})}\big|x'\rangle \\
 G^{(2)}(x,x') &=& \langle x\big|\frac{1}{\partial^2+g^4k^2/4\pi^2}\big|x'\rangle.	
\end{eqnarray}
Both terms in Eq.~(\ref{Seff}) can be evaluated by transforming to momentum space.  One can show, as we do in Appendix~\ref{CS}, that the first term Eq.~(\ref{Seff}) contributes a braiding transformation $e^{i \Phi}$, with the phase
\begin{equation}
\Phi = \frac{2\pi a b}{k}\left( 1-\sqrt{\frac{\pi g^2 k R}{4\pi}}e^{-g^2 k R/2\pi}\right)+\mathcal{O}\left( e^{-g^2 k R/2\pi}\right),
\end{equation}
which has finite-$R$ corrections, but is independent of the braiding speed.
Evaluation of the second term in the action shows that it grows linearly in ${t_\text{op}}$ and is the {\it same for all braiding processes}, i.e. it is independent of the charge of the second anyon at the origin, as is expected for a dynamical phase. If there are diabatic corrections to braiding, they must arise from effects not allowed in this simple theory.

We now modify our theory such that anyon $a$ is dynamical. Its position is no longer a classical
parameter but is, instead, controlled by a pinning potential. We move the pinning potential in order
to transport anyon $a$ around static anyon $b$.
We again set $b$ to have fixed position. The effective action reads
\begin{equation}
	\begin{split}
		{\mathcal S}= \int dt\bigg[\int d^2\vr   \,\Big( & \frac{k}{4\pi}\epsilon_{\mu \nu \lambda}a^\mu \partial^\nu a^\lambda-j^\mu a_\mu\Big) \\
&+\frac{1}{2}m\Big( \frac{d{\bf q}}{dt}\Big)^2 -V_a({\bf q}-\mathbf{R}(t))\bigg].
	\end{split}
\end{equation}
Here ${\bf q}$ is the coordinate of the particle, which is now a dynamical variable. $\mathbf{R}(t)$ parameterizes the trajectory of the pinning potential $V_q$.

To proceed, we first integrate out $a_\mu$. As before, this will generate a Hopf term for the worldlines and, in the present configuration, this term is just the winding number of $\mathbf{q}(t)$ around the origin.

We can simplify this problem further by ignoring the radial motion of particle $a$, which is an inessential complication,
so we only need to keep the polar angle $\theta$.  The above action now can be reduced to the problem of a particle on a ring with a flux tube in the center. However, we still have the external ``driving'' force that moves the anyon, which is given by the time dependent pinning potential $V_a(\mathbf{q}(t)-\mathbf{R}(t))$:
\begin{equation}
 {\mathcal S} =\int_0^{t_\text{op}} dt \,\Big[\frac{1}{2} I \dot{\theta}^2 +\frac{a b}{k}\dot{\theta}-V_a\Big(\theta-\frac{2\pi t}{t_\text{op}}\Big)\Big].
\end{equation}
Here $I$ is the effective rotational inertia. In the following, we assume that the pinning potential is moving with a constant angular velocity and that the pinning potential completes one circuit and returns to $\theta=0$ after time ${t_\text{op}}$. The path integral representation of the transition amplitude is
\begin{equation}
\langle \theta\equiv 0|U(t_{\text{op}},0)|\theta\equiv 0\rangle= \sum_{n=-\infty}^\infty \int_{\theta(0)=0}^{\theta({t_\text{op}})=2\pi n} \mathcal{D}\theta(t) e^{i \mathcal{S}}.
\end{equation}

Notice that we need to sum over different winding number sectors for $\theta(t)$.
Let us make the change of variable $\theta=\tilde{\theta}+\frac{2\pi t}{t_\text{op}}$, so that $\tilde{\theta}(0)=0$ and $\tilde{\theta}({t_\text{op}})={2\pi(n-1)}$. This yields
\begin{equation}
\begin{split}
{\mathcal S} = &\Big(\frac{2\pi I}{t_\text{op}}+\frac{ab}{k}\Big)\Big[\tilde{\theta}({t_\text{op}})-\tilde{\theta}(0)\Big]+\frac{2\pi^2 I}{t_\text{op}}+\frac{2\pi ab}{k}
\\ &+\int_0^{t_\text{op}} dt \,\Big[\frac{1}{2}I\dot{\tilde{\theta}}^2-V_a(\tilde{\theta})\Big].
\end{split}
\end{equation}
Let us denote
\begin{equation}
	{\mathcal S}_m=\int_{\theta(0)=0}^{\theta({t_\text{op}})=2\pi m}\mathcal{D}\theta(t)\exp\left\{ i\int_0^{t_\text{op}}\di t \left[\frac{1}{2}I\dot{\theta}^2 - V_a(\theta)\right]\right\}.
	\label{}
\end{equation}
The transition amplitude is then
\begin{equation}
\langle \theta \equiv 0|U(t_{\text{op}},0)|\theta\equiv 0 \rangle
= e^{i\frac{2\pi ab}{k}}\sum_{n=-\infty}^\infty e^{i\frac{4\pi^2 I}{t_\text{op}}(n+\frac{1}{2})+i\frac{2\pi ab n}{k}} \mathcal{S}_{n}.
\end{equation}

In order to find the braiding transformation, we need to compare the above transition amplitude with the case there is no anyon $b$
sitting at the origin. If we let $H_0 (t)$ denote the Hamiltonian in this case, we find
\begin{equation}
\langle \theta \equiv 0 | U_{0}( t_\text{op} , 0)|\theta\equiv 0 \rangle=\sum_{n=-\infty}^\infty e^{i\frac{4\pi^2 I}{t_\text{op}}(n+\frac{1}{2})} \mathcal{S}_n.
\end{equation}

The braiding transformation is, thus, given by the ratio of these two amplitudes, resulting in the phase factor:
\begin{equation}
e^{i\Phi}=	\frac{e^{i\frac{2\pi ab}{k}}\sum\limits_{n=-\infty}^{\infty}e^{i\frac{4\pi^2I}{t_\text{op}}n} e^{i\frac{2\pi ab n}{k}} \mathcal{S}_n}{\sum\limits_{n=-\infty}^{\infty} e^{i\frac{4\pi^2I}{t_\text{op}}n}\mathcal{S}_n}.
\label{relphase}
\end{equation}
The quantization of $\Phi$ is destroyed in general because the moving anyon now has some amplitude $\mathcal{S}_{n\neq 0}$ of escaping the pinning potential and tunneling around the static anyon an additional
$n$ times. In the adiabatic limit, the system remains in the instantaneous ground state at all times,
so the moving anyon remains trapped in the pinning potential. In this limit, $\mathcal{S}_n = 0$ for all $n\neq 0$, and the braiding phase is quantized to $\Phi=2\pi ab/k$.

We note that Eq.~\eqref{relphase} ignores coupling to an environment. Realistically, the environment will detect the sectors associated with distinct winding numbers $n$, since these are macroscopically different trajectories.
This ``which-path'' information will remove the interference between $n$-sectors in Eq.~\eqref{relphase}, resulting in a decohered state.
Presumably the bath can help to the extent it relaxes the escaped anyon back into the moving trap before it is left behind.

Clearly a theory that does not fix anyon number will also have diabatic corrections to the braiding transformation.  A pair of anyons with nontrivial topological charge could be created.  If one of the anyons circles $a$ or $b$ before annhilating with its antiparticle, the braiding transformation will be affected.  If we braid two anyons with a fixed fusion channel in a non-Abelian Chern-Simons theory, we can reduce the calculation to a calculation in Abelian Chern-Simons theory, since the result must be a phase. So long as we do not allow any type of quasiparticle creation (real or virtual), the fusion channel will remain fixed, so the preceding calculation is, in fact, completely general and pinpoints the source of diabatic errors in the general case.

We have seen that both sources of diabatic corrections to the braiding tranformation arise from transitions out of the ground state subspace that result in the uncontrolled motion
of anyons -- either the anyon $a$ winds around $b$ too many or too few times, or else an anyon pair is created and one of the new anyons winds around $a$ and/or $b$. We are now in a position to understand the power law behavior of corrections to the braiding transformation shown in Fig.~\ref{fig:Berry-correction-with dissipation}. Corrections to the braiding transformation must be the result of two transitions: a transition out of the ground state, causing the error, and a transition back into the ground state allowing us to define an operation within the ground state subspace.  As shown in Refs.~\onlinecite{Lidar09, Wiebe12, Lidar15}, for $\mathcal{C}^k$ smooth time evolution, the transition amplitude is $\mathcal{O}\Big( t_{\text{op}}^{-k-1}\Big)$, therefore corrections to the braiding transformations are $\mathcal{O}\Big( t_{\text{op}}^{-2k-2}\Big).$


\section{A Correction Scheme for Diabatic Errors to the Braiding of MZMs in $T$-junctions}
\label{sec:MZM-trijunction}

 In the previous section, we found that errors in the braiding transformation caused by diabatic effects originate from the uncontrolled creation or motion of anyons.  We now use this result to devise a correction scheme for such diabatic errors.  In this section, we focus on the particular example of braiding MZMs in a $T$-junction and provide concrete proposals in this context. In Section~\ref{sec:general-braiding}, we will generalize our diabatic error correction scheme to systems supporting arbitrary types of non-Abelian anyons or defects.

\subsection{Relation Between Two-Level Systems and Braiding MZMs at $T$-junctions}
\label{sec:MZM-LZ}

Section~\ref{sec:general-theory} focused on the adiabatic evolution of two-level systems.
Since our main interest in this paper is the braiding of quasiparticles in a topological phase, in particular the braiding of MZMs,  we pause now to map the braiding and two-level problems onto each other. With such a mapping in hand, we can translate the results discussed in Section~\ref{sec:general-theory}
to the context of quasiparticle braiding in a topological phase. More specifically, we consider braiding of MZMs in a network of topological superconducting wires. The essential building block of the network is a so-called $T$-junction.

\begin{figure}
	\includegraphics[width=1.8\columnwidth]{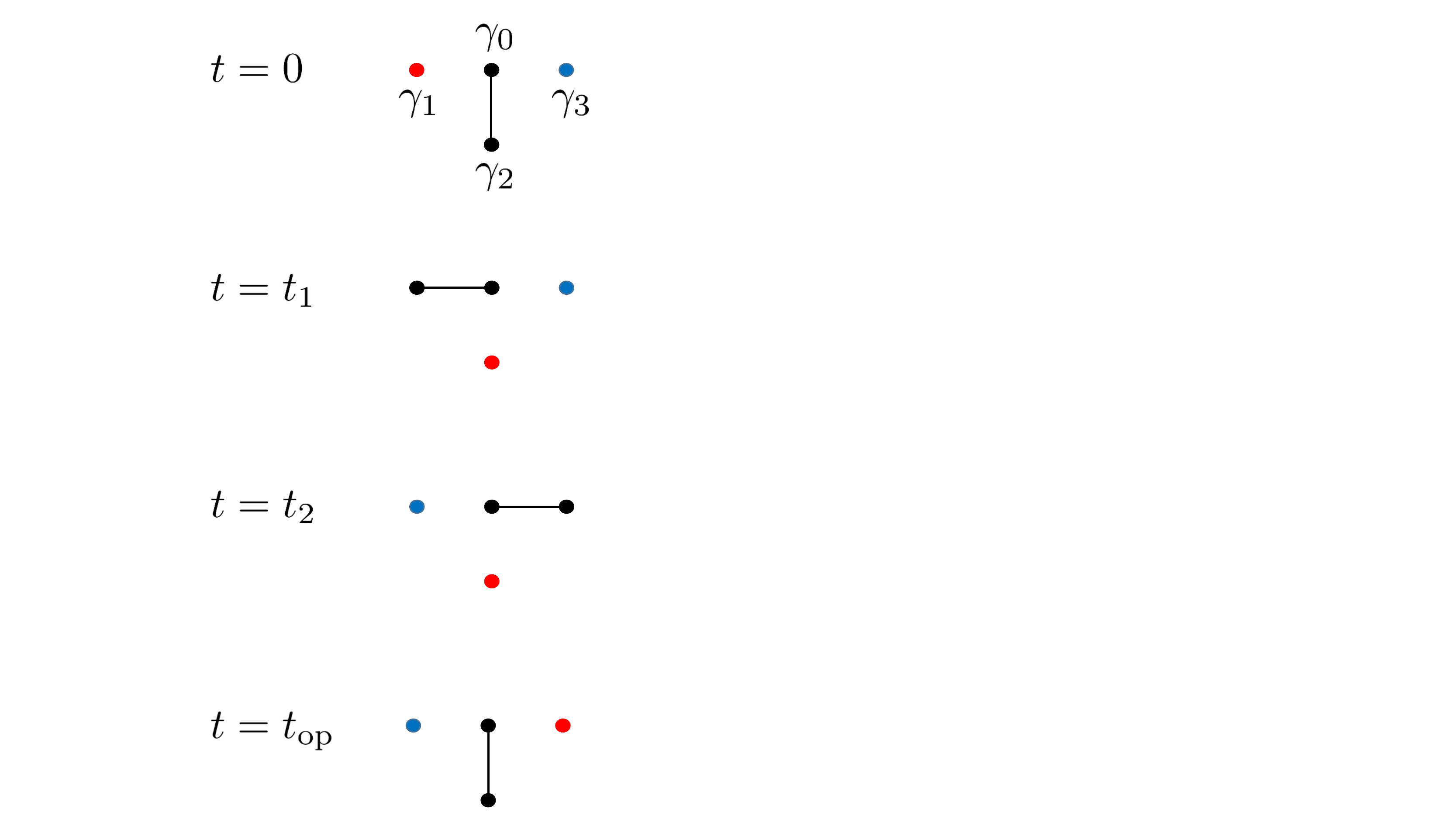}
\caption{Schematic of braiding process at a $T$-junction. Each dot represents a MZM and the lines connecting dots indicates which MZMs
are in a definite fusion channel at a given time. This sequence of states can be obtained as the ground states of a Hamiltonian with nonzero coupling between the pair connected by a line at each step and by adiabatically tuning the Hamiltonian from one step to the next. This sequence effectively braids the MZMs labeled by red and blue dots.}
\label{fig:T-junction-braiding}
\end{figure}

A $T$-junction is composed of four MZMs. At the initial and final configurations, two of these MZMs are completely decoupled (up to exponentially suppressed corrections) and will, in part, comprise the topological qubit, while the other two MZMs are an ancillary pair that are coupled to each other. (Eventually, it will be convenient to have three MZMs replacing the one in the middle, following Ref.~\onlinecite{Hyart13}, but we will focus on the simpler situation here.) The braiding operation is partitioned into three steps, that end at time $t_1$, $t_2$, and $t_3$, respectively. (For simplicity, we will typically let $t_1={t_\text{op}}/3$, $t_2=2{t_\text{op}}/3$, $t_3={t_\text{op}}$.) Each step changes which MZMs are decoupled (and correspond to the topological qubit pair) and which MZMs are coupled (and correspond to the ancillary pair). We call the configurations at the end of each step a ``turning point.'' This sequence is depicted in Fig.~\ref{fig:T-junction-braiding}.

The Hamiltonian for these MZMs takes the form
\begin{equation}
H = -\sum_{j=1}^{3} {\Delta_j}(t)\, i \gamma_j \gamma_0
\label{eqn:MZM-braiding-Hamiltonian}
\end{equation}
where $\{{\gamma_i},{\gamma_j}\} = 2\delta_{ij}$ for $i,j = 0,1,2,3$ and $\Delta_j(t)$ ranges between $0$ and $\Delta>0$.
In each panel of Fig.~\ref{fig:T-junction-braiding}, the dots represent MZMs and the line connecting two dots represents a Hamiltonian of the form of Eq.~ (\ref{eqn:MZM-braiding-Hamiltonian}) with the corresponding ${\Delta_i}=\Delta$ and all other ${\Delta_i}=0$. By changing which MZM is coupled to the central MZM in an adiabatic manner, the topological state information is teleported between MZMs, so as to always be encoded in the uncoupled MZMs. Following the indicated sequence of such teleportations results in a braiding transformation of the topological qubit pair of MZMs.

We assign the overall fermion parity of these four MZMs to be even when $\gamma_0 \gamma_1 \gamma_2 \gamma_3 =-1$ and odd when $\gamma_0 \gamma_1 \gamma_2 \gamma_3 =+1$.
If we fix the overall fermion parity of these four MZMs, they share a two-dimensional topological state space, which we map to a spin-$1/2$ system according to the representation of the Pauli operators $\sigma_j = i\gamma_0 \gamma_j$ for overall parity even, and $\sigma_j = -i\gamma_0 \gamma_j$ for overall parity odd.

This representation reveals the equivalence between the MZM Hamiltonian of Eq.~(\ref{eqn:MZM-braiding-Hamiltonian}) and the spin Hamiltonian of Eq.~(\ref{eqn:spin-hamiltonian-with-diss}) without the bath coupling. In particular, the even and odd overall parity sectors of the four-MZMs Hamiltonian are mapped to the $\tau_z = +1$ and $-1$ sectors of the two-spin Hamiltonian, respectively. The difference between the holonomies in the sectors of the two-spin model is mapped to the difference between the holonomies in the even and odd fermion parity sectors of the MZMs, giving the relative phase of the braiding transformation.

Let us focus in more detail on the first step of this process, which transfers the state information initially encoded in $\gamma_1$ to $\gamma_2$,
and occurs between $t=0$ and $t=t_1$, as shown in Fig.~\ref{fig:T-junction-braiding}.
Consider varying the couplings linearly during this time segment:
\begin{eqnarray}
\Delta_1 (t) &=& \Delta\frac{t}{t_1}  \\
\Delta_2 (t) &=& \Delta\left(1-\frac{t}{t_1}\right) \\
\Delta_3 (t) &=&0
,
\end{eqnarray}
so that the $\tau_z = +1$ sector of the spin Hamiltonian (corresponding even fermion parity) takes the following form for $0 \leq t \leq t_1$:
\begin{equation}
H = \Delta
\left[\frac{t}{t_1} \sigma_x +
\Bigl(1-\frac{t}{t_1}\Bigr)\sigma_y
\right].
\end{equation}
If we define the following unitary transformation
\begin{equation}
M=\frac{1}{2\sqrt{2+\sqrt{2}}}\Big[i(\sigma_z+\sigma_y)(1+\sqrt{2})-(i\sigma_x+\openone)\Big],
\end{equation}
then
\begin{equation}
M H M^\dagger =
 \frac{\Delta}{\sqrt{2}}
\left[ h(t) \,\sigma_z  - \sigma_x
\right]
,
\end{equation}
where $h(t) =1-\frac{2 t}{t_{1}}$. Thus, we obtain $MHM^\dagger$ to be in the same form as the Landau-Zener Hamiltonian in Eq.~(\ref{eqn:piece-wise-model}).
As we show in Appendix \ref{sec:MZM-LZ-details}, the other steps in the braiding protocol can also
be mapped to Landau-Zener problems that can be pieced together.

The relation between a MZM $T$-junction and a two-level system implies that the diabatic errors that
we encountered in the latter case will also arise in the former. Consequently, if braiding is not done infinitely-slowly,
the resulting unitary transformation will generically differ from the expected adiabatic result by $\mathcal{O}(1/{t_\text{op}})$ errors. This can be improved to $\mathcal{O}(1/{t^{k+1}_\text{op}})$ if the time-dependence of the control parameters of the Hamiltonian is $\mathcal{C}^{k}$, which requires fine-tuning the time-dependence by setting $k$ derivatives of the Hamiltonian to zero at the initial and final times.
On the other hand, Section~\ref{sec:CS-theory} leads us to anticipate that
errors in the braiding transformation must be due to the creation or uncontrolled movement of topological quasiparticles.
In the next section, we show that this is, indeed, the case: if a sequence of measurements shows that no quasiparticles
have been created at intermediate steps of the evolution, then the braiding phase is fixed to its topologically-protected value. Moreover, this fact allows us to specify a protocol for detecting and correcting diabatic error that would affect the braiding transformation.

\subsection{Error Correction through Measurement}
\label{sec:Measurement=Correction}

In this section, we show that projecting the system into the ground state at the turning points during the $T$-junction
braiding process is sufficient to fix all diabatic errors within the MZM system.
This suggests an error correction scheme for braiding MZMs, based on a repeat-until-success
protocol, that produces topologically-protected gates.   For now, we focus on errors occurring within the low energy subspace of the four MZMs, because we expect these errors to be the most prevalent. We address diabatic transitions out of this subspace in Section~\ref{sec:DiabaticErrorTopTransmon}.

We consider the time evolution depicted in Fig.~\ref{fig:T-junction-braiding}.
At any point in the system's time evolution, the energy levels in the even parity sector
$\gamma_0 \gamma_1 \gamma_2 \gamma_3 = -1$
are the same as those of the odd parity sector $\gamma_0 \gamma_1 \gamma_2 \gamma_3 = +1$.
This follows from the fact that there is always a pair of MZMs that is decoupled from the Hamiltonian
(the one that is unaffected during that step of the protocol and a linear combination of the other three),
and switching the parity of this pair does not affect the energy. This correspondence between the spectra in the two sectors implies the dynamical phase is identical for both sectors and, thus, does not affect the braiding transformation.

At each turning point of the braiding process, there are two decoupled MZMs which sit at the endpoints of the $T$-junction: at $t=0$, $\gamma_1$ and $\gamma_3$ are decoupled; at $t_1$, $\gamma_2$ and $\gamma_3$ are decoupled; at $t_2$, $\gamma_1$ and  $\gamma_2$ are decoupled; and, at $t_3$, $\gamma_1$ and  $\gamma_3$ are decoupled.  We can consider the unitary time evolution of each step between turning points, which we denote as $U_{ij}$, to indicate the Hamiltonian starts with $\gamma_{j}$ coupled to $\gamma_0$ and $\gamma_{i}$ decoupled, and ends with $\gamma_{i}$ coupled to $\gamma_0$ and $\gamma_{j}$ decoupled. In this notation, ${U}_{12}$ is the evolution from time $t=0$ to $t_1$, ${U}_{31}$ is from $t_1$ to $t_2$, and ${U}_{23}$ is from $t_2$ to $t_3$.
We emphasize that $\gamma_k$ for $k\neq 0,i,j$ remains decoupled throughout the step corresponding to $U_{ij}$, as this fact is crucial for the topological protection of the braiding, and we will utilize it to analyze the diabatic error.  (By decoupled, we mean up to the residual, exponentially suppressed couplings due to nonzero correlation length. Such exponentially suppressed corrections can easily be made arbitrarily small and so are left implicit throughout this paper.)

Let us first choose a basis for the Hilbert space of the four MZMs. For calculational purposes, it will be useful to employ the basis $\ket{-\gamma_0 \gamma_1 \gamma_2 \gamma_3 =\pm 1, i \gamma_{2} \gamma_{0} = \pm 1}$, specified by the overall fermion parity of the four MZMs and the parity of the initial/final ancillary pair of MZM.
In this basis, the four MZMs have the following matrix representations
\begin{eqnarray}
\gamma_0 &=& -\sigma_y \otimes \sigma_y \\
\gamma_1 &=& \sigma_x \otimes \mathbb{1} \\
\gamma_2 &=& \sigma_y \otimes \sigma_x \\
\gamma_3 &=& \sigma_y \otimes \sigma_z.
\end{eqnarray}
Since the total fermion parity must be conserved (as these four MZMs only interact with each other for the specified Hamiltonian), the unitary evolution operators ${U}_{ij}$ are block diagonalized into $2\times2$ blocks $U_{ij}^{\mathrm{e}}$ and $U_{ij}^{\mathrm{o}}$ corresponding to even and odd fermion parity sectors, respectively:
\begin{equation}
U_{ij}=
\left[
\begin{matrix}
		U_{ij}^{\mathrm{e}} & 0\\
		0 & U_{ij}^{\mathrm{o}}
\end{matrix}
\right]
.
\end{equation}
The property $[U_{ij}, \gamma_k]=0$ for $k\neq 0, i, j$ yields the relations between even and odd overall parity sectors
\begin{eqnarray}
\label{eqn:conditionsonUa}
U_{12}^{\mathrm{o}} &=& \sigma_z U_{12}^{\mathrm{e}}\sigma_z \\
\label{eqn:conditionsonUb}
U_{31}^{\mathrm{o}} &=& \sigma_x U_{31}^{\mathrm{e}}\sigma_x \\
\label{eqn:conditionsonUc}
U_{23}^{\mathrm{o}} &=& U_{23}^{\mathrm{e}}.
\end{eqnarray}

We now consider what happens if we apply a projective measurement of the fermion parity eigenstates of the ancillary pair of MZMs at each turning point (which are also their energy eigenstates at those points). Later, we will discuss how to do this in a physical setup; for now, we will simply analyze what happens when such projections are applied at the turning points of the braiding process. We define the projection operators
\begin{equation}
\Pi_s^{(ij)}=\frac{1+is\gamma_i\gamma_j }{2},
\end{equation}
which projects to the state with definite fermion parity $i \gamma_i \gamma_j = s = \pm 1$ for the pair of MZMs $\gamma_i$ and $\gamma_j$. For the above representation of MZM operators, the projectors of interest are given by
\begin{eqnarray}
\Pi_{s_0}^{(20)} &=& \frac{1+s_0\mathbb{1}\otimes\sigma_z}{2} \\
\Pi_{s_1}^{(10)} &=& \frac{1+s_1\sigma_z\otimes\sigma_y}{2} \\
\Pi_{s_2}^{(30)} &=& \frac{1-s_2\mathbb{1}\otimes\sigma_x}{2}
.
\end{eqnarray}

The total evolution operator for the braiding process with fermion parity measurements of the ancillary pairs at the turning points given by
\begin{equation}
W_\text{Total} = \Pi_{s_3}^{(20)} {U}_{23} \Pi_{s_2}^{(30)}{U}_{31}\Pi_{s_1}^{(10)} {U}_{12}\Pi_{s_0}^{(20)},
	\label{}
\end{equation}
where $s_j$ is the measurement outcome at the $j$th turning point. Clearly this operator is not unitary, in general, since it involves projective measurements.
In order to for this operator to represent a braiding operation, the initial and final configurations of the ancillary pair must match, that is, we must have $s_3 = s_0$.

Substituting Eqs.~\eqref{eqn:conditionsonUa}-(\ref{eqn:conditionsonUc}) and assuming $s_3 = s_0$, we find
\begin{equation}
W_\text{Total}=
\left[
\begin{matrix}
		1 & 0\\
		0 & is_0s_1s_2
\end{matrix}
\right]
\otimes W',
\label{eqn:Etot}
\end{equation}
where $W'$ is given by
\begin{equation}
W'= \frac{1+s_3\sigma_z}{2}{U}^{\text{e}}_{23} \frac{1-s_2\sigma_x}{2}{U}^{\text{e}}_{31}\frac{1+s_1\sigma_y}{2} {U}^{\text{e}}_{12}\frac{1+s_0\sigma_z}{2}.
	\label{}
\end{equation}
We notice that $W'$ takes the form $w' \Pi_{s_0}^{(20)}$ for a scalar $w'$ that depends on the precise details of the unitary evolution operators and measurement outcomes. (This scalar encodes the probability of the measurement outcomes, but is otherwise unimportant, since the quantum state is normalized after each measurement.)

In order to obtain the effect of this operation on the topological qubit, it is useful to convert to the more relevant basis given by $\ket{i \gamma_1 \gamma_3 =\pm 1, i \gamma_{2} \gamma_{0} = \pm 1}$ (which is obtained by a simple permutation of basis states). In this basis, the total evolution operator is
\begin{equation}
W_\text{Total} = \left[ R_{13} \right]^{s_1 s_2} \otimes w \, \Pi_{s_0}^{(20)} ,
	\label{}
\end{equation}
where
\begin{equation}
R_{13}=
\left[
\begin{matrix}
		1 & 0\\
		0 & i
\end{matrix}
\right]
\label{eqn:Etot}
\end{equation}
is the (projective) braiding transformation for exchanging the MZMs $\gamma_1$ and $\gamma_3$ in a counterclockwise fashion. Once again, $w = -i s_1 s_2 w'$ is an unimportant overall scalar. The parity of the exponent $s_1 s_2 = \pm 1$, i.e. the measurements outcomes at the $t_1$ and $t_2$ turning points, determines whether $W_\text{Total}$ acts as a counterclockwise or clockwise braiding transformation.

The preceding argument shows that the braiding process with fermion parity measurements of the ancillary pairs at the turning points acts on the topological qubit pair of MZMs in the same way as the topologically protected braiding transformation $R_{13}$, so long as a neutral fermion is not created (paying its concomitant energy penalty) throughout the process. When precisely one of the intermediate measurements finds the ancillary pair to have odd parity, this means that a fermion is transferred from the qubit pair to the ancillary pair during the preceding time step and then transferred back during the following time step.

\begin{figure}[t]
	\includegraphics[width=0.5\columnwidth]{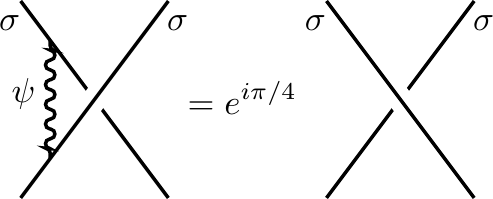}
\caption{The effect of a diabatic error, which transfers a neutral fermion from the qubit to the ancillas, on the braiding. For Ising anyons, it turns a counterclockwise braiding into a clockwise one, up to an overall phase.}
\label{fig:DL}
\end{figure}

\begin{figure}[t!]
	\centering
	\includegraphics[width=0.7\columnwidth]{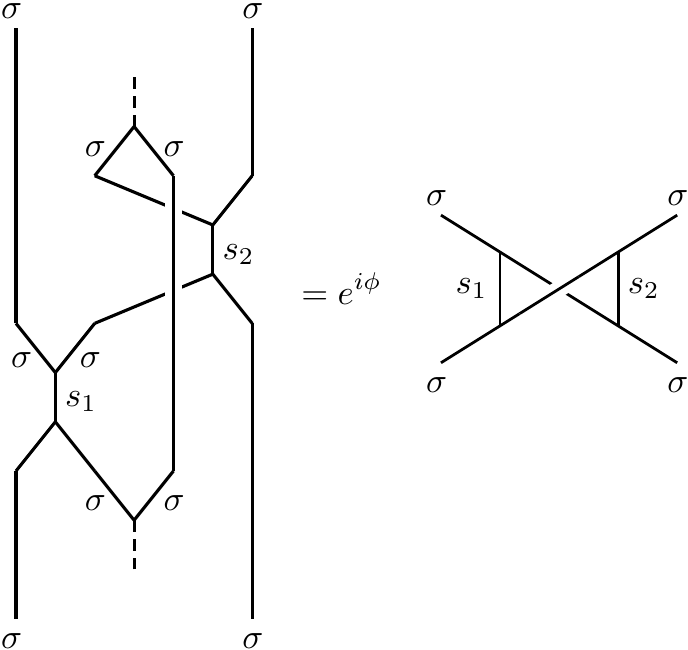}
	\caption{The measurement-only protocol for braiding. The resulting operation depends on the fusion channels $s_1$ and $s_2$ of the intermediate measurements (which can take the values $I$ or $\psi$). If $s_1=s_2$, the result is a counterclockwise braid, otherwise it is the inverse braid.}
	\label{fig:measure}
\end{figure}

This can be understood diagramatically from the arguments of Refs.~\onlinecite{Bonderson08b,Bonderson08c,BondersonPRB2013}, as summarized in Figs.~\ref{fig:DL} and \ref{fig:measure}. (These are shown with labels from the Ising anyon theory, but the same essential arguments hold for MZMs.) It follows from the properties of the Ising anyon model that a braiding exchange of two Ising $\sigma$ non-Abelian anyons with a neutral fermion transferred between them is equivalent to their inverse braid with no fermion transfer, up to an overall phase, as shown in Fig.~\ref{fig:DL}. (The same property is true for MZMs.)

At a $T$-junction governed by the Hamiltonian of Eq.~(\ref{eqn:MZM-braiding-Hamiltonian}), the emitted fermion can
only be transferred to one place, the ancillary pair of MZMs, since the Hamiltonian does not couple any other degrees of freedom.
A pair of such transfers of fermions, which corresponds to the measurements finding the ancillas in their excited state at both $t_1$ and $t_2$, essentially cancel each other. In this case, we have $s_1 s_2=1$ and the braiding transformation is still $R_{13}$.

In summary, we can understand the correction of diabatic errors via measurement from the general viewpoint of measurement-only protocol for braiding~\cite{Bonderson08b,Bonderson08c,BondersonPRB2013}, as depicted in Fig.~\ref{fig:measure}. One can clearly see that the resulting operation effected by the protocol depends on the outcomes of the two intermediate fusion channel measurements, in agreement with the result we found Eq.~\eqref{eqn:Etot}. This analysis reveals that diabatic transition errors that occur between the turning points can be corrected and topological protection of the resulting operation will be recovered if we introduce measurements at the turning points of the braiding process. If the measurements do not produce the desired outcomes, the resulting operation, though topologically protected, may not be the intended braiding transformation. Hence, we would like to impose a protocol that guarantees that we obtain the desired outcomes and braiding operation.
For this, we now devise a generalization of the ``forced measurement'' scheme introduced in Ref.~\onlinecite{Bonderson08b}.

First, let us recall the original forced measurement protocol in the context of this braiding process. Suppose that the first measurement of the fermion parity $i \gamma_1 \gamma_0$ returns the undesired outcome $s_1=-1$ with probability $p_0$. We can recover from such an undesired outcome by measuring the fermion parity of $i \gamma_2 \gamma_0$, which now has equal probability of measurement outcomes $i \gamma_2 \gamma_0 = s'_0 = \pm 1$ (projecting with $\Pi_{s'_0}^{(20)}$), and then repeating the measurement of $i \gamma_1 \gamma_0$, which now also has equal probability of measurement outcomes $i \gamma_1 \gamma_0 = s'_1 = \pm 1$ (projecting with $\Pi_{s'_1}^{(10)}$). This process can be repeated as many times as necessary until we obtain the desired measurement outcome $s_1 =1$. Each recovery attempt has probability $1/2$ of succeeding or failing, so the probability of needing $n$ recovery attempts for the forced measurement process in order to obtain the desired outcome of $s_1=+1$ is $p_n = p_0 2^{-n}$ and the average number of recovery attempts needed for this will be $\langle n \rangle = 2 p_0$. A similar protocol can be used for each of the three segments of the braiding process if the corresponding measurements do not initially yield the desired outcome.

The original forced measurement protocol may be less efficient than is desirable if the measurement times are relatively long, as each recovery attempt only has $1/2$ probability of success. In this case, it may be preferable to utilize a hybrid approach that combines the use of nearly-adiabatic evolution with the forced measurement scheme in order to generate a high probability of success for each recovery attempt. Consider, again, the situation where we reach the first turning point with Hamiltonian $H=-i\Delta\gamma_1\gamma_0$ (we assume $\Delta>0$), and perform a measurement of the fermion parity $i \gamma_1 \gamma_0$ and obtain the undesired outcome $s_1=-1$ with probability $p_0$. We can now follow the hybrid adiabatic-measurement recovery protocol:

\begin{enumerate}
\item Change the sign of the coupling between $\gamma_0$ and $\gamma_1$, so that the Hamiltonian goes from $H=-i\Delta\gamma_1\gamma_0$ to $H=i\Delta\gamma_1\gamma_0$.

\item Nearly-adiabatically tune the Hamiltonian from $H=i\Delta\gamma_1\gamma_0$ to $H=-i\Delta \gamma_2 \gamma_0$, and then to $H=-i\Delta\gamma_1\gamma_0$.

\item Measure the fermion parity $i \gamma_1 \gamma_0$. If the outcome is $s_1=-1$, go to step 1. If the outcome is $s_1 =+1$, stop.
\end{enumerate}

In step 1, we emphasize that the Hamiltonian only involves the MZMs $\gamma_0$ and $\gamma_1$, so the process of changing the sign of the coupling does not change the state, i.e. the state remains in the $i \gamma_1 \gamma_0 = -1$ state during this process, due to conservation of fermion parity. It just goes from being an excited state to being a ground state. Note that ancillary MZMs' states $i\gamma_1\gamma_0=\pm 1$ will temporarily become degenerate in this step when the Hamiltonian passes through zero. Clearly, this means that this step will not be adiabatic (nor nearly-adiabatic) with respect to the energy difference between the $i\gamma_1\gamma_0=\pm 1$ states, but we also want to make sure that it is fast with respect to any of the exponentially suppressed energy splittings between topologically degenerate states.  Of course, if the MZMs are embedded in a superconductor, then we must also ensure that the process is slow enough not to excite states above the superconducting gap. In other words, if this process is carried out in time $t_{\text{flip}}$, then we require $\Delta_{\text{SC}}^{-1} \ll  t_{\text{flip}} \ll 1/ \delta E$. 

In step 2, we really just want any adiabatic path from $H=i\Delta\gamma_1\gamma_0$ to $H=-i\Delta\gamma_1\gamma_0$. Taking a path that passes through $H=-i\Delta \gamma_2 \gamma_0$ and which only involves $\gamma_0$, $\gamma_1$, and $\gamma_2$ is most convenient, because it limits the possible diabatic errors to transitions involving the three MZMs that we are already manipulating and measuring in this segment of the braiding process. Moreover, as we will discuss later, we may need to pause at $H=-i\Delta \gamma_2 \gamma_0$ during this step in order to flip the sign of the possible coupling between $\gamma_0$ and $\gamma_1$, while its magnitude is at zero and can be done without affecting the state. As long as the Hamiltonian is changed slowly and smoothly (near adiabatically) during this step, the system will remain in the ground state with high probability. In this case, the subsequent measurement in step 3 will have a high probability of obtaining the desired measurement outcome $s_1 =+1$. If the probability of obtaining the undesired outcome $s_1 = -1$ after one such hybrid recovery attempt is $p$, the probability of needing $n$ recovery attempts in order to obtain the desired outcome $s_1=+1$ is $p_n = p_{0} p^{n-1}(1-p)$ and the average number of recovery attempts needed for this will be $\langle n \rangle = \frac{p_0}{1-p}$. In this hybrid scheme, $p$ can be made arbitrarily small by making the nearly-adiabatic evolution take a longer amount of time and by making the Hamiltonian time dependence smoother.

If the system is coupled to a dissipative bath of the type described in Section~\ref{sec:dissipation}, there is yet another generalization of forced measurement protocol. There is some rate $\Gamma$ for relaxation to the ground state which vanishes at the turning points and is largest midway between two turning points.  In this case, after performing a measurement of the fermion parity $i \gamma_1 \gamma_0$ with the undesired outcome $s_1=-1$, we can follow the dissipation-assisted hybrid recovery protocol:
\begin{enumerate}
\item Nearly-adiabatically tune the Hamiltonian from $H=-i\Delta\gamma_1\gamma_0$ to $H=-i\Delta \frac{1}{2} \left( \gamma_1\gamma_0 + \gamma_2 \gamma_0 \right)$.

\item Pause for an amount of time approximately equal to $\Gamma^{-1}$.

\item Nearly-adiabatically tune the Hamiltonian from $H=-i\Delta \frac{1}{2} \left( \gamma_1\gamma_0 + \gamma_2 \gamma_0 \right)$ to $H=-i\Delta\gamma_1\gamma_0$.

\item Measure the fermion parity $i \gamma_1 \gamma_0$. If the outcome is $s_1=-1$, go to step 1. If the outcome is $s_1 =+1$, stop.
\end{enumerate}

The effectiveness of this strategy strongly depends on the system-bath coupling. It has the advantage over the previously described hybrid strategy that it does not require the ability to change the sign of couplings.

\begin{figure}[t!]
\includegraphics[width=1.5\columnwidth]{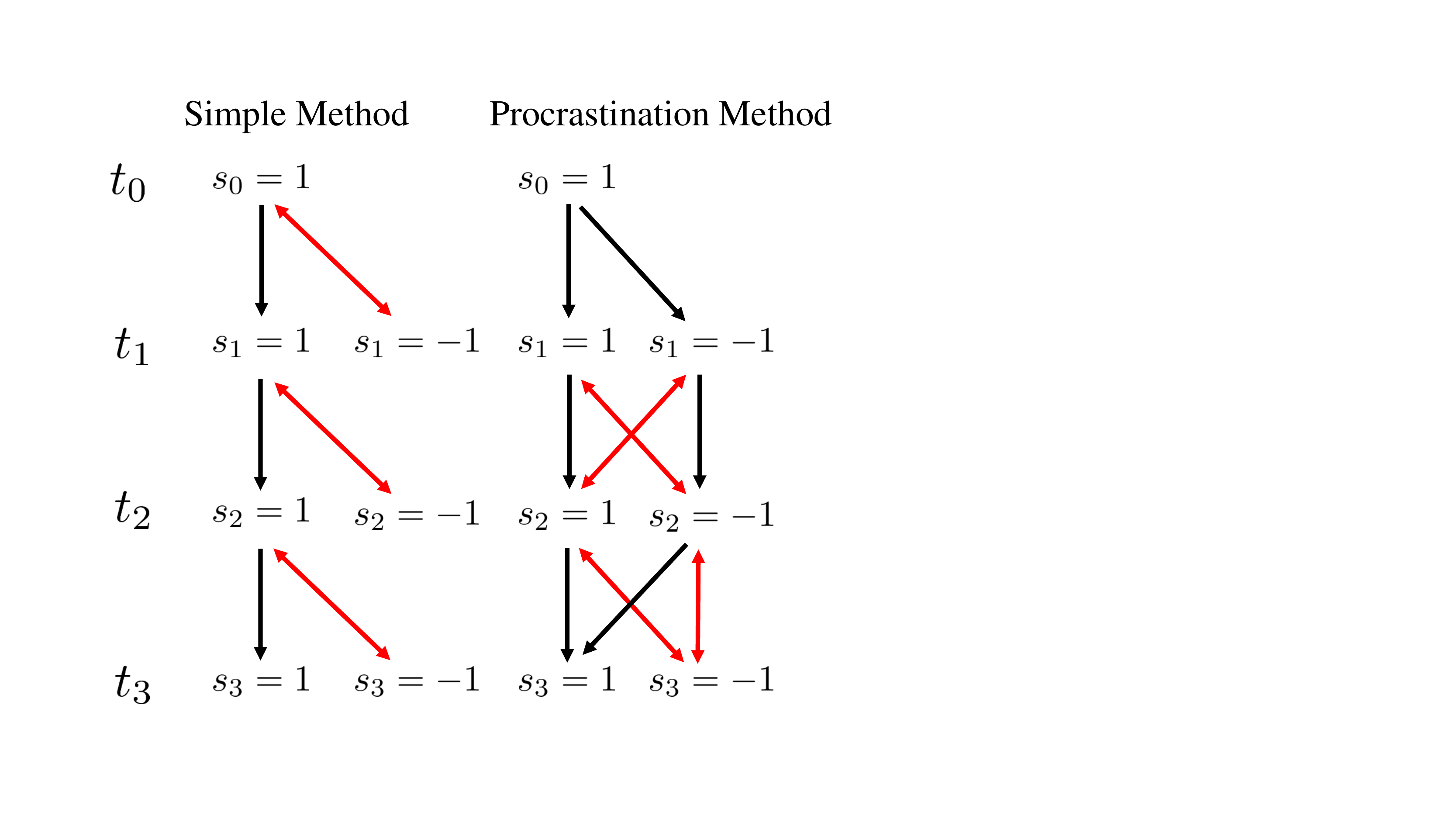}
\caption{The above flow chart outlines the two methods of using forced measurement or its generalizations for recovery protocols, as discussed in the text. The one directional arrows (black) indicate a process that yields a desired or acceptable outcome for which we do not apply a recovery protocol. The bidirectional arrows (red) indicate a process that yields an undesired or unacceptable outcome for which we apply a recovery protocol. We can schematically think of the recovery protocol as backtracking and trying the process again, with a new probability of obtaining a desired outcome. The simple method applies a recovery protocol whenever a turning point measurement outcome indicates a diabatic transition occurred to an excited state. The procrastination method will accept either measurement outcome at the first turning point. However, when the first measurement outcome is $s_1 =-1$, if we procrastinate, we must require the second turning point to have measurement outcome $s_2 = -1$, because we need two wrongs to make a right.}
\label{CorrectionMethods}
\end{figure}

We have outlined three approaches to correcting diabatic errors at each turning point: the forced measurement, hybrid, and dissipation-assisted hybrid protocols. As described above, we can employ one of these recovery schemes as soon as we measure the system in its excited state at each turning point of the braiding process.  This is outlined in the left panel of Fig.~\ref{CorrectionMethods}.  A slightly more efficient method is to procrastinate correcting certain errors.  If we measure $s_1=s_2=-1$, then as long as we measure $s_3=s_0$, we will obtain the correct braiding transformation. In other words, two wrongs make a right. Thus if we measure $s_1=-1$, there is some chance that, if we continue to evolve, we will find $s_2=-1$ and $s_3=s_0$, in which case the system has made the right number of errors to correct itself. The likelihood of such self-correction can be increased by changing the sign of the coupling between $\gamma_0$ and $\gamma_3$, so that the Hamiltonian is taken from $H=-i\Delta\gamma_1\gamma_0$ at time $t_1$ to $H=i \Delta \gamma_3 \gamma_0$ at time $t_2$. In this way, if there is no diabatic transition during the second braiding segment, the system will stay in the excited state, and yield $s_1=s_2=-1$. If a diabatic error does occur during this segment, yielding the measurement outcome $s_2=+1$, then we apply a recovery protocol. This procrastination method is shown in the right panel of Fig.~\ref{CorrectionMethods}.


\section{A Correction Scheme for Diabatic Errors to the Braiding of Anyons}
\label{sec:general-braiding}

We now explain how the previous section's correction scheme for diabatic errors to braiding MZMs can be generalized to the braiding of generic non-Abelian anyons.  We will first consider braiding transformations generated using a $T$-junction type setup with tunable couplings between non-Abelian anyons at fixed locations, as described in Ref.~\onlinecite{BondersonPRB2013}. At the end of this section, we will explain how to correct for diabatic errors in the more general scenario of transporting anyons through a two-dimensional space.

\begin{figure}[t!]
\includegraphics[width=.65\columnwidth]{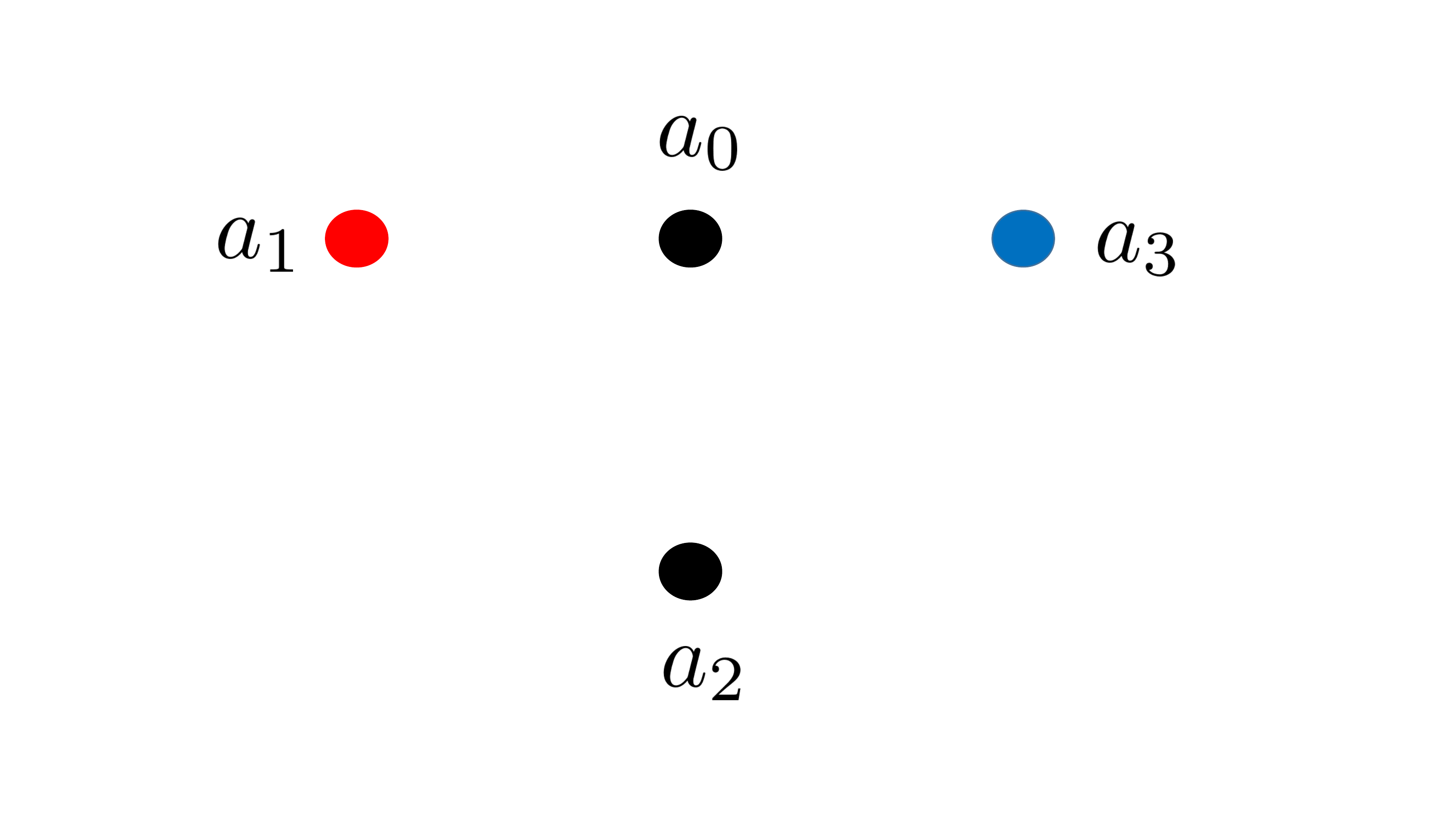}
\caption{In the $T$-junction shown here, anyons $a_1, a_2,$ and $a_3$ have topological charge $a$ and anyon $a_0$ has topological charge $\bar{a}$.  Similar to the protocol for braiding MZMs, at times $t=0$ and $t=t_{\text{op}}$ anyons $a_0$ and $a_2$ (both in black) form the ancillary pair of anyons.  We exchange the positions of the red and blue anyons using a protocol identical to that shown in Fig.~\ref{fig:T-junction-braiding}, with the labels $\gamma_i$ replaced by $a_i$.}
\label{fig:general-T}
\end{figure}

It is straightforward to generalize the MZM braiding protocol depicted in Fig.~\ref{fig:T-junction-braiding} to the braiding of two non-Abelian anyons of topological charge $a$ in the $T$-junction shown in Fig.~\ref{fig:general-T}.
\footnote{The braiding protocol can also be applied to anyons of different topological charge values $a$ and $b$, provided that they share an Abelian fusion channel~\cite{BondersonPRB2013}, but we restrict out attention to the simpler case.} 
For the sake of simplicity, we assume that the anyons $a$ and $\bar{a}$ obey the fusion rule
\begin{equation}
\label{eq:fusion}
a \times \bar{a}= 0 + c,
\end{equation}
where $0$ is the vacuum topological charge and $c$ is some nontrivial topological charge.

As with the example of braiding MZMs, we partition the braiding operation into three steps that end at times $t_{\text{op}}/3$, $2t_{\text{op}}/3$, and $t_{\text{op}}$, which we call the turning points. At each turning point, two anyons are coupled with each other and the other two anyons are decoupled. Each step interpolates between the turning points, changing which two anyons are coupled (or decoupled). The sequence is identical to that in Fig.~\ref{fig:T-junction-braiding}, with the labels $\gamma_i$ replaced by $a_i$.  The corresponding Hamiltonian governing this (sub)system of four anyons can be written as
\begin{eqnarray}
H&=&-\sum_{j=1}^3 \Delta_{j}(t) Z_{j} \\
Z_{j} &=&  \ket{a_j,a_0;0}\bra{a_j,a_0;0} - \ket{a_j,a_0;c}\bra{a_j,a_0;c}
\end{eqnarray}
where $\ket{a_j,a_0; 0}$ and $\ket{a_j,a_0; c}$ correspond to the states in which the anyons $a_j$ and $a_0$ are in the $0$ and $c$ fusion channels (i.e. have collective topological charge of the corresponding values), respectively. The energy splittings given by $\Delta_j(t)$ reflect which pairs of anyons are coupled, as was the case for MZMs. For the (near) adiabatic braiding process, the nonzero values of $\Delta_{j}(t)$ at the turning points are $\Delta_{2}(0)$, $\Delta_{1}(t_{\text{op}}/3)$, $\Delta_{3}(2 t_{\text{op}}/3)$, and $\Delta_{2}(t_{\text{op}})$.

We assume these energy scales are much smaller than the bulk gap of the system, $|\Delta_j(t)| \ll \Delta_{\text{bulk}}$, so that the dominant diabatic errors will be transitions to excited states within the fusion state space of these four anyons, rather than to states with additional bulk quasiparticle excitations. Assuming only such dominant diabatic errors, the discussion follows that of Section~\ref{sec:Measurement=Correction}. In particular, when a diabatic error occurs in a given step of the braiding process, the ancillary (decoupled) pair of anyons at the end of the step will be in the $c$ fusion channel, rather than the $0$ fusion channel corresponding to the ground state. In accord with Section~\ref{sec:CS-theory}, where we demonstrated that diabatic errors result from uncontrolled creation and movement of anyons, we can interpret such errors as corresponding to an unintended transfer of topological charge $c$ between the anyon being transported and the ancillary pair. (There is nowhere else for the topological charge to come from or go, in the state level of approximation, since the Hamiltonian does not couple to any other degrees of freedom.) If we project the ancillary pair of anyons to their vacuum fusion channel after each step, we recover the braiding transformation for adiabatic evolution.  Thus, as we saw with MZMs, a measurement-based error correction protocol can correct all diabatic errors within the four anyon subspace.

Let us focus on the situation where we are tuning between the initial configuration at $t=0$ with $H=- \Delta Z_2$ and the first turning point at $t=t_{\text{op}}/3$ with $H=-\Delta Z_1$. When we reach the first turning point, we perform a measurement of fusion channel of the pair of anyons $a_1$ and $a_0$ and obtain outcome $s_1$. (The precise method of measurement will depend on the details of the system in which the anyons exist.) The desired measurement outcome, corresponding to no diabatic error occuring, is $s_1 =0$. Let us assume the outcome $s_1 = c$, corresponding to a diabatic error, occurs with probability $p_0$. In the event of this diabatic error, we can apply the following hybrid adiabatic-measurement diabatic error correction protocol:
\begin{enumerate}
\item Change the sign of the coupling between $a_0$ and $a_1$, so that the Hamiltonian goes from $H=-\Delta Z_1 $ to $H=\Delta Z_1$.

\item Nearly-adiabatically tune the Hamiltonian from $H=\Delta Z_1$ to $H=-\Delta Z_2$, and then to $H=-\Delta Z_1 $.

\item Measure the fusion channel of $a_0$ and $a_1$. If the outcome is $s_1=c$, go to step 1. If the outcome is $s_1 =0$, stop.
\end{enumerate}

The above steps are identical to the hybrid adiabatic-measurement recovery protocol outlined for MZMs in Section~\ref{sec:Measurement=Correction}.  In step 1, the Hamiltonian only involves anyons $a_0$ and $a_1$, so the process of changing from $H=-\Delta Z_1 $ to $H=\Delta Z_1$ does not change the state.  It just takes it from being an excited state to being a ground state. In doing so, the fusion channels $0$ and $c$ will temporarily become degenerate, thus this step will not be adiabatic (nor nearly-adiabatic) within the four anyon subspace.

Step 2 really just requires any nearly adiabatic path from $H=\Delta Z_1$ to $H=-\Delta Z_1 $. The path described limits the possible diabatic errors to involving the three anyons that we are already manipulating and measuring in this segment of the braiding process.  As long as the Hamiltonian is changed nearly adiabatically during this step, the system will remain in the ground state with high probability.  In this case, the measurement in step 3 will have a high probability of obtaining the desired measurement outcome of $s_1=0$.  If the probability of an undesired measurement outcome $s_1=c$ after one such hybrid recovery attempt is $p$, the probability of needing $n$ recovery attempts in order to correct the diabatic error is $p_n=p_0 p^{n-1}(1-p)$ and the average number of recovery attempts needed for this will be $\langle n \rangle=\frac{p_0}{1-p}$.  In this hybrid scheme, $p$ can be made arbitrarily small by making the nearly-adiabatic evolution take a longer amount of time and by making the Hamiltonian time dependence smoother.

Similarly, one can also adapt the dissipation-assisted hybrid recovery protocol of Section~\ref{sec:Measurement=Correction} to apply to non-Abelian anyons, but we will not repeat the details. Of course, we could alternatively use other methods, such as a measurement-only protocol, if they provide preferable time costs.

It is straightforward to generalize the above discussion to the case of general fusion rules $a \times \bar{a}= \sum_{c} N_{a \bar{a}}^{c} c$ (note that we always have $N_{a \bar{a}}^{0}=1$, by definition), as it it simply involves keeping track of additional energies levels corresponding to the additional fusion channels and multiplicities. It does, however, require having greater control over the system parameters, because errors corresponding to the different undesired fusion channel measurement outcomes ($s_{j} \neq 0$) will require tuning the couplings in a manner that is specific to the particular fusion channel.

We note that, for general non-Abelian anyons, one cannot always use the procrastination method, described in Section~\ref{sec:Measurement=Correction}, for reducing the number of diabatic error correction protocols applied during a complete braiding operation. It can only be used when the undesired fusion channel measurement outcomes at intermediate turning points $s_1 = s_2 = c$ is an Abelian topological charge.

\begin{figure}[t!]
\includegraphics[width=.65\columnwidth]{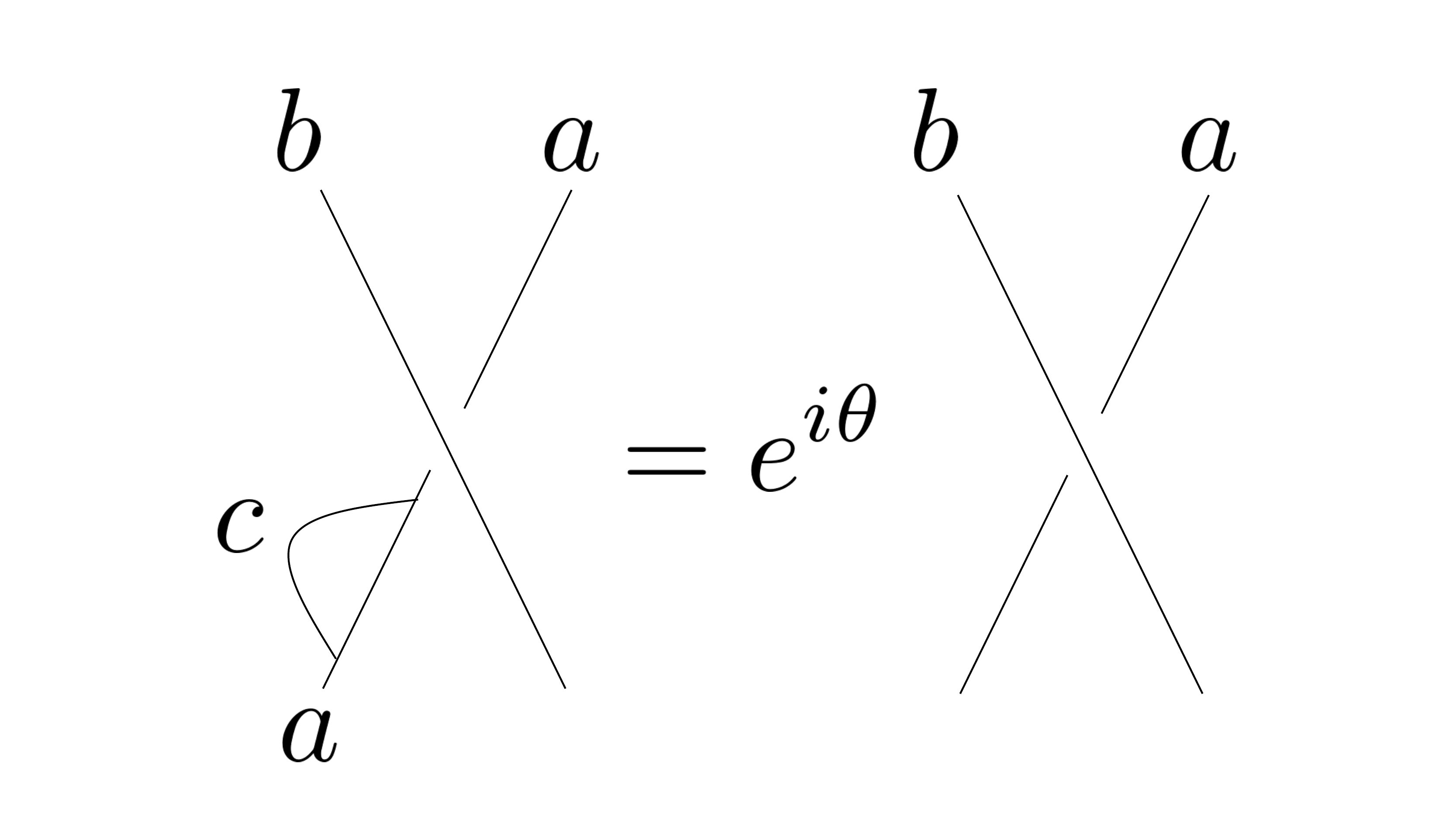}
\caption{The left side shows the braiding diagram corresponding to the diabatic error correction protocol addressing the creation of a bulk quasiparticle.  The anyon $a$ can emit an anyon $c$, which we need to detect, trap, then fuse back together with $a$.  The right side shows that this process is equivalent to the intended braid, up to an unimportant overall phase. Note that the $b$ line is not actually neccessary for this statement.}
\label{fig:braid-emit-refuse}
\end{figure}

If the diabatic errors associated with creation of quasiparticles in the bulk, i.e. transitions to states above the bulk gap, are not negligible (as we have previously assumed), then we require additional machinery to correct such errors. By locality, such diabatic errors will create quasiparticles in the vicinity of the ``transport path,'' which is to say along the two legs of the $T$-junction connecting the three anyons involved in a given step. We must monitor the bulk region along this path to detect whether there is an unintentional creation of a bulk quasiparticle that leaves the $T$-junction. (If the unintentionally created quasiparticle remains in the $T$-junction, it will be dealt with by the previous diabatic error correction protocols.) In the event that an emitted quasiparticle is detected, it must be trapped and fused back into the anyons involved in the corresponding transport process. 

This protocol also applies more generally to the case where an anyon is being physically transported through the 2D system by some arbitrary method, e.g. being dragged around by a moving pinning potential. This can be understood schematically from the diagrams shown in Fig.~\ref{fig:braid-emit-refuse}, where we show a moving anyon of topological charge $a$ that emits an anyon of topological charge $c$. If we trap the anyon $c$ and fuse it back into anyon $a$, the process is equivalent (in the topological state space) to the process where anyon $a$ is moved without emitting anyons, up to unimportant overall phase factors.


\section{Implementation of Measurement-Based Correction in a Flux-Controlled Architecture
for Manipulating MZMs}
\label{sec:Hyart-system}

\subsection{Review of the Top-Transmon}

We now adapt the diabatic error correction scheme of Section~\ref{sec:Measurement=Correction} to a particular MZM device in which the Hamiltonian parameters are tuned by varying the magnetic flux.  The idea is to embed a MZM $T$-junction~\cite{Alicea11,Heck12} or $\pi$-junction~\cite{Hyart13} inside a system of superconductors, coupled to each other with split Josephson junctions.  Changing the magnetic flux through a junction changes the Coulomb couplings between MZMs on the same island.  This proposal is appealing both because it does not require careful control over local parameters, as would be necessary to move topological domain walls, and because the sophistication of superconducting qubit technology can be easily transferred to a combined superconductor-topological qubit system.  In particular, superconducting qubit experiments are able to carefully control the time evolution of the magnetic flux through a split Josephson junction, so it is feasible to set multiple time derivatives of the flux time dependence to zero at the beginning and end of the evolution.

\begin{figure}[t!]
\includegraphics[width=\columnwidth]{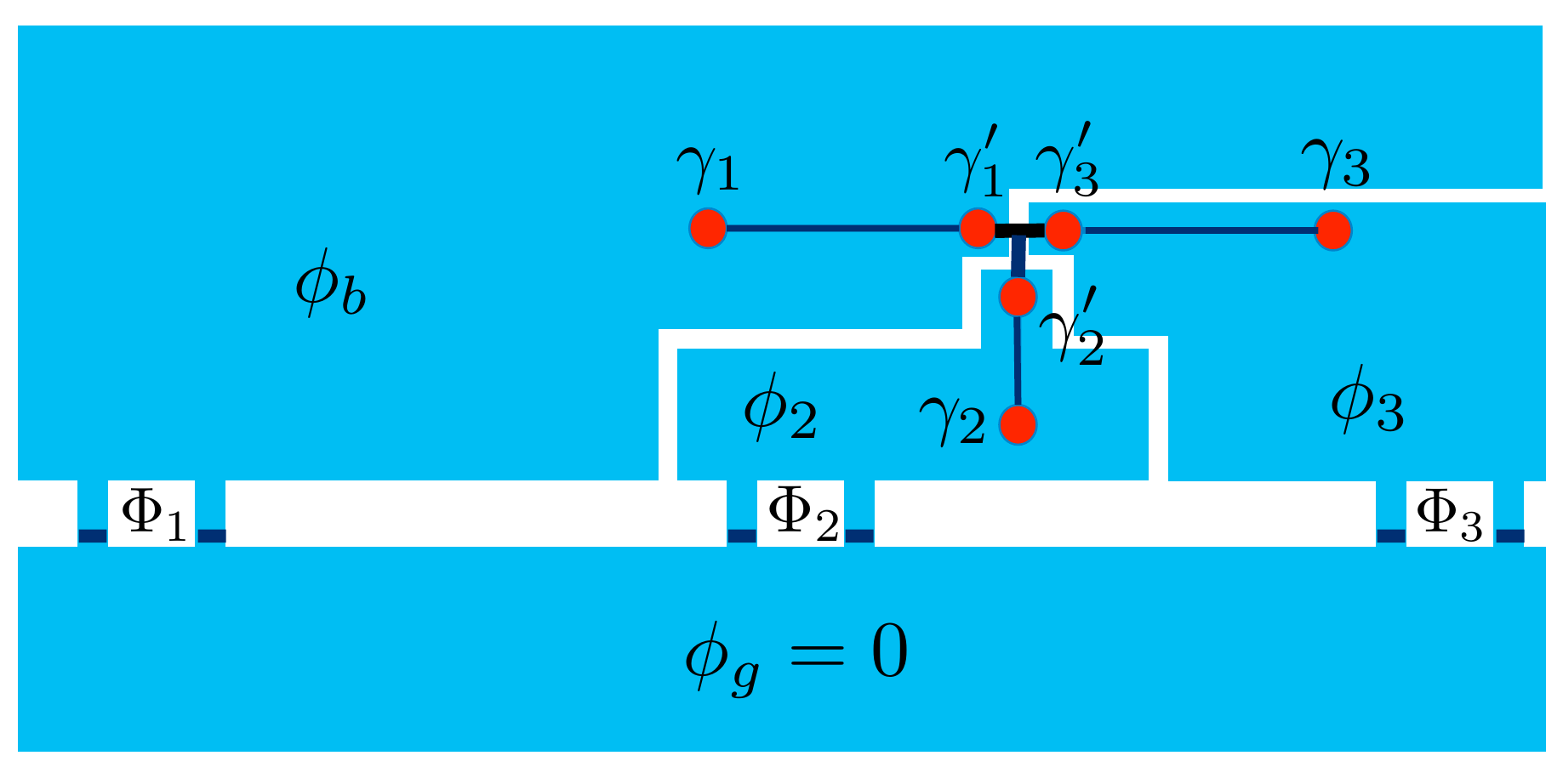}
\begin{tabular}{|c|c |c|c|}
\hline
time & $|\Phi_1|$ & $|\Phi_2|$ & $|\Phi_3|$ \\ \hline
0 & 0 & $ \Phi_{\text{max}} $& 0 \\ \hline
$t_1 $ &$\Phi_{\text{max}} $& 0 & 0 \\ \hline
$t_2$ & 0 & 0 &$ \Phi_{\text{max}} $\\ \hline
$t_3 $ & 0 & $\Phi_{\text{max}}$ & 0 \\ \hline
\end{tabular}
\caption{Top: The simplest flux-tunable MZM braiding setup, following Refs.~\onlinecite{Heck12,Hyart13}. The black lines are nanowires hosting MZMs (red dots) at their endpoints.
We tune the Josephson junctions' flux values $|\Phi_i|$ between $0$ and $\Phi_{\text{max}}<\frac{1}{2}\Phi_0$ to change the strength of the Coulomb coupling between the MZM pairs $\gamma_i$ and $\gamma_i'$.
Bottom: Flux values at the four turning points.  When $|\Phi_i|=\Phi_{\text{max}}$,  $\Delta_i$ is maximized and $\gamma_i$ is coupled to the center MZM $\gamma_0$, formed out of a linear combination of $\gamma_1', \gamma_2',$ and $\gamma_3'$. }
\label{SimpleHyart13}
\end{figure}

\begin{figure}[t!]
\includegraphics[scale=0.6]{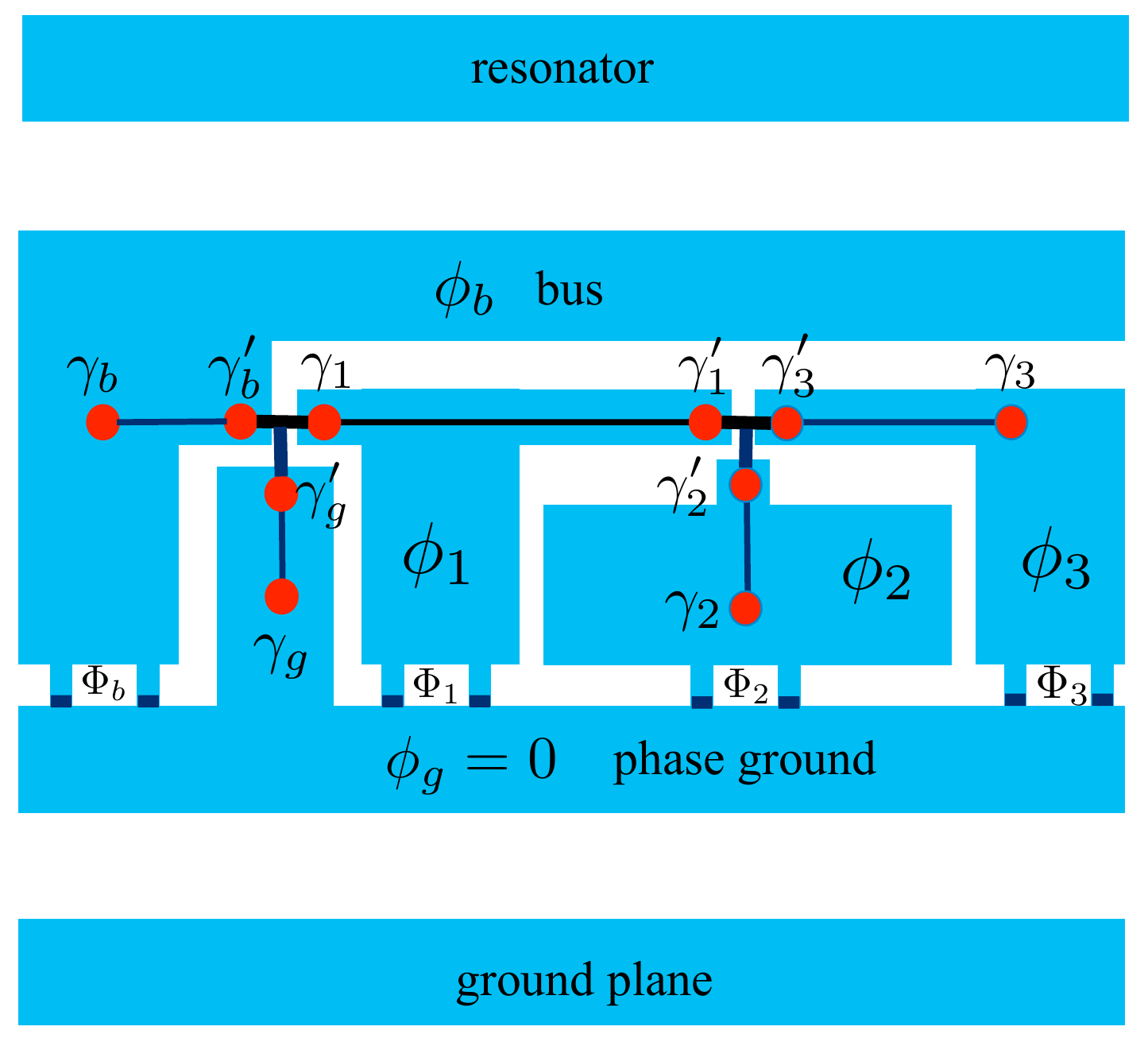}
\caption{The $\pi$-junction proposed by Ref.~\onlinecite{Hyart13}.  The MZM system sits inside a superconducting qubit formed by the bus and ground islands.  The topological qubit is embedded into a transmission line resonator to allow read-out of the qubit state.}
\label{HyartFig}
\end{figure}

The minimal braiding setup is the $T$-junction shown in Fig.~\ref{SimpleHyart13}.  The minimal setup that encodes a topological qubit is the $\pi$-junction in Fig.~\ref{HyartFig}, but most of the underlying physics is already captured by the $T$-junction.
We review the braiding scheme for the $T$-junction and discuss the diabatic errors to which this setup is susceptible.
We then propose a modification to the superconducting system that allows for correction of the most common
diabatic errors.  Finally, we show how this modification can be easily extended to the $\pi$-junction.

Fig.~\ref{SimpleHyart13} shows three superconducting islands, each hosting a semiconductor nanowire tuned to have a MZM on either end, connected via split Josephson junctions to a superconducting phase ground.  The island hosting MZMs $\gamma_1$ and $\gamma_1'$ is referred to as the ``bus" and is assumed to be much larger than islands 2 and 3.  The nanowires form a $T$-junction, with $\gamma_1$, $\gamma_2$, and $\gamma_3$ located at the endpoints of the $T$ and $\gamma_1'$, $\gamma_2'$, and $\gamma_3'$ situated at the center of the $T$.

The $\gamma_{j}'$s are coupled to each other through a Majorana-Josephson potential of strength $E_M$, which couples the three MZMs in the low-energy subspace, leaving behind a single MZM that we denote as $\gamma_0$, which is a linear combination of $\gamma_1'$, $\gamma_2'$, and $\gamma_3'$
that commutes with the six MZM Hamiltonian, given in Eq.~(\ref{HyartHeff}).

If we ignore the excited states associated with $\gamma_1'$, $\gamma_2'$, and $\gamma_3'$,
then the low-energy Hamiltonian (up to small corrections that we will for now assume to be negligible) is
\begin{equation}
H_\text{eff} =- \sum_j \Delta_j i\gamma_j \gamma_0.
\label{eqn:MZM-braiding-effective}
\end{equation}
Hence, the low-energy effective Hamiltonian of this system is the Hamiltonian of Eq.~(\ref{eqn:MZM-braiding-Hamiltonian}) that we analyzed in Section~\ref{sec:MZM-trijunction}.
The couplings  $\Delta_i$ are~\cite{Hyart13}
\begin{multline}
\Delta_i =
16 \left( \frac{E_{C,i}E_{J,i}(\Phi_i)^3}{2\pi^2}\right)^{1/4}\\
\times e^{-\sqrt{8E_{J,i}(\Phi_i)/E_{C,i}}} \cos(q_i \pi/e) f(\alpha)
 \end{multline}
where $f(\alpha)$ is a function depending on the Aharanov-Bohm phase shifts which is $\mathcal{O}(1)$ during the braiding process and $q_i$ is the induced charge on island $i$, controlled through electrostatic gates.  The Josephson energy associated with junction $i$ is
\begin{equation}
\label{eq:JJ_energy}
E_{J,i}(\Phi_i) =E_{J,i}(0)\cos(\pi\frac{\Phi_i}{\Phi_0})
,
\end{equation}
and $E_{C,i}$ is the single electron charging energy of junction $i$.   The system is operated in the regime $E_{J,i}\gg E_{C,i}$.  Thus, when we tune $\Phi_i\approx 0$, the ratio $E_{J,i}(\Phi_i)/E_{C,i}$ is maximized and the coupling $\Delta_i$ is exponentially suppressed.  A reasonable parameter choice is $E_{J,i}(0)/E_{C,i}\sim 50$~\cite{Koch07}, indicating that $\Delta_i$ can be tuned to a minimum value $\sim e^{-20}$.  This justifies our approximation in Section~\ref{sec:Measurement=Correction}  that the system Hamiltonian commutes with $\gamma_k$ when $\Delta_k$ is tuned to its minimum value.    When we tune flux $\Phi_i=\Phi_{\text{max}} \lesssim \frac{\Phi_0}{2}$, $E_{J,i}(\Phi_i)/E_{C,i}$ is minimized and $\Delta_i$ reaches its maximum value. Note that the sign of $\Delta_i$, which determines which ancilla parity state corresponds to the ground state, depends on the induced charge $q_i$.  By tuning the fluxes according to the schedule shown in Fig.~\ref{SimpleHyart13},
we can vary the Hamiltonian with time in the manner considered in Section~\ref{sec:MZM-trijunction}.

To enable measurements of the system, the setup in Fig.~\ref{SimpleHyart13} is capacitively coupled to a transmission line resonator, as shown schematically for the $\pi$-junction in Fig.~\ref{HyartFig}.  The frequency of the resonator is shifted by the state of the superconducting-MZM system.  This results in an energy-dependent transmission amplitude of a microwave sent down the transmission line, which can be used to extract the state of the superconducter-MZM system~\cite{Sank14}.  A system of a superconducting bus and ground coupled to each other through a split Josephson junction and capacitively coupled to a transmission line resonator is a particular type of superconducting qubit, known as a ``transmon" when operated in the regime $E_J\gg E_C$~\cite{Koch07}.  The system described here embeds a topological qubit within a transmon.  When this system is tuned such that all islands are either phase locked to the bus or the ground it forms a ``top-transmon"~\cite{Hassler11, Heck12, Hyart13}.   The top-transmon proposal and the measurement scheme are discussed in further detail in Appendices~\ref{Details} and~\ref{Measurement}, respectively.

\subsection{Diabatic Errors in a Top-Transmon}
\label{sec:DiabaticErrorTopTransmon}

We now consider diabatic errors that could occur in the device of Fig.~\ref{SimpleHyart13}.
The errors of the type analyzed in Section~\ref{sec:MZM-trijunction} can occur: the system remains within the
low-energy subspace governed by Eq.~(\ref{eqn:MZM-braiding-effective})  and errors can be identified and
corrected by the protocol of Section~\ref{sec:Measurement=Correction}. In Section~\ref{sec:SC-measurement},
we discuss how the necessary measurements can be carried out.

It is also possible for the system to transition out of the low energy subspace described by Eq.~(\ref{eqn:MZM-braiding-effective}).   The low energy subspace includes the ground and first excited states, separated by energy gap $\mathcal{O}\left(\Delta\right)$. The gap separating these two lowest lying states from the higher excited states is $\mathcal{O}\left( E_M \right)$, where $E_M$ is the energy scale of the Majorana-Josephson coupling between $\gamma_1'$, $\gamma_2'$, and $\gamma_3'$ that splits their shared degeneracy.  As discussed in Section~\ref{sec:general-theory}, the probability of diabatic transitions to excited states of energy $E_{\text{gap}}$ scales with the operational time as $\mathcal{O} \left(\frac{1}{( {t_\text{op}} E_{\text{gap}})^{2k+2} } \right)$, for $\mathcal{C}^{k}$ smoothness of the time evolution. Hence, the relative likelihood of errors due to diabatic transition to the second or third excited states [at energies $\mathcal{O}\left( E_M \right)$] compared to errors due to diabatic transition to the first excited states [at energies $\mathcal{O}\left(\Delta\right)$], which are correctible by the protocol of Section~\ref{sec:Measurement=Correction}, will scale as $ \left( \Delta/E_M\right)^{2k+2}$.

Other errors can occur from transitioning to even higher energy levels on the order of the Josephson
energy $E_J$ or the bulk superconducting gap $\Delta_{\text{SC}}$.  One possibility is diabatic-induced quasiparticle
poisoning, which can be understood as follows. Tuning the flux at a Josephson junction decreases the energy gap to the continuum for the Andreev bound state (ABS)
at that junction.  If the ABS transitions to the continuum and travels into the superconducting island $i$, it changes the induced
charge of that island, flipping the sign of $\Delta_i$ and thereby interchanging the ground and first excited states.
However, we expect the probability of such errors to be suppressed due to both the larger energy scale and the fact
that the time evolution of each Josephson junction is ``more adiabatic" than the evolution of the six MZM system.
This second point is because the Josephson energies in Eq.~(\ref{eq:JJ_energy}) depend less sensitively on changes in the flux
than the Coulomb couplings, which depend roughly as $e^{-\sqrt{8E_{J,i}(\Phi_i)/E_{C,i}}}$.

One might also worry that despite the larger gap and the less dramatic time dependence, diabatic transitions above the superconducting gap would be significant due to the continuum of available states.  The following argument suggests that the continuum of states above the gap does not introduce significant errors. The system's time dependence arises from changing the Josepshon energy, which is a local quantity. Thus, we expect a diabatic transition to excite a quasiparticle state localized near the junction.  Provided the junction is spatially separated from the MZMs, the effect of this excited state on the MZM subspace is very small. In other words, the spectral weight for the local density of states near the MZM wires is small and finite, which implies that the matrix elements between excitations above the superconducting gap and the state of the system are small.  Integrating these matrix elements over the continuum of available states will still be small. Combined with the suppressed probability of such a transition, we, thus, expect diabatic transitions above the superconducting gap to be much less significant than errors within the MZM system. Hence, we will satisfy ourselves by correcting the latter.  We note that, if we were to braid the MZMs by physically moving them around each other, we would need to take into account diabatic transitions to the continuum of states above the superconducting gap as such excitations would be localized near the MZMs.
Such errors could be dealt with using the quasiparticle trap method described schematically at the end of Section~\ref{sec:general-braiding}.

Finally, we note that the magnetic fluxes threaded through the split Josephson junctions control the time evolution.  Each flux is tuned independently from the others. Thus, noise introduced in one junction will be uncorrelated with noise associated with the flux in a different junction.  This justifies our choice of system-bath coupling in Eq.~\ref{eqn:spin-hamiltonian-with-diss} and demonstrates that the analysis of Section~\ref{sec:dissipation} applies to a top-transmon.

\subsection{Extension to the $\pi$-junction}
\label{sec:flux-control-error-correction}

The same analysis holds for $\pi$-junctions, with the modification that there are now always four decoupled MZMs.  For instance, in Fig.~\ref{HyartFig}, $\gamma_b$ and $\gamma_g$ are always decoupled, as are two of the remaining four MZMs ($\gamma_1'$, $\gamma_2'$, and $\gamma_3'$ are Majorana-Josephson coupled and so effectively comprise one MZM, as is the case with $\gamma_1$, $\gamma_b'$, and $\gamma_g'$).  In the effective six MZM picture, the two MZMs appearing in the Hamiltonian form the ancilla and the four decoupled MZMs comprise the topological qubit.  The two energy levels are determined by the parity of the ancilla.  If we fix total parity, each energy level is two-fold degenerate; e.g. for total parity even, at the first turning point the ground state corresponds to ancilla parity even and two degenerate qubit states,  ${\ket{0}=\ket{i\gamma_g\gamma_b=+1,i\gamma_2\gamma_3=+1}}$ and ${\ket{1}=\ket{i\gamma_g\gamma_b=-1,i\gamma_2\gamma_3=-1}}$.

\subsection{Error Detection through Projective Measurement}
\label{sec:SC-measurement}

\begin{figure}
\includegraphics[width=\columnwidth]{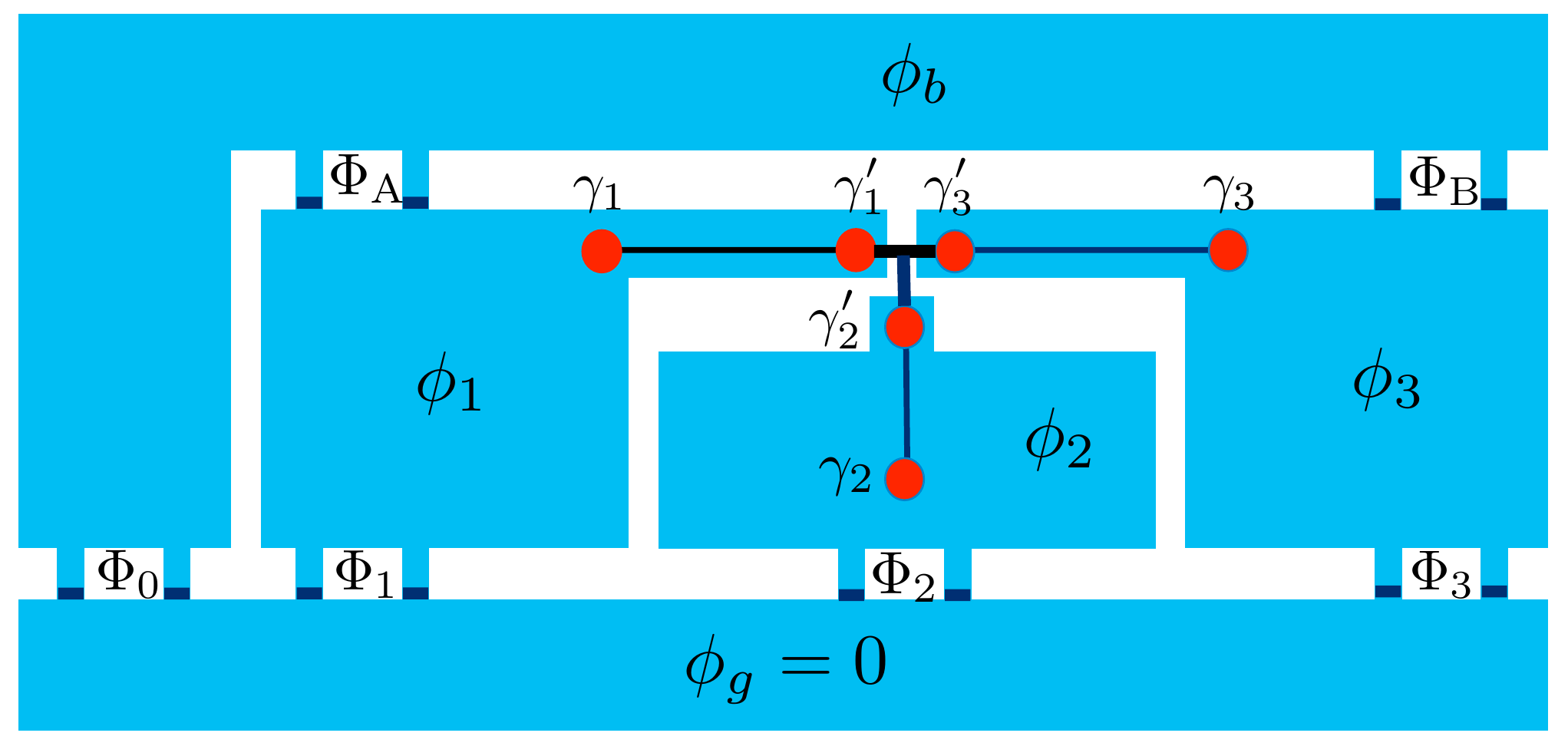}
\begin{tabular}{|c|c|c|}
\hline
time & Junctions with $\Phi=0$ & Junctions with $|\Phi|=\Phi_{\text{max}}$ \\ \hline
0 & 2, A, B & 0, 1, 3 \\ \hline
$t_1$ & 2, 3, A & 0, 1, B\\ \hline
$t_2$ & 1, 2, B & 0, 3, A \\ \hline
$t_3$ & 2, A, B  & 0, 1, 3 \\ \hline
\end{tabular}
\caption{Top: Modified $T$-junction structure designed to allow for parity measurements at each turning point.
Bottom: Flux values for measurement at the turning points.  Note that a junction with $\Phi=0$ maximizes the Josephson energy and phase locks its neighboring superconductors, while a junction with $|\Phi|=\Phi_{\text{max}}$ minimizes its Josephson energy and essentially decouples the neighboring superconductors.  At time $t_1$, islands 2 and 3 are phase locked to the ground and decoupled from the bus, while island 1 is phase locked to the bus and decoupled from the ground. }
\label{HyartMod}
\end{figure}

We now explain how to carry out the projective measurements needed for our error correction protocol.  We modify the experimental architecture from that shown in Fig.~\ref{SimpleHyart13} to that of Fig.~\ref{HyartMod}.  The braiding protocol is the same up to the minor change that the Coulomb couplings of islands 1 and 3 now depend on the magnetic flux tuned through two junctions: $\Delta_1(\Phi_A,\Phi_1),~\Delta_3(\Phi_B,\Phi_3)$  (recall that $\Delta_i$ couples the MZMs $\gamma_i$ and $\gamma_i'$ in Fig.~\ref{HyartMod}).  The essential feature that each $\Delta_i$ can be independently tuned between exponentially separated minimum and maximum values is unchanged (see Appendix~\ref{Modified} for more detail).   As before, we will write the maximum and minimum values of $|\Delta_i|$ as $\Delta$ and $0$, respectively.

The benefit of the geometry of Fig.~\ref{HyartMod} is that at each turning point the system can be turned into a top-transmon~\cite{Hassler11}, allowing for measurement of the parity of the ancillary pair of MZMs.  This is accomplished by decoupling the bus and the ground and connecting each MZM to either the bus or the ground.  Measurement returns the parity of the MZMs connected to the bus.

The table in Fig.~\ref{HyartMod} shows the necessary flux values that one must have to perform the measurements at each turning point.  In order to measure $i\gamma_1\gamma_0$, we couple island 1 to the bus and islands 2 and 3 to the ground.  Similarly, to measure $i\gamma_3\gamma_0$, we connect island 3 to the bus and islands 1 and 2 to the ground.  To measure $i\gamma_2\gamma_0$, we connect islands 1 and 3 to the bus and island 2 to the ground.  Assuming the total parity of the system is fixed, one can infer the parity of $i\gamma_2\gamma_0$ from this measurement. In Appendix~\ref{Modified}, we explain how this assumption can be relaxed and one can explicitly check the total parity by introducing additional superconducting islands, as shown in the structure of Fig.~\ref{fig:ModifiedTPCheck}.

In Section~\ref{sec:Measurement=Correction}, we discussed several approaches for correcting diabatic error by utilizing measurements: the forced measurement, hybrid, and dissipation-assisted hybrid protocols.   We emphasize that utilizing the forced measurement protocol in the architecture of Fig.~\ref{HyartMod} involves tuning fluxes in order to isolate different pairs of MZMs for subsequent measurements. We can implement the hybrid approach if we are able to flip the sign of $\Delta_i$ for each island independently.  This can be done if the induced charge on each island is independently controlled by external electrostatic gates, as flipping the sign of $\Delta_i$ corresponds to changing $q_i\to q_i\pm e$. Note that this swaps the ground and first excited states of the MZMs, but does not introduce electrons into the system and, therefore, does not affect the total fermion parity of the system.  We can also use the dissipation-assisted hybrid protocol, which does not require tuning the induced charge, if the system is coupled to a dissipative bath.

Consider the recovery step for the hybrid protocol for the architecture of Fig.~\ref{HyartMod}.  If we measure $s_1=-1$, we first need to change the sign of $\Delta_1$.  This is done by tuning $q_1\to q_1'=q_1\pm e$.  We then reverse the time evolution back to $H(0)$.  At this point, $\gamma_1$ is decoupled from the other MZMs, so that when we tune $q_1'\to q_1'\pm e$ it has no effect on the energy levels of the system.  This is in contrast to the initial change of $q_1$ to $q_1'$, which is intentionally done while $\gamma_1$ and $\gamma_1'$ are coupled, in order to swap the energy level of the occupied level from an excited state to a ground state.  We then evolve back to $H(t_1)$ and remeasure $s_1$.

Once the appropriate islands are coupled to the bus or ground, one measures the state of the system with the transmission line resonator (see Appendix~\ref{Measurement} for more detail).  The system with fixed parity, say even, has four energy levels: the ground and first excited state, separated by energy $\mathcal{O}(\Delta)$, and the second and third excited state, with energy $\mathcal{O}(E_M)$ above the ground state (energy subspaces are discussed in Appendix~\ref{Subspaces}).  For $\Delta>0$, at the first turning point the ground state and the second excited state correspond $i\gamma_1\gamma_0=+1$, while the first and third excited states correspond to $i\gamma_1\gamma_0=-1$~\cite{Hyart13}.  By sending a microwave through the transmission line resonator and measuring the shift in the resonant frequency, one can infer the parity of $i\gamma_1\gamma_0$.  If the system has remained in the lowest two energy states, one can also infer the parity of the qubit.

 Noise broadens the effective frequency of the resonator into a normal distribution, thus measurement will only distinguish the different parity states of $i\gamma_1\gamma_0$ provided the peak spacing is sufficiently larger than the width of the distributions. We obtain a rough estimate of the measurement time as follows: the difference in the resonator's effective frequencies determines the peak spacing of the distributions, which in turn sets an upper bound on the width (in frequency) of each distribution.  The uncertainty principle allows us to translate an upper bound on the width of the distribution to a lower bound on the measurement time.  

For the system under consideration, 
the measurement must resolve a frequency splitting $\mathcal{O}(\frac{g^2 \delta_+}{ \delta\omega^2})$, where $g$ is the coupling strength of the transmon to the resonator, $\delta\omega=\Omega_0-\omega_0$ is the detuning, $\Omega_0$ is the transmon frequency, and $\omega_0$ is the bare resonator frequency.  $\delta_+$ is the average dispersion of the transmon energy levels, see Appendix~\ref{Details} for an explicit definition. For the frequency estimates given in Ref.~\onlinecite{Hyart13}, this frequency splitting corresponds to a lower bound on the measurement time of
\begin{equation}
\label{eq:mst-time-bound}
t_{\text{meas}}\gg 20\,\text{ns}.
\end{equation}

Provided the experimental details of the resonator, we could calculate the photon transmission probability, $T_\pm$ corresponding to the parity states $i\gamma_1\gamma_0=\pm1$.  Let the probability that $N$ photons pass through the resonator during a measurement time $t_{\text{meas}}$ when the system is in the state $i\gamma_1\gamma_0=\pm 1$ be denoted $P(N, t_{\text{meas}}|i\gamma_1\gamma_0=\pm 1)$.  As described in Ref.~\onlinecite{Hyart13},  this probability distribution is Poissonian, and at long measurement times approaches a normal distribution:
\begin{equation}
P(N,t_{\text{meas}}|i\gamma_1\gamma_0=\pm1)=\text{Pois}(N,\lambda_{\pm})\approx \frac{e^{-\frac{(N-\lambda_{\pm})^2}{2\lambda_{\pm}}}}{\sqrt{2\pi \lambda_{\pm}}},
\end{equation}
where $\lambda_\pm \propto T_\pm t_{\text{meas}}$.  We see that the peak spacing between the distributions grows linearly in time, while the width of each distribution grows as a square-root in time.
Due to the finite overlap of the two possible distributions, there is some probability of incorrectly interpreting a measurement outcome; this separation error decreases exponentially with increasing measurement time.  Thus, we expect that a measurement time of $100$\,ns is sufficient to satisfy the bound in Eq.~(\ref{eq:mst-time-bound}).

Recall that $\gamma_0$ is a linear combination of $\gamma_1',\gamma_2',\gamma_3'$.  There are two other Majorana operators, composed of different linearly-independent combinations of $\gamma_1',\gamma_2',\gamma_3'$.  It is the parity of these additional two Majorana operators that determines whether or not the MZMs are in the low energy subspace.  These Majorana operators couple less strongly to the resonator and, thus, greater resolution is necessary to determine their parity.  In order to detect a transition to the second or third excited states, the measurement needs to resolve a frequency splitting $\mathcal{O}(\frac{g^2 \delta_+\delta_-}{ \delta\omega^2 E_M})$, where $\delta_-$ is half the difference of the dispersion of the transmon energy levels, see Appendix~\ref{Details}.  For this resolution, the lower bound on the measurement time increases to
\begin{equation}
t_{\text{meas}}\gg 1\,\mu\text{s}.
\end{equation}
Measurement details are discussed in Appendix~\ref{Measurement} and the bounds on the measurement time are derived in Appendix~\ref{Reality}.  As discussed in Appendix~\ref{Subspaces}, coupling the system to a cold bath can relax the second and third excited states  to the first excited state and the ground state respectively.  Thus, in the presence of a dissipative bath and with sufficiently slow evolution, the system will relax into the lowest two energy levels and the more precise measurement resolution is unnecessary.

\begin{figure*}
\centering
\includegraphics[scale=.6]{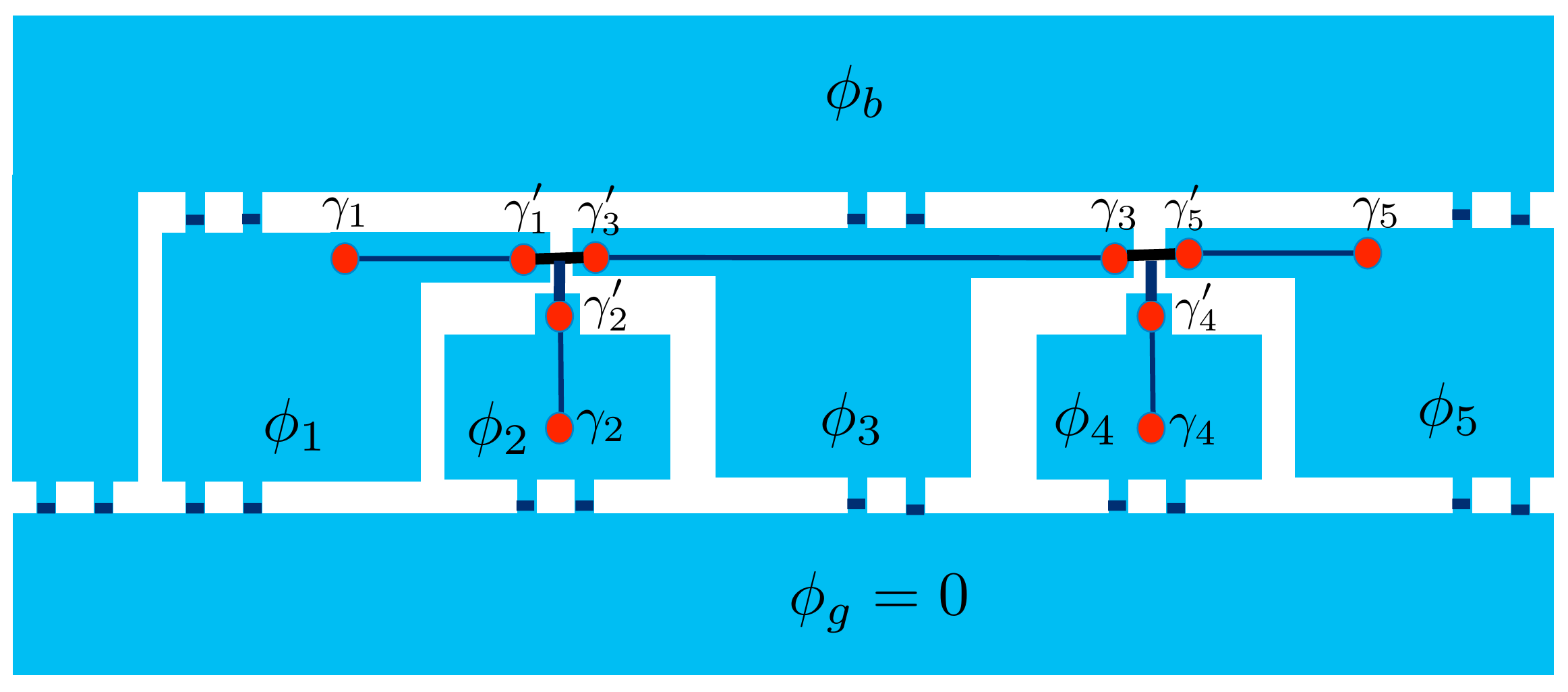}
\caption{A $\pi$-junction designed to allow fermion parity measurements at each turning point.}
\label{GG}
\end{figure*}

Finally, we note that we can easily generalize from a $T$-junction~\cite{Heck12}) to a $\pi$-junction~\cite{Hyart13} while still maintaining the ability to measure the parity at each corner of the braiding process.  This generalization is shown in Fig.~\ref{GG}.

\section{Feasibility Estimates}
\label{sec:feasibility}

The proposed platform for demonstrating non-Abelian braiding takes advantage of some well-established methods in superconducting qubit experiments.  In particular, careful control over the time evolution of the system can reduce diabatic errors and the measurement scheme used to read out the collective fermionic parity of MZMs can be used to detect and subsequently correct diabatic errors.  Moreover, the usual decoherence times, i.e. the relaxation time $T_1$ and decoherence time $T_2$, that plague superconducting qubits do not apply to the MZM based qubits and operations considered in this paper, since we simply want the transmon to remain in its ground state. However, the modifications presented here and in Ref.~\onlinecite{Hyart13} introduce new challenges, which we now address.

The time dependence of the combined MZM-transmon system enters entirely through the magnetic flux threaded through the split Josephson junctions.   As mentioned in Section~\ref{sec:general-theory}, setting time derivatives of the Hamiltonian to zero at the beginning and end of the evolution significantly decreases the diabatic error.  A benefit of using the transmon architecture is that control over the time evolution of the flux is excellent and current experiments can easily set $\dot{\Phi}(t)=0$ at the beginning and end of each time step~\cite{Barends14, Martinis14}.

Transmon experiments do not control the bias flux directly, but rather set the value of a digitally controlled voltage source for an external circuit, which induces a flux through the split Josephson juction via the mutual inductance~\cite{Koch07}.  Each additional split Josephson junction complicates the experiment due to unwanted cross-talk between the wrong bias circuit and junction.  There exist clever schemes to minimize the off-diagonal terms in the mutual inductance matrix through the geometry of the system.  It might be possible to achieve the topology of Fig.~\ref{HyartMod} using external circuits. However, doing so while maintaining the ability to independently tune the strength of each junction would undoubtedly be challenging.  A possible solution is to use a qubit which relies on voltage rather than magnetic field to tune the ratio of $E_J/E_C$~\cite{Larsen15, Lange15}.  This would eliminate the need for bias circuits while still retaining careful control over the time evolution.

The necessary energy resolution for the top-transmon is $\mathcal{O}(\frac{g^2 \delta_+}{\delta\omega^2}),$ while for the transmon it is $\mathcal{O}(\frac{g^2}{\delta\omega})$.  When operated strictly in the transmon regime, $E_J\gg E_C$, the required resolution for the top-transmon is orders of magnitude larger than for the transmon.  However, during measurement the top-transmon is tuned out of the transmon regime, and $\delta_+$ can be comparable to $\delta\omega$.  We expect measurement times for the top-transmon to be comparable to those of the transmon.

The hybrid approach for error correction relies on independently tuning the induced charge for each MZM island.  Such control can be achieved by gating each island and changing the gate voltage.  One can also avoid the additional complication of adding electrostatic gates by using other error correction schemes, such as the dissipation-assisted hybrid protocol, as described in Section~\ref{sec:Measurement=Correction}.

\section{Discussion}
\label{sec:discussion}

With the preceding analysis in hand, we are now in a position to answer the question posed in the title
of this paper. As we have shown, diabatic errors occur when anyons are unintentionally
created or move in an uncontrolled way. Such errors can be suppressed by making the time dependence of the Hamiltonian
as smooth as possible and by coupling to a dissipative bath.
They can be further reduced by measuring and correcting for the unwanted creation or motion
of anyons, which can be done without measuring the encoded quantum information that we wish to manipulate.  Let us suppose that we can tolerate a  probability $\varepsilon_0$ of a diabatic error per braiding step. The value of $\varepsilon_0$ will depend on the task we wish to accomplish and whether or not we hope to carry out a computation without additional error correction.  This error probability (assuming, for the moment, that there are no other sources of error, apart from diabatic errors) can be achieved by performing the unitary evolution slowly and smoothly. However, if the time required by this strategy exceeds the time needed for a measurement, then it may be advantageous to utilize a hybrid strategy that involves a faster ``nearly adiabatic'' evolution together with measurements that detect the occurrence of errors from diabatic transitions.

We assume that the process of nearly adiabatically tuning the Hamiltonian between any two turning points is carried out with the same time $t_u$. We denote the diabatic transition error  probability associated with each segment of nearly adiabatic unitary evolution as $\varepsilon[t_u]$. In the case of no dissipation,
\begin{equation}
\label{eqn:error-rate}
\varepsilon[t_uf]=\frac{c[k]}{(\Delta t_u)^{2k+2}},\quad \text{for}\quad t_u > t_{\text{th}}
\end{equation}
where $t_{\text{th}}=(10 c[k])^{\frac{1}{2k+2}}/\Delta$ is the threshold time above which the transition probability is bounded by a power-law, see Fig.~\ref{fig:transition-rate-dissipation}.  Recall that $\Delta$ is the maximum Coulomb coupling between MZMs on the same island and that the gap separating the ground and first excited state is $\mathcal{O}(\Delta)$.    Here $k$ is the number of vanishing time derivatives at the beginning and end of the unitary evolution and $c[k]$ is some $k$-dependent constant.  If we can tolerate an error probability of $\varepsilon_0$ for one-third of the braiding process, the time needed for a braid with unitary evolution is
\begin{equation}
\label{eqn:braid-time-unitary}
t_{\text{op}}=\frac{3}{\Delta}\left( \frac{c[k]}{\varepsilon_0}\right)^{\frac{1}{2k+2}}.
\end{equation}
For $\varepsilon_0$ very small, this will become a slow process.  We could, instead, perform the evolution faster and correct errors using, for instance, the hybrid protocol discussed in Section~\ref{sec:Measurement=Correction}.  There is no benefit to performing the unitary evolution arbitrarily fast, since the likelihood of diabatic errors will be high and several measurements will be necessary.  If we perform the unitary evolution much more slowly than the measurement time, $t_{\text{meas}}$, then we fail to take advantage of the benefits of performing measurements.  This can be made more quantitative as follows.

Let $t_{\text{op}}/3$ be the total time needed to evolve the system between two turning points (one third of the total time for a braid), including possible diabatic transition error recovery steps.  As discussed in Section~\ref{sec:SC-measurement} this time will be divided between unitary evolution, measurement, and flipping the sign of the couplings by changing the induced charge on the MZM island of interest. We will assume that each segment of nearly adiabatic unitary evolution (from one corner point to the next) takes the same amount of time $t_u$ and that the time needed to flip the sign of the coupling on any given island takes the same amount of time $t_{\text{flip}}$. We also assume that measurement is a relatively slow process, i.e. $t_{\text{meas}} \gg t_u, t_{\text{flip}}$. For the moment we will ignore errors associated with measurement, transitions to the second and third MZM excited states, and flipping the sign of the couplings, we will address these concerns later.

One diabatic transition error recovery step involves the following sequence of processes, to be performed at the desired turning point, following a syndrome measurement of the ancillary pair of MZMs that detected a diabatic transition error: (1) flip the induced charge on the ancillas' island (say island $i$), (2) near adiabatically tune the Hamiltonian to the previous turning point, (3) flip the induced charge on island $i$, (4) near adiabatically tune the Hamiltonian to the desired turning point, and (5) perform a syndrome measurement on the ancillary pair of MZMs. Consequently, the time required to perform one recovery step is
\begin{equation}
t_{\text{rec}} = 2t_u+t_{\text{meas}}+2t_{\text{flip}}.
\end{equation}

The corresponding probability that process of evolving between two turning points will be completed with $n$ recovery steps (i.e. that the initial near adiabatic evolution and subsequent $n-1$ recovery attempts had a diabatic transition error, but the $n$th recovery process was successful), for $n\ge 1$, is
\begin{equation}
\label{eqn:prob-n}
p_n[t_u]=\varepsilon[t_u] \left[ 2\varepsilon[t_u] (1-\varepsilon[t_u])\right]^{n-1} (1-2\varepsilon[t_u]+2\varepsilon[t_u]^2),
\end{equation}
and, clearly, $p_0 = 1 - \varepsilon[t_u]$.  In Eq.~(\ref{eqn:prob-n}), the first factor of $\varepsilon[t_u]$ is the probability of a diabatic transition error on the initial attempt; each factor of $2\varepsilon[t_u] (1-\varepsilon[t_u])$ is the probability of a diabatic transition error occurring on one of the two near adiabatic evolution segments associated with a recovery step; and the final factor of $1-2\varepsilon[t_u]+2\varepsilon[t_u]^2$ is the probability of successfully completing one recovery step without a diabatic transition error (i.e. with either zero or two diabatic transitions occurring during the two near adiabatic evolution segments).

The average number of recovery steps needed to evolve between two turning points without error is thus
\begin{equation}
\label{eqn:avg-n}
\langle n[t_u] \rangle = \sum_{n=0}^{\infty} n  p_n[t_u] = \frac{\varepsilon[t_u]}{ 1-2\varepsilon[t_u]+2\varepsilon[t_u]^2}.
\end{equation}
Hence, the average time needed to evolve between two turning points with the diabatic transition errors corrected is
\begin{eqnarray}
\frac{ \langle t_{\text{op}}[t_u] \rangle }{3}&=& t_u +t_{\text{meas}}+\langle n[t_u] \rangle t_{\text{rec}} \notag \\
&=& t_u +t_{\text{meas}} + \frac{(2 t_u +t_{\text{meas}}+2t_{\text{flip}}) \varepsilon[t_u] }{ 1-2\varepsilon[t_u]+2\varepsilon[t_u]^2}
.
\label{eqn:avg-time-meas}
\end{eqnarray}
This average operation time is minimized by some optimal choice of the time $t_u$, subject to the constraint that $t_u>t_{\text{th}}$, which is straightforward to compute when the other quantities are specified.

We now apply this to the system discussed in Section~\ref{sec:Hyart-system}. Ref.~\onlinecite{Hyart13} estimates $\Delta\sim 10$\,GHz, for which we find $t_\text{meas} \gg 20$\,ns.  We satisfy this inequality by setting $t_{\text{meas}}=100$\,ns  (see discussion in Appendix~\ref{Reality}).  We use $c[0]=2.2$ and $c[1]=162.5$, obtained from the data shown in Fig.~\ref{fig:transition-rate-dissipation}.  As a rough approximation, we set $E_M=50$\,GHz and $t_{\text{flip}}=10/E_M=.2$\,ns. In Tables~\ref{table:hybrid} and \ref{table:unitary}, we compare the average operation time for a braid with error-correction to the time for a braid with nearly adiabatic unitary evolution and target error probabilities $\varepsilon_0=10^{-4}$, $10^{-6},$ and $10^{-8}$.

We can also consider the effects of dissipation, as discussed in Section~\ref{sec:dissipation}.  Fitting to the data shown in Fig.~\ref{fig:transition-rate-dissipation} with system-bath coupling $\lambda=0.01\Delta$, we see that the error probability for unitary evolution for $k=0$ is
\begin{equation}
\varepsilon_{k=0}[t_u]=\frac{0.52}{(\Delta t_u)^{1.97}}, \quad \text{ for }\varepsilon_{k=0}[t_u]<10^{-3}
\end{equation}
and for $k=1$ it is
\begin{equation}
\varepsilon_{k=1}[t_u] = \left \{ \begin{array}{cc} \frac{162.5}{(\Delta t_u)^{4}}, & 10^{-2}> \varepsilon_{k=1}>10^{-5}
\\  \frac{0.05}{(\Delta t_u)^{1.95}},  &\varepsilon_{k=1}<10^{-5}\end{array} \right .
\end{equation}
Using these expressions, we estimate the braiding times with dissipation for unitary evolution and for the hybrid error correction scheme in Tables~\ref{table:hybrid} and \ref{table:unitary}.

\begin{table}
\begin{centering}
\begin{tabular}{|c|c|c|c|c|}
\hline $k$ & $\lambda$ (diss.) & $\langle t_{\text{op}}\rangle$ & $t_u$ & $\langle n(t_u)\rangle$ \\ \hline
0 & 0 & 308\,ns & 1.7\,ns & 0.008 \\ \hline
1 & 0 & 305\,ns & 1.5\,ns & 0.004 \\ \hline
0 & 0.1\,GHz & 308\,ns & 2.4\,ns & 0.001 \\ \hline
1 & 0.1\,GHz & 306\,ns & 1.5\,ns & 0.004 \\ \hline
\end{tabular}
\par\end{centering}
\caption{Braiding time using the hybrid protocol for the system discussed in Section~\ref{sec:Hyart-system} with $\Delta$=10\,GHz  and temperature $T=0.001\Delta$.  The columns label: the smoothness of the time evolution of the system Hamiltonian ($k=0,1$); the system-bath coupling $\lambda=0$ (no dissipation) or $\lambda=0.01 \Delta$ (dissipation); the average braiding time, $\langle t_{\text{op}}\rangle$; the corresponding unitary time, $t_u$; and the average number of recovery steps needed to complete the braid, $\langle n(t_u)\rangle$.  The above values assume no measurement error and no error from transitioning to excited states with energy $\mathcal{O}(E_M)$ above the ground state.  We use the estimates $t_{\text{flip}}=0.2$\,ns and $t_{\text{meas}}=100$\,ns, and then choose $t_u$ to minimize Eq.~(\ref{eqn:avg-time-meas}), subject to the constraint that $t_u>t_{\text{th}}$.  }
\label{table:hybrid}
\end{table}

\begin{table}
\begin{centering}
\begin{tabular}{|c|c|c|c|c|}
\hline
$k$ & $\lambda$ (diss.) & $t_{\text{op}},~\varepsilon_0=10^{-4}$ & $t_{\text{op}},~\varepsilon_0=10^{-6}$& $t_{\text{op}},~\varepsilon_0=10^{-8}$ \\ \hline
0 & 0 & 45 ns & 450 ns & 4.5 $\mu$s\\ \hline
1 & 0 & 11 ns & 34 ns & 110 ns \\ \hline
0 & 0.1 GHz & 23 ns & 240 ns & 2.5 $\mu$s\\ \hline
1 & 0.1 GHz & 11 ns & 77 ns & 820 ns \\ \hline
\end{tabular}
\par\end{centering}
\caption{Braiding time using nearly adiabatic unitary evolution for the system discussed in Section~\ref{sec:Hyart-system} with $\Delta$=10 GHz and temperature $T=0.001\Delta$.  The first two columns label the smoothness of the time evolution of the system Hamiltonian ($k=0,1$) and whether the system-bath coupling is $\lambda=0$ (no dissipation) or $\lambda=0.01 \Delta$ (dissipation). The third, fourth, and fifth columns list the braiding time to reach a target error probability of $\varepsilon_0=10^{-4},$ $10^{-6}$, and $10^{-8}$ respectively, between two turning points.  A smaller target error probability increases the corresponding braiding time.}
\label{table:unitary}
\end{table}

Tables~\ref{table:hybrid} and \ref{table:unitary} give rough estimates of the braiding times for MZMs in a flux-tunable architecture.  We see that, if we use an error correcting protocol involving measurements, our braiding operation time is limited by the measurement time. When $t_u>t_{\text{th}}$, the initial syndrome measurement at each turning point has a high probability of finding the desired outcome and projecting the system into its ground state, so we only rarely need to implement the recovery procedure.  With error-correction, the times do not depend strongly on $k$ nor on whether the system is coupled to a dissipative bath.  For nearly adiabatic unitary evolution, there is a significant improvement in braiding time for $k=1$ compared to $k=0$.  As discussed in Section~\ref{sec:dissipation}, for small error probabilities, dissipation reduces the braiding time for unitary evolution if $k=0$, but not if $k=1$. Our analysis suggests that, for a target error probability of $\varepsilon_0=10^{-6}$, the braiding time for the hybrid protocol is comparable to the braiding time for unitary evolution when $k=0$, and is faster when the system is not coupled to a dissipative bath.  When $k=1$, unitary evolution is significantly faster than correcting error through measurement, both with and without a bath.  For a target error probability of $\varepsilon_0=10^{-8}$, the hybrid protocol is faster than unitary evolution unless $k=1$ and there is no system-bath coupling.  These comparisons neglect measurement error and diabatic transitions to the second and third MZM excited states.  Taking into account these errors could shift the crossover point at which the hybrid protocol becomes better than unitary evolution.

The above analysis applies when we restrict our attention to the low energy subspace.  Let $\varepsilon_M[t_u]$ be the error probability associated with transitions out of this subspace to excited states of energy $\mathcal{O}(E_M)$, associated with the states supported by the triples of MZMs at $T$-junction intersections, whose degeneracies are lifted by Majorana-Josephson coupling. As $E_M\gg \Delta$, $\varepsilon_{M}[t_u]$ is expected to be much smaller than $\varepsilon[t_u]$.  While including $\varepsilon_{M}[t_u]$ could increase the braiding time for unitary evolution, it will not greatly affect the average braiding time with error-correction (even if $\varepsilon_{M}[t_u]\approx \varepsilon[t_u]$, $\langle n[t_u] \rangle$ would remain close to zero and the dominant contribution to the braiding time would still be $t_{\text{meas}}$).  For our choice of $t_{\text{meas}}=100$\,ns, measurement does not distinguish the ground and third excited state, thus such a transition results in an error.  Increasing the measurement time would allow us to detect, and correct, such a transition.

The values in Tables~\ref{table:hybrid} and \ref{table:unitary} are subject to change given the experimental implementation.  In particular, $\Delta, E_M, t_{\text{meas}}, \lambda$, and $c[k]$ will depend significantly upon system details.  ($E_M$ is exponentially sensitive to the separation of MZMs at the center of the $T$-junction.)  We chose $t_{\text{flip}}=10/E_M$ to justify ignoring errors associated with flipping the sign of the couplings.  (Recall that when we exchange the ground state and first excited state, the only transitions that conserve total parity are between states whose energies are separated by a gap $\mathcal{O}(E_M)$.)  With more information about the physical system, $t_{\text{flip}}$ could be optimized to be as short as possible without inducing diabatic transitions.

Measurement error is another potential issue.  Generally, there will be some probability of the measurement projecting the ancillary pair onto an excited state (odd parity), while providing an erroneous readout indicating that the outcome is a ground state (even parity), or vice-versa. Such errors can typically be reduced by repeating the measurement to increase the level of confidence of the measurement, as we discuss in Appendix~\ref{Measurement}.  Nonetheless, it is useful to know how small measurement errors must be in order to safely ignore them in the preceding analysis. In Appendix~\ref{MeasurementError}, we show that we can ignore a measurement error probability of $\varepsilon_{\text{meas}}$ at the $n$th recovery step, provided that
\begin{equation}
\label{eqn:meas-error-ineq}
\varepsilon_{\text{meas}}\ll \text{min}_{n\in\mathbb{N}}\left(\varepsilon[t_u],\left( 2\varepsilon[t_u](1-\varepsilon[t_u])\right)^{ n[t_u]}\right).
\end{equation}

It is important to remember that while braiding MZMs can realize single-qubit Clifford gates, universal quantum computation requires additional gates, such as the two-qubit entangling gate CNOT and the single qubit $\pi/8$ phase gate. There are a number of proposals for how one might implement such additional gates for MZM systems that may be incorporated in the Majorana nanowire (and other) systems considered in this paper~\cite{Bravyi06,Bonderson10c,Sau10c,Clarke10,Jiang11,Bonderson11b,Hyart13,Clarke15,Karzig15a}. Since these implementations of the additional gates will likely possess undesirable error rates and utilize significantly different methods from those of braiding, they will require the use of different error correction protocols, such as magic-state distillation~\cite{Bravyi05}. We do not focus on this matter here and the errors introduced by these additional (non-braiding) gates are not taken into account in our analysis and Tables~\ref{table:hybrid} and \ref{table:unitary}. Ref.~\onlinecite{Aasen15b} discusses milestone experiments leading to MZM based quantum computing, including fusion rule detection, which is simpler to execute than braiding.  These experiments are susceptible to the same diabatic errors discussed in the present paper.  An interesting future direction is to extend our analysis to the systems discussed in these papers, thereby better understanding the role diabatic errors play in topological quantum computation.

Our measurement-based correction protocol focuses on diabatic transitions from the ground state to the first excited state of the MZM system.  For longer measurement time, it is also possible to detect transitions to the second and third MZM excited states, and one could generalize the hybrid protocol to correct these errors as well.  We do not take into account transitions above the superconducting gap.  Such excitations are especially dangerous as quasiparticles could braid with the MZMs in an uncontrolled manner.  Quasiparticle traps could potentially help with these errors, although perhaps the best strategy is to optimize parameters such that these excitations are extremely rare.  As our interest in this paper has been on diabatic effects, we do not address errors arising from thermally-excited quasiparticles. Such errors (analyzed, for instance, in Ref.~\onlinecite{Pedrocchi15}) can be reduced by maximizing $\beta\Delta$ and, possibly, by variations on the ideas discussed in the present paper.

The hybrid error-correction protocol, introduced in Section~\ref{sec:MZM-trijunction} for MZMs and in Section~\ref{sec:general-braiding} for general non-Abelian anyons, interpolates between braiding via adiabatic tuning of the couplings~\cite{Alicea11,Heck12, Hyart13} and measurement-only topological quantum computation (MOTQC)~\cite{Bonderson08b,Bonderson08c}. It uses nearly-adiabatic tuning of the couplings to generate a very high probability of the state being the desired (ground) state at each topological charge/fermion parity measurement step, subject to the constraint that this does not take too long.
If measurement returns the excited state, the hybrid scheme is used to converge exponentially to the desired result, albeit with the cost of slowing braiding down to the speed of a measurement, in addition to introducing energy dissipation and heating associated with measurement.

If the braiding operation time $t_{\text{op}}$ becomes too long when using nearly adiabatic evolution or our hybrid protocol, one might consider simply using the MOTQC scheme.  For the Majorana network discussed in this paper, we must tune the couplings between subsequent measurements in order to isolate different pairs of MZMs for measurement.  This tuning should be done as fast as possible without inducing transitions to higher excited states of energy $\mathcal{O}(E_M)$. Let $t_{\text{tun}}=10/E_M$ be the required time to tune couplings between subsequent measurements (note that while $t_{\text{tun}}$ applies to a different process than $t_{\text{flip}}$, both times are subject to the same constraints).  Then the average braiding operation time would be $\langle t_{\text{op}}\rangle_{\text{MOTQC}}=9 t_{\text{meas}}+6t_{\text{tun}}=901$ ns for our energy estimates.  This is slower than the hybrid protocol for the systems considered in detail in this paper, and hence not the preferred protocol. However, one might envision other system designs for which the MOTQC scheme yields the faster protocol.

 In analyzing the diabatic errors in anyon braiding, we have mainly focused on satisfying the lower bound on the operational time. However, as mentioned in the introduction, it is of crucial importance that the braiding time is sufficiently fast that the system does not resolve the ground state degeneracy splitting, which are inevitably present due to nonzero correlation length. The resulting upper bound on braiding time depends on the details of the system. For the system discussed in Section~\ref{sec:Hyart-system}, the wires hosting MZMs must be sufficiently long compared to the correlation length (coherence length) and we must be able to tune the magnetic fluxes sufficiently close to $\Phi_0/2$.  It is worth noting that, to the best of our knowledge, the degeneracy splitting of MZM wires in current experiments is too large for the time estimates given in Tables~\ref{table:hybrid} and \ref{table:unitary}. However, the exponential suppression of the degeneracy splitting as a function of $L/\xi$ indicates that only modest increases in the length of the wires and/or the energy gap (which decreases the correlation length) is necessary to obtain an upper time limit much larger than the braiding times estimated in this paper.  For the system of Ref.~\onlinecite{Albrecht16}, tripling the length of the longest wire to $4.5$\,$\mu$m is sufficient.


\acknowledgements

We thank A. Dunsworth, R. Lutchyn, D. Sank, and N. Wiebe for useful discussions and J. Alicea, D. Clarke, S. Flammia, and T. Karzig for comments on a draft of this paper.
C.K. acknowledges support by the National Science Foundation Graduate Research Fellowship Program under Grant No. DGE 1144085.

\appendix

\section{Review of the Landau-Zener Effect}
\label{sec:Landau-Zener}

We now derive Eq.~(\ref{eqn:S1-LZ-sudden}) in the main text.  The Hamiltonian of Eq.~(\ref{eqn:piece-wise-model}) between $-\tau<t<\tau$ is
\begin{equation}
H=ct\sigma_z-\lambda\sigma_x.
\end{equation}
We assume $c>0$.  Write the wave function as $\psi=(a(t),b(t))^T$, the Schroedinger equation reads
\begin{equation}
\begin{split}
i\dot{a}&=ct a-\lambda b
\\ i\dot{b}&=-\lambda a -ct b.
\end{split}
\end{equation}
We can eliminate $a$ to obtain the differential equation for the evolution of $b$:
\begin{equation}
\ddot{b}=icb-(\lambda^2+c^2 t^2)b.
\label{eqn:b-diff}
\end{equation}

Define
\begin{equation}
z=\sqrt{2c}e^{-i\frac{\pi}{4}}t, \quad n =\frac{i\lambda^2}{2c}, \quad \Lambda=\frac{\lambda^2}{c}.
\end{equation}
Eq. (\ref{eqn:b-diff}) becomes the Weber equation
\begin{equation}
\frac{d^2 b}{dz^2}+\left( n+\frac{1}{2}-\frac{z^2}{4}\right)b=0,
\end{equation}
with linearly-independent solutions $D_n(-z)$ and $D_{-n-1}(-iz)$, where $D$ is the parabolic cylinder function.  Therefore, for $t>-\tau$ the solution can be written as
\begin{equation}
b(t)=\alpha D_n(-z)+\beta D_{-n-1}(-iz).
\end{equation}

We consider $\tau$ large, which means $\sqrt{c}\tau\gg 1$.  $D_\mu(z)$ has the following asymptotics for $|z|\gg 1$ \cite{Suominen90}:
\begin{widetext}
\begin{eqnarray}
\begin{split}
D_\mu (z) &\sim e^{-\frac{z^2}{4}}z^\mu\left[ 1+\mathcal{O}(z^{-2}) \right], &|\arg z|<\frac{3\pi}{4}
\\ D_\mu (z) & \sim e^{-\frac{z^2}{4}}z^\mu\left[ 1+\mathcal{O}(z^{-2})  \right]
+\frac{\sqrt{2\pi}}{\Gamma(-\mu)}e^{i\mu \pi}e^{\frac{z^2}{4}} z^{-\mu-1}\left[ 1+\mathcal{O}(z^{-2}) \right],
& \frac{\pi}{4}<\arg z<\frac{5\pi}{4}
\\ D_\mu (z) & \sim e^{-\frac{z^2}{4}}z^\mu\left[ 1+\mathcal{O}(z^{-2})  \right]
-\frac{\sqrt{2\pi}}{\Gamma(-\mu)}e^{-i\mu \pi}e^{\frac{z^2}{4}} z^{-\mu-1}\left[ 1+\mathcal{O}(z^{-2}) \right],
&  -\frac{\pi}{4}>\arg z>-\frac{5\pi}{4}.
\end{split}
\label{eqn:asymptotics}
\end{eqnarray}
\end{widetext}
Define $\Phi(t)=\frac{c|t|^2}{2}+\frac{\Lambda}{2}\ln|\sqrt{2c}t|$.
Note that
\begin{equation}
\begin{split}
\arg(- z(t<0))&= -\frac{\pi}{4}
\\ \arg( -iz(t<0))&=\frac{\pi}{4}
\\ \arg( -z(t>0))&= \frac{3\pi}{4}
\\ \arg(-iz(t>0))&= -\frac{3\pi}{4}.
\end{split}
\end{equation}
Therefore, using the appropriate expression in Eq. (\ref{eqn:asymptotics}) we find for $t<0$
\begin{equation}
\begin{split}
D_n(-z) &\sim e^{\frac{\pi \Lambda}{8}+i\Phi}
\\ D_{-n-1}(-iz) & \sim \frac{e^{\frac{\pi \Lambda}{8}-i\frac{\pi}{4} -i\Phi}}{\sqrt{2c} |t|}
\\ \dot{D}_n(-z) &\sim ic e^{\frac{\pi \Lambda}{8}+i\Phi}t
\\ \dot{D}_{-n-1}(-iz) &\sim \sqrt{\frac{c}{2}}e^{\frac{\pi \Lambda}{8}+i\frac{\pi}{4}-i\Phi}
\end{split}
\end{equation}
and for $t>0$
\begin{equation}
\begin{split}
D_n(-z) &\sim e^{-\frac{3\pi \Lambda}{8}+i\Phi}-\sqrt{\frac{\pi}{c}}\frac{1}{\Gamma(-n)t}e^{-\frac{\pi \Lambda}{8}-i\Phi+i\frac{\pi}{4}}
\\ D_{-n-1}(-iz) &\sim \frac{e^{-\frac{3\pi \Lambda}{8}+\frac{3\pi i }{4} -i\Phi}}{\sqrt{2c}t}+\frac{\sqrt{2\pi}}{\Gamma(n+1)}e^{-\frac{\pi \Lambda}{8}} e^{i\Phi}.
\end{split}
\end{equation}

By matching initial conditions at $t=-\tau$ we find that
\begin{equation}
\begin{split}
\alpha &= e^{-i\Phi-\frac{\pi \Lambda}{8}}\left( b_0 -\frac{\lambda}{2c\tau}a_0\right),
\\ \beta &= \frac{\lambda a_0}{\sqrt{2c}}e^{-\frac{\pi \Lambda}{8}+i\frac{\pi}{4}+i\Phi}.
\end{split}
\end{equation}
Therefore, at a much later time $t=\Top>0$ we have
\begin{equation}
\begin{split}
b(T) &\sim \left( b_0 -\frac{\lambda}{2c\tau}a_0\right)\left( e^{-\frac{\pi\Lambda}{2}}-\sqrt{\frac{\pi}{c}}\frac{e^{-\frac{\pi \Lambda}{4}-2i\Phi+i\frac{\pi}{4}}}{\Gamma(-n)T}\right)
\\ &~+\lambda a_0 \left( \sqrt{\frac{\pi}{c}} \frac{e^{-\frac{\pi \Lambda}{4}+i\frac{\pi}{4}+2i\Phi}}{\Gamma(n+1)}-\frac{e^{-\frac{\pi\Lambda}{2}}}{2cT}\right).
\end{split}
\end{equation}

Similarly, the differential equation for $a$ can be written as a Weber differential equation
\begin{equation}
\frac{d^2 a}{dw^2}+\left( m+\frac{1}{2}-\frac{w^2}{4}\right)a =0
\end{equation}
for $w=\sqrt{2c} e^{i\frac{\pi}{4}}t, ~ m=-\frac{i\Lambda}{2}$.  We can write $a(t)=\alpha' D_m(-w)+\beta' D_{-m-1}(iw)$ (note that $D_m(\pm w)$ satisfy the same Weber's equation and are linearly independents from $D_{-m-1}(\pm iw)$, which also satisfy the same Weber's equation).
We see $\arg -w=-\pi/4$ and $\arg iw=\pi/4$, therefore for $t<0$
\begin{equation}
\begin{split}
D_m(-w)&\sim e^{\frac{\pi\Lambda}{8}-i\Phi}
\\ D_{-m-1}(iw) &\sim \frac{e^{\frac{\pi \Lambda}{8}+i\frac{\pi}{4}+i\Phi}}{\sqrt{2c}|t|}
\\ \dot{D}_m(-w) &\sim -ict e^{\frac{\pi \Lambda}{8}-i\Phi}
\\ \dot{D}_{-m-1}(iw) &\sim \sqrt{\frac{c}{2}}e^{\frac{\pi \Lambda}{8}-i\frac{\pi}{4}+i\Phi}.
\end{split}
\end{equation}
As $\arg(-w(t>0))=\frac{-3\pi}{4}$ and $\arg(iw(t>0))=\frac{3\pi}{4}$, we have for $t>0$
\begin{equation}
\begin{split}
D_m(-w) &\sim e^{-i\Phi}e^{-3\pi\Lambda/8}-\sqrt{\frac{\pi}{c}}\frac{e^{-\pi\Lambda/8+i3\pi/4}e^{i\Phi}}{\Gamma(-m)t}
\\ D_{-m-1}(iw) &\sim \frac{e^{-\frac{3\pi\Lambda}{8}-\frac{3\pi i}{4}+i\Phi}}{\sqrt{2c}t}-\frac{\sqrt{2\pi}}{\Gamma(m+1)}e^{-\frac{\pi \Lambda}{8}}e^{-i\Phi}.
\end{split}
\end{equation}
By matching boundary conditions at $t=-\tau$ we find
\begin{equation}
\begin{split}
e^{\pi \Lambda/8} (e^{-i\Phi}\alpha'+\frac{e^{i\pi/4+i\Phi} }{\sqrt{2c}\tau}\beta') &= a_0
\\ e^{\pi \Lambda/8}(e^{-i\Phi}\alpha' -\frac{e^{i\pi/4+i\Phi}}{\sqrt{2c}\tau}\beta') &= a_0 +\frac{\lambda}{c\tau}b_0
\end{split}
\end{equation}
thus we find
\begin{equation}
\begin{split}
\alpha'&= e^{-\pi\Lambda/8+i\Phi}(a_0+\frac{\lambda}{2c\tau}b_0)
\\ \beta' &= -e^{-\pi \Lambda/8-i\pi/4-i\Phi}\frac{\lambda}{\sqrt{2c}}b_0.
\end{split}
\end{equation}

Therefore, we have
\begin{equation}
\begin{split}
&a(\Top) = \alpha' D_m(-w)+\beta' D_{-m-1}(iw)
\\ &= a_0  e^{-\pi \Lambda/8+i\Phi}D_m(-w )
\\ &+b_0 \frac{e^{-\pi \Lambda/8}\lambda }{\sqrt{2c}} \left(e^{-i\pi/4-i\Phi} D_{-m-1}(iw) -\frac{e^{i\Phi}}{\sqrt{2c}\tau}D_m(-w)\right)
\\ &b(\Top) = \alpha D_n(-z)+\beta D_{-n-1}(-iz)
\\ &=a_0 \frac{e^{-\pi\Lambda/8}\lambda}{\sqrt{2c}}\left( \frac{e^{-i\Phi}}{\sqrt{2c}\tau}D_n(-z)+e^{-i\pi/4+i\Phi} D_{-n-1}(-iz)\right)
\\ &+ b_0 e^{-\pi \Lambda/8-i\Phi}D_n(-z)
\end{split}
\end{equation}
where in the above every $w, z,$ and $\Phi$ is evaluated at $t=\Top>0$.
It follows that at time $\Top$
\begin{equation}
\begin{split}
S_{1} &= e^{-\pi\Lambda /8+i\Phi}D_m(-w)
\\ &= e^{-\pi\Lambda/2}-\sqrt{\frac{\pi}{c}}\frac{e^{-\pi\Lambda/4+3\pi i /4+2i\Phi}}{\Gamma(-m)\Top}
\\ S_2 &=- \frac{e^{-\pi \Lambda/8}\lambda}{\sqrt{2c}} \left(e^{-i\pi/4-i\Phi} D_{-m-1}(iw) -\frac{e^{i\Phi}}{\sqrt{2c}\tau}D_m(-w)\right)
\\ &=\frac{e^{-\pi\Lambda/2}\lambda}{c}\left( \frac{1}{\Top}+\frac{1}{\tau}\right)
\\ &+\sqrt{\frac{\pi}{c}}e^{-\pi\Lambda/4-i\pi/4}\lambda\left( \frac{e^{i2\Phi}}{2c\tau \Top\Gamma(-m)}+\frac{e^{-2i\Phi}}{\Gamma(m+1)}\right)
\\ &\approx \sqrt{\frac{\pi}{c}}\frac{e^{-i\pi/4-\pi\Lambda/4-2i\Phi}\lambda}{\Gamma(m+1)}.
\end{split}
\end{equation}
In the last line we keep only the leading order term, noting that $ \tau\gg 1$ and discarding the term $\mathcal{O}(e^{-\pi\Lambda/2})$.

The transition amplitude is given by $|S_1|^2$:
\begin{equation}
|S_1|^2 =\frac{\pi}{c} \frac{e^{-\pi\Lambda/2}}{|\Gamma(-m) t_{\text{op}}|^2}.
\end{equation}
Note that
\begin{equation}
|\Gamma(m)|=\sqrt{\frac{\pi}{|m|\text{sinh}(\pi|m|)}},
\end{equation}
therefore
\begin{equation}
|S_1|^2 =\frac{\Lambda}{4c t_{\text{op}}^2} +\mathcal{O}\Big(e^{-\pi\Lambda},\frac{e^{-\pi\Lambda/2}}{\sqrt{c t_{\text{op}}}}\Big).
\end{equation}

\section{Mapping of the Braiding of MZMs to the Landau-Zener Problem}
\label{sec:MZM-LZ-details}

In Section~\ref{sec:MZM-LZ}, we showed how the first step in the braiding protocol
at a $T$-junction could be mapped onto the Landau-Zener problem with sudden turn-on/off. Here, we perform this mapping for the other two steps.

Consider the time period $t_1<t\leq t_2$. We find that:
\begin{equation}
M_2 H M_2^\dagger = \frac{\Delta}{\sqrt{2}}\left( h_2(t)\sigma_z-\sigma_x\right)
\label{Map2}
\end{equation}
where $h_2(t)=\frac{6 t}{t_{\text{op}}}-3$ and $M_2$ is the unitary matrix
\begin{equation}
	M_2=\frac{1}{2}\Big[ -(2+\sqrt{2})^{1/2}\sigma_z +(2-\sqrt{2})^{1/2}\sigma_x\Big].
\end{equation}

Meanwhile, for $t_2<t\leq t_3$
\begin{equation}
M_3 H M_3^\dagger = \frac{\Delta}{\sqrt{2}}\left(h_3(t)\sigma_z-\sigma_x \right)
\label{Map3}
\end{equation}
where $h_3(t)=\frac{6 t}{t_{\text{op}}}-5$ and $M_3$ is the unitary matrix
\begin{equation}
	M_3= \frac{1}{2^{5/4}}(\sigma_x+\sigma_y)\big[(\sqrt{2}-1)^{1/2}\sigma_y -(\sqrt{2}+1)^{1/2}\sigma_z\big].
\end{equation}

Eqs.~(\ref{Map2}) and (\ref{Map3}) are both of the form of the Landau-Zener hamiltonian, therefore we see that each step of the MZM braiding process with linear couplings can be mapped to the Landau-Zener Hamiltonian with sudden turn-on/off of off diagonal couplings.


\section{Master Equation Formalism for Time-Dependent Hamiltonians Coupled to a Bath}
\label{sec:master-equation}

In this Appendix, we give a derivation of the master equation for a time-dependent Hamiltonian
coupled to a bath. We begin with a general system-bath Hamiltonian
\begin{equation}
H(t)=H_{S}(t)+H_{B}+H_{I}(t),
\end{equation}
where $H_{S}(t)$ is the time-dependent system Hamiltonian and $H_{B}$ is
the bath Hamiltonian (e.g. a set of harmonic oscillators).  The interaction between the system and the bath can be written in the general form
\begin{equation}
H_{I}(t)=\sum_{\alpha}g_{\alpha}(t)\, A_{\alpha}\otimes B_{\alpha},
\end{equation}
where the operator $A_{\alpha}$ is a Hermitian operator acting only on the
degrees of freedom of the system and $B_{\alpha}$ is a Hermitian operator acting only on the
degrees of freedom of the bath. For
a time-independent system-bath coupling, $g(t)=g$, a derivation of the master
equation is given in Ref.~\onlinecite{Albash12}.
Here, we generalize the formalism to a time-dependent system-bath
coupling in order to derive the results in Section~\ref{sec:dissipation}.

Consider the reduced density matrix $\tilde{\rho}_{S}(t)={\rm Tr_{B}}\tilde{\rho}(t)$
in the interaction picture:
\begin{equation}
\tilde{\rho}(t)=U_{0}^{\dagger}(t,0)\rho(t)U_{0}(t,0),
\end{equation}
with $U_{0}(t,t')=U_{S}(t,t')\otimes U_{B}(t,t')$ and
\begin{equation}
\begin{split}
U_{S}(t,t')&=\mathcal{T}\exp[-i\int_{t'}^{t}d\tau H_{S}(\tau)]
\\ U_{B}(t,t')&=\exp[-i(t-t')H_{B}].
\end{split}
\end{equation}
The system-bath Hamiltonian
can be written in the interaction picture as
\begin{equation}
\begin{split}
\tilde{H}_{I}(t) & =  U_{0}^{\dagger}(t,0)H_{I}U_{0}(t,0)\\
 & =  \sum_{\alpha}g_{\alpha}(t)\, U_{S}^{\dagger}(t,0)A_{\alpha}U_{S}(t,0)\otimes U_{B}^{\dagger}(t,0)B_{\alpha}U_{B}(t,0)\\
 & =  \sum_{\alpha}g_{\alpha}(t)\, A_{\alpha}(t)\otimes B_{\alpha}(t).
\end{split}
\end{equation}
Applying the standard Born approximation,
\begin{equation}
 \rho(t)\approx\rho_{S}(t)\otimes\rho_{B},
\end{equation}
which assumes that the system-bath coupling is small and the influence of
the system on the bath is weak, we obtain
\begin{equation}
\frac{d\tilde{\rho}_{S}(t)}{dt}=-\int_{0}^{t}d\tau \,\text{Tr}_{B}\Big[\tilde{H}_{I}(t),\;\Big[\tilde{H}_{I}(t-\tau),\;\tilde{\rho}_{S}(t-\tau)\otimes\rho_{B}\Big]\Big].
\end{equation}
We next use the Markov approximation, that is, we replace $\tilde{\rho}_{S}(t-\tau)$
by $\tilde{\rho}_{S}(t)$ and let the upper limit of the integral go to infinity.  This is valid for $g_{\alpha}\ll1/\tau_{B}$.
Here, $\tau_{B}$ is the correlation time of the bath:
\begin{equation}
\begin{split}
\mathcal{B}_{\alpha\beta}(t)={\rm Tr}\big[B_{\alpha}(t)B_{\beta}(0)\rho_{B}\big]
\sim\exp(-t/\tau_{B}).
\end{split}
\end{equation}
The master equation becomes
\begin{multline}
\label{eqn:early-master-eqn}
\frac{d\tilde{\rho}_{S}(t)}{dt}=\sum_{\alpha\beta}\int_{0}^{\infty}d\tau\, g_{\alpha}(t)g_{\beta}(t-\tau)\,\times\\
\Big[\Big(A_{\beta}(t-\tau)\tilde{\rho}_{S}(t)A_{\alpha}(t)-A_{\alpha}(t)A_{\beta}(t-\tau)\tilde{\rho}_{S}(t)\Big)\mathcal{B}_{\alpha\beta}(\tau)\\
+h.c.\Big].
\end{multline}
In this expression, $A_{\beta}(t)$ is given by:
\begin{equation}
A_{\alpha}(t) =  U_{S}(t,0)^{\dagger}A_{\beta}U_{S}(t,0).
\end{equation}

In general, the right-hand-side of the master
equation includes: (1) the unitary evolution superoperator (which, in
the Schr\"odinger picture, takes the form $-i[H_{S}(t)+H_{LS},*]$ where
$H_{LS}$ is the Lamb shift and $*$ refers to any operator), (2) a dissipative superoperator due to purely adiabatic
processes, and (3) a dissipative superoperator due to diabatic
corrections.

Eq.~(\ref{eqn:early-master-eqn}) requires us to perform an integral over $\tau$ for the time-evolution operator
$U_{S}(t-\tau,0)$, which is very difficult. In order
to avoid this, we make a further simplifying approximation \cite{Albash12},
\begin{equation}
\label{eqn:AlbashApprox}
U_{S}(t-\tau,0)=U_{S}^{\dagger}(t,t-\tau)U_{S}(t,0)\approx e^{i\tau H_{S}(t)}U_{S}(t,0).
\end{equation}
Eq.~(\ref{eqn:AlbashApprox}) is justified by the smallness of the bath correlation time $\tau$: $H_S(t)$ is almost a constant over a time $\tau$ due to the rapid decay of $\mathcal{B}_{\alpha\beta}(\tau)$ as a function of $\tau$.
Using this approximation we find
\begin{multline}
\int_{0}^{\infty}d\tau\, g_{\alpha}(t)g_{\beta}(t-\tau)A_{\beta}(t-\tau)\tilde{\rho}_{S}(t)A_{\alpha}(t)\mathcal{B}_{\alpha\beta}(\tau) =\\
\int_{0}^{\infty}d\tau\, g_{\alpha}(t)g_{\beta}(t-\tau)U_{S}(t,0)^{\dagger}e^{-i\tau H_{S}(t)}A_{\beta}e^{i\tau H_{S}(t)}U_{S}(t,0)\\
\times \tilde{\rho}_{S}(t)U_{S}(t,0)^{\dagger}A_{\beta}U_{S}(t,0)\mathcal{B}_{\alpha\beta}(\tau).
\end{multline}
Returning to the Schr\"odinger picture density matrix, $\rho(t)$, via
$\tilde{\rho}_{S}(t)=U_{S}(t,0)^{\dagger}\rho(t)U_{S}(t,0)$,
all the $U_{S}(t,0)$ terms will cancel.
We insert a resolution of the identity
$\mathbb{1}=\sum_{\alpha}|\epsilon_{\alpha}(t)\rangle\langle\epsilon_{\alpha}(t)|$
in the instantaneous eigenbasis ${|\epsilon_{a}(t)\rangle}$ of $H_{S}(t)$,
which has  instantaneous eigenvalues ${\epsilon_{a}(t)}$ according to
$H_{S}(t)|\epsilon_{\alpha}(t)\rangle=\epsilon_{\alpha}(t)|\epsilon_{\alpha}(t)\rangle$.
We obtain
\begin{multline}
\label{eqn:MasterEquation}
\frac{d\rho_{S}(t)}{dt}=-i[H_{S}(t)+H_{LS}(t),\rho_{S}(t)]+\\
\sum_{\alpha\beta}\sum_{\omega}\gamma_{\alpha\beta}(t,\omega)\Big[A_{\beta,\omega}(t)\rho_{S}(t)A_{\alpha,\omega}(t)^{\dagger} \\-\frac{1}{2}\{A_{\alpha,\omega}(t)^{\dagger}A_{\beta,\omega}(t),\;\rho_{S}(t)\}_{+}\Big].
\end{multline}
The Lamb shift is
\begin{equation}
H_{LS}(t)=\sum_{\alpha\beta}\sum_{\omega}A_{\alpha,\omega}(t)^{\dagger}A_{\beta,\omega}(t)S_{\alpha\beta}(t,\omega).
\end{equation}
It is conventional to combine $\gamma_{\alpha\beta}(t,\omega)$ and $S_{\alpha\beta}(t,\omega)$ into
the noise function $\Gamma_{\alpha\beta}(t,\omega) = \frac{1}{2}\gamma_{\alpha\beta}(t,\omega)+iS_{\alpha\beta}(t,\omega)$.  In the present calculation,
\begin{equation}
\Gamma_{\alpha\beta}(t,\omega) =
\int_{0}^{\infty}d\tau\, g_{\alpha}(t)g_{\beta}(t-\tau)e^{i\tau\omega}\mathcal{B}_{\alpha\beta}(\tau).
\end{equation}
$A_{\beta,\omega}(t)$ is
\begin{equation}
A_{\beta,\omega}(t)=\sum_{\omega_{ba}(t)=\omega}\langle\epsilon_{a}(t)|A_{\beta}|\epsilon_{b}(t)\rangle|\epsilon_{a}(t)\rangle\langle\epsilon_{b}(t)|,
\end{equation}
where $\omega_{ba}(t)=\epsilon_{b}(t)-\epsilon_{a}(t)$.

\section{Numerical Solution of the Master Equation for a $T$-junction Coupled to a Dissipative Bath}
\label{sec:dissipation-details}

Let us first consider diabatic corrections without dissipation, which in the language of Appendix \ref{sec:master-equation} means we set $g_\alpha=0$.  After solving the Heisenberg equation of motion, we obtain the reduced density matrix for the system at the end of the braiding process, $\rho_S(t_{\text{op}})$.
We focus on two quantities.  The first is the transition probability from the ground state (computational basis) to the excited state (non-computational subspace) for the even parity sector, $P_{G\to E}=2\rho_S^{22}(t_{\text{op}})$.  The second quantity, $||\rho_{\text{G}}(t_{\text{op}})-\rho_A||$, quantifies the deviation of the relative phase phase from its adiabatic value of $\pi/2$ if the system has remained in the computational subspace.  That is, we project the system into the computational subspace and find the trace norm, denoted $||...||$, of the difference between the projected density matrix, $\rho_{\text{G}}(t_{\text{op}})$, and the adiabatic density matrix, $\rho_A$.
As the two $\tau_z=\pm 1$ sectors are decoupled (fermion parity is conserved), measuring the system at the end of the braiding process to be in the computational subspace is equivalent to applying a phase gate to this subspace.

We consider time-dependent Hamiltonians with $k=0,1$ derivatives vanishing at $t=0, t_1, t_2, t_3$.  Diabatic corrections should vanish as  ${\cal O}({t_\text{op}}^{-2k-2})$, given the results of Refs.~\onlinecite{Lidar09,Wiebe12,Lidar15},

\begin{figure}
\begin{centering}
\includegraphics[scale=0.3]{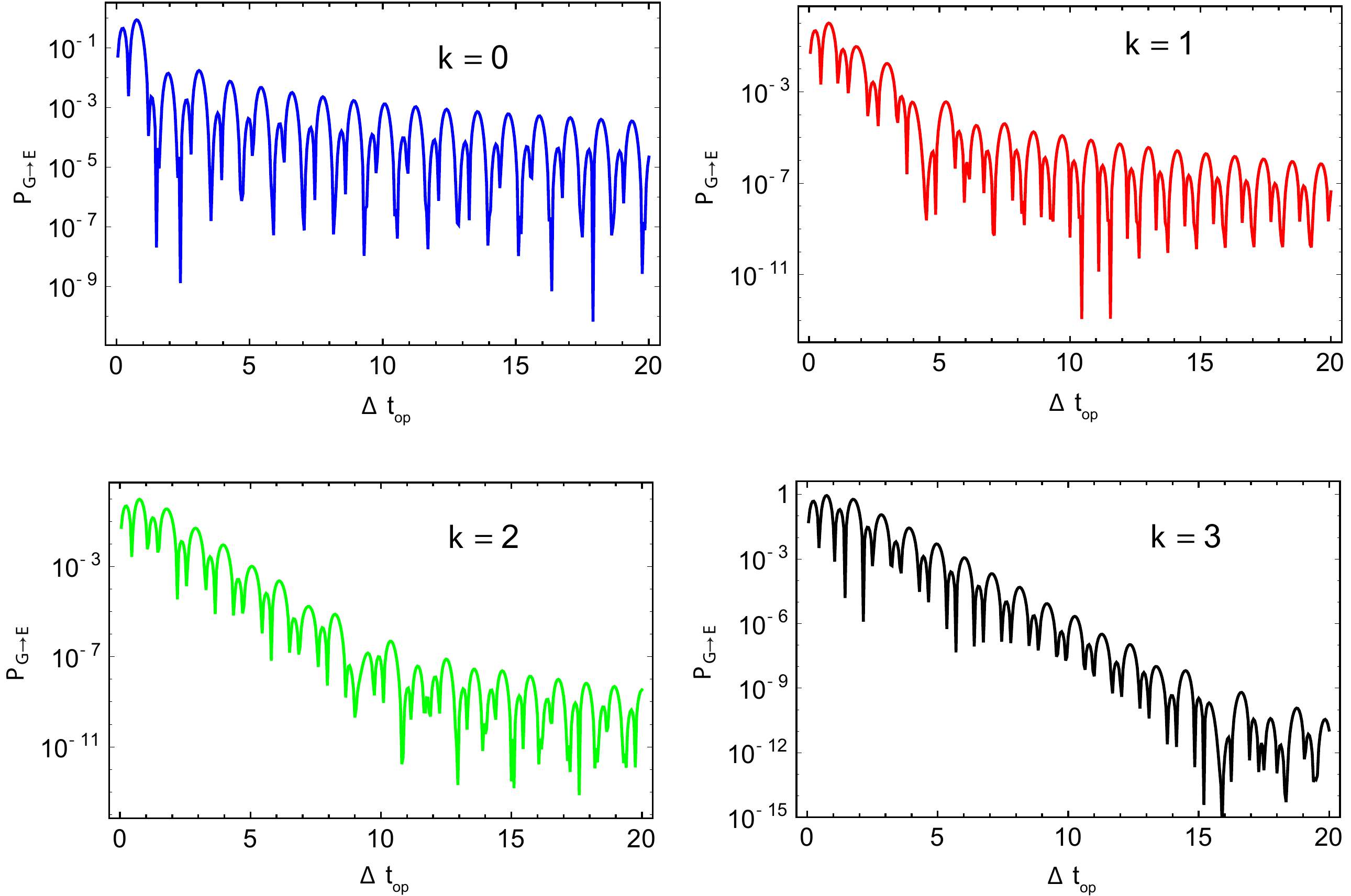}
\end{centering}
\caption{Transition probability, $P_{G\rightarrow E}$, versus braiding period, $t_{\text{op}}$, without dissipation.  $k=0,1,2,3$ refers to the number of vanishing time derivatives of the Hamiltonian at each turning point.}
\label{fig:transprob-no-diss}
\end{figure}

\begin{figure}
\begin{centering}
\includegraphics[scale=0.37]{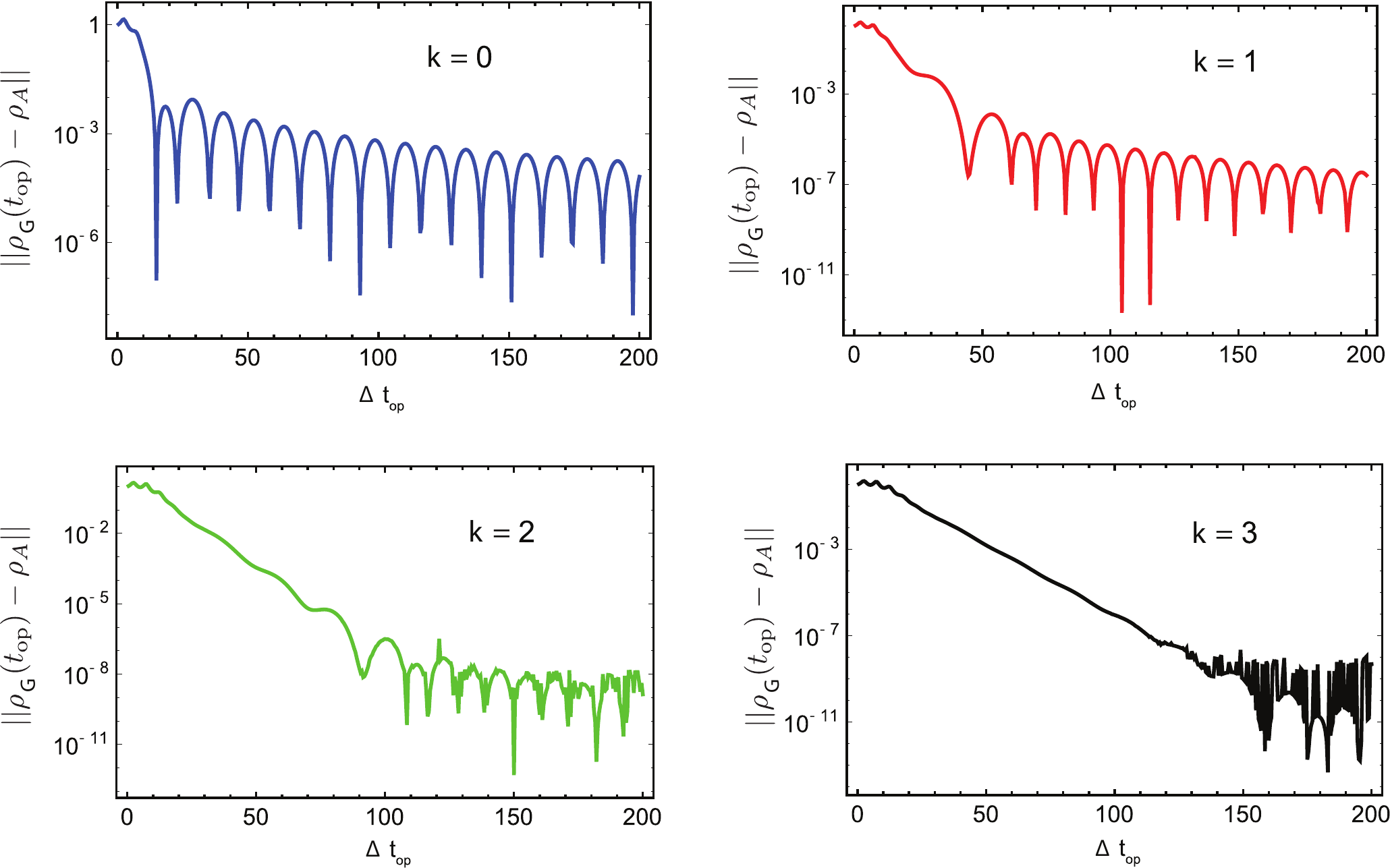}
\end{centering}
\caption{Phase error, $||\rho_{\text{G}}(t_{\text{op}})-\rho_A||$, against braiding period, $t_{\text{op}}$, without dissipation.  $k=0,1,2,3,4$ refers to the number of vanishing time derivatives of the Hamiltonian at each turning point. }
\label{fig:phaseerror-no-diss}
\end{figure}

We choose the following braiding protocol (see Fig.~\ref{fig:T-junction-braiding}):
\begin{equation}
\Delta_{1}(k,\; t)=\begin{cases}
\Delta\theta\left(k,\frac{t}{t_1}\right), & 0\leq t<{t_1}\\
\Delta\left(1-\text{\ensuremath{\theta}}\left(k,\;\frac{\left(t-t_{1}\right)}{t_1}\right)\right), & t_{1}\leq t< t_{2}\\
0. &{t_2}\leq t\leq {t_3}
\end{cases}
\end{equation}
\begin{equation}
\Delta_{2}(k,\; t)=\begin{cases}
\Delta\left(1-\text{\ensuremath{\theta}}\left(k,\frac{t}{t_1}\right)\right), &
0\leq t<{t_1}\\
0, &  {t_1}\leq t< {t_2}\\
\Delta\theta\left(k,\frac{t-t_2}{t_3 - t_2}\right), & {t_2}\leq t\leq {t_3}
\end{cases}
\end{equation}
\begin{equation}
\Delta_{3}(k,\; t)=\begin{cases}
0, & 0\leq t< {t_1}\\
\Delta\theta\left(k,\frac{t-{t_1}}{t_1}\right), & {t_1}\leq t<{t_2}\\
\Delta\left(1-\text{\ensuremath{\theta}}\left(k,\frac{t-{t_2}}{t_3 - t_2}\right)\right), &  {t_2}\leq t\leq {t_3}
\end{cases}
\end{equation}
where $\theta(k,\tau)$ is the regularized incomplete beta function:
\begin{equation}
\begin{split}
\theta(k,\tau)&=\frac{B(\tau,1+k,1+k)}{B(1,1+k,1+k)}
\\  B(\tau,a,b)&=\int_{0}^{\tau}dy\, y^{a-1}(1-y)^{b-1},
\end{split}
\end{equation}
and ${\rm Re}(a)>0,$ ${\rm Re}(b)>0$, and $|\tau|\leq1$. In this
protocol, the first $k$ derivatives vanish at the turning points  $t=0,{t_1},{t_2}, {t_3}$.
Figs.~\ref{fig:transprob-no-diss} and \ref{fig:phaseerror-no-diss} plot the transition probability $P_{G\rightarrow E}$ and phase error $||\rho_\text{G}(t_\text{op}) - \rho_A ||$ as a function of braiding
period ${t_\text{op}}$. As expected, we see that diabatic-induced excitations and Berry
phase error can be reduced by making the time dependence smoother.

We now consider the effect of dissipation.  To obtain the master equation, Eq.~(\ref{eqn:MasterEquation}), we used the standard Born and Markovian approximations.  We also replaced $U_{S}^\dagger (t,t-\tau)$ with $e^{i\tau H_S(t)}$,
which is justified when the bath has short correlation time.  We now apply a fourth approximation:
\begin{equation}
\begin{split}
\label{eqn:approx4}
\Gamma_{\alpha\beta}(t,\omega) &=\int_{0}^{\infty}d\tau\, g_{\alpha}(t)g_{\beta}(t-\tau)e^{i\tau\omega}\mathcal{B}_{\alpha\beta}(\tau)\\
&\approx g_{\alpha}(t)g_{\beta}(t)\int_{0}^{\infty}d\tau\, e^{i\tau\omega}\mathcal{B}_{\alpha\beta}(\tau)
\end{split}
\end{equation}
which significantly simplifies the numerical calculation. Eq.~(\ref{eqn:approx4})
generally overestimates the effect of dissipation, and is justified in the limit of
small $\tau_B /{t_\text{op}}$ (again $\tau_B$ is bath correlation time).
The comparisons between the transition probability $P_{G\rightarrow E}$
and $||\rho_\text{G}(t_\text{op}) - \rho_A ||$
with and without dissipation are shown in Figs.~\ref{fig:transition-rate-dissipation} and \ref{fig:Berry-correction-with dissipation}.  The choice of parameters for Figs.~\ref{fig:transition-rate-dissipation} and \ref{fig:Berry-correction-with dissipation} 
are valid for the first-order weak coupling expansion and the Born-Markov approximation used in the previous Appendix to derive the master equation.  
That is, the parameters satisfy the constraints that the bath correlation time ($\sim \hbar/T$) is much smaller than the system relaxation time ($\sim \hbar\Delta/\lambda^2$), i.e. $\lambda^2/\Delta\ll T$, and that the timescale associated with the system dynamics is much less than the system relaxation time (in terms of energies, $\lambda \ll \Delta$).  We see
that dissipation washes out the oscillations in the transition probability.

\section{Chern Simons Calculation}\label{CS}

We fill in the details of Section~\ref{sec:CS-theory}. In the following, we will replace $k/2\pi$ with $\theta$ to avoid confusion with the momentum variable $k$.   Beginning with the action of Eq.~(\ref{eqn:CS-action}), integrating out $a_\mu$ yields the effective action
\begin{align}
S_{\text{eff}} &= \int d^3xd^3y \, j_\mu(x) G^{\mu \nu }(x, y) j_\nu (y)
\end{align}
 where  the propagator
\begin{align}
 G^{\mu \nu}(x,y) &=\langle x | \frac{ m^2}{\theta}\frac{\epsilon^{\mu \nu \lambda}\partial_\lambda}{\partial^2(\partial^2+m^2)}-\frac{ m}{\theta}\frac{g^{\mu \nu}}{\partial^2-m^2}|y \rangle
\end{align}
and  $m= g^2 \theta$.  We fix the gauge such that $\partial_\mu a^\mu=0$ and work with the signature $g^{00}=1, g^{ii}=-1$.

One approach to finding the topological contribution to the phase is to find the magnetic field due to the stationary $b$ particle and calculate the flux enclosed by the trajectory of the $a$ particle.  Note that the vector potential due to the $b$ particle is
\begin{align}
	a_{\mu,b}(x) &=\int d^3y\, G_{\mu \nu }(x,y) j^\nu_{ b}(y),
\end{align}
and to single out the part contributing to the braiding phase we can replace $G$ with $G^{(1)}$.
The magnetic field is then
\begin{align}
B_b = \epsilon^{0ij}\partial_i a_{j,b}
\end{align}
Rotational symmetry implies $B_b(\vr)\equiv B_b(r)$.

 Noting that $j_b^0(k) = 2\pi\delta(\omega)$ we find
\begin{equation}
\begin{split}
B_b(r) &= \frac{bm^2}{(2\pi)^2\theta}\int_0^\infty d|k| \int_0^{2\pi}d\theta_k \frac{|k| e^{i|k|r\cos(\theta-\theta_k)}}{|k|^2+m^2}
\\ &= \frac{bm^2}{\theta}\frac{1}{2\pi}\int_0^\infty d|k| \frac{|k|}{|k|^2+m^2}J_0(|k|r)
\end{split}
\end{equation}
where the last line follows from the identity
\begin{equation}
J_n(x)=\frac{1}{2\pi}\int_{-\pi}^\pi d\theta e^{i(n\theta-x\sin\theta)}.
\end{equation}
The integral evaluates to
\begin{align}
B_b(r)&= \frac{bm^2}{2\pi \theta}K_0(mr).
\end{align}

Particle $a$ sweeps out a circular area of radius $R$, enclosing flux
\begin{equation}
\begin{split}
\Phi &=a \int_0^{2\pi}d\theta \int_0^R dr B_b(r)
\\ &= \frac{ab}{\theta}\left[1-m R K_1(mR)\right].
\end{split}
\end{equation}
Working in the limit $mR=g^2 \theta R\gg1$ we are interested in the asymptotic form of $K_1(x)\sim \sqrt{\frac{\pi}{2x}}e^{-x}(1+\mathcal{O}(\frac{1}{z}))$, therefore the braiding phase is given by
\begin{align}
\Phi&= \frac{ab}{\theta}\left( 1-\sqrt{\frac{ \pi mR }{2}}e^{-mR}+...\right).
\end{align}
As we claimed, this has no dependence on the braiding time ${t_\text{op}}$.  The exponential suppression in $m$ and $R$ originates from the Maxwell term: the flux attached to each particle is no longer an infinitely thin solenoid but rather has finite width and so has an exponential decay away from the particle.

The second term in the propagator $\langle x |\frac{m}{\theta (k^2-m^2)}g^{\mu \nu}|y \rangle$ contributes an overall phase that grows linearly in time and is thus reminiscent of a dynamical phase.  We are interested in the phase resulting from the braiding process and thus only keep the term involving both currents.  Transforming to momentum space we find this term is
\begin{equation}
\begin{split}
&\frac{m}{\theta}\frac{1}{(2\pi)^3} \int d^2 k\int d\omega\, j_a^\mu(-k) \frac{g^{\mu \nu}}{k^2-m^2}j_b^\nu(k)
\\ &= \frac{bm}{\theta(2\pi)^2}\int d|k|\, \frac{|k|}{|k|^2+m^2}\int d\theta_k\, j^0_a(-\mathbf{k},\omega=0).
\end{split}
\end{equation}

The $a$ particle current is
\begin{equation}
\begin{split}
j^0_{a}(k)
&= \int d^2 r \int_0^{t_\text{op}} dt\, e^{ikr}j^0_a(r)
\\ &=a \int_0^{t_\text{op}} dt\,  e^{-i|k|R\cos(\theta_k-\frac{2\pi t}{t_\text{op}})+i\omega t}
\\ &=a \int_0^{t_\text{op}} dt\, \sum_n i^n e^{-in\theta_k} J_n(|k|R) e^{i(\omega-\frac{2\pi n}{t_\text{op}})t}
\\ &=a 2{t_\text{op}} \sum_n i^n e^{-in\theta_k}J_n(|k|R)e^{i(\omega {t_\text{op}}-2\pi nm)}
\\ &\quad\quad\quad\quad\times\text{sinc}(\omega {t_\text{op}}  -2\pi n m)
\end{split}
\end{equation}
and when we plug this into the $\theta_k$ integral we find:
\begin{align}
\int_0^{2\pi}d\theta_k\, j_{a}^0(-\mathbf{k},\omega=0)
= a2{t_\text{op}} J_0(|k|R).
\end{align}
At this point the dependence on the number of times the particle circles the origin has vanished, thus this term does not contribute to the braiding phase.

\section{Top-Transmon Details}\label{Details}

\subsection{Deriving the Effective Hamiltonian}

In this section we use the notation of Ref.~\onlinecite{Hyart13} to facilitate comparison with their results.  The microscopic Lagrangian of Fig.~\ref{HyartFig} is~\cite{Hyart13}
\begin{equation}\label{Lagrangian}
\mathcal{L}=\Omega^\dagger\left( T-V_J-V_M\right)\Omega
\end{equation}
where $T$ is the charging energy, $V_J$ is the Josephson potential, $V_M$ is the Majorana-Josephson potential, and $\Omega$ is a gauge transformation that enforces a constraint between the charge in a superconducting island and the Majorana parity of that island:
\begin{equation}\label{Omega}
\Omega=e^{\frac{i}{4}(1-i\gamma_b\gamma_b')\phi_b}\prod_{k=1}^3 e^{\frac{i}{4}(1-i\gamma_k\gamma_k')\phi_k},
\end{equation}
where $\phi_k$ is the superconducting phase of island $k$.

Making the assumption that the cross-capacitance between Majorana islands $i$ and $j$, $C_{ij}$, is negligible in comparison with the capacitances involving the bus and the ground, $C_{B,i}$, $C_{G,i}$, $C_{B,G}$, the charging energy is:
\begin{equation}\label{THyart13}
\begin{split}
\Omega^\dagger T \Omega &=\frac{\hbar^2}{8e^2}C_b \dot{\phi}_b^2+\frac{\hbar^2}{8 e^2}\sum_{k=1}^3\big[C_{G,k}\dot{\phi}_k^2+C_{B,k}(\dot{\phi}_k-\dot{\phi}_b)^2\big]
\\ &~~+\frac{\hbar}{2e}\Big[q_b\dot{\phi}_b+\sum_{k=1}^3\big[ q_k+\frac{e}{2}(1-i\gamma_k\gamma_k')\big] \dot{\phi}_k\Big].
\end{split}
\end{equation}
In the above, $q_k$ is the induced charge of island $k$.

The Josephson potential takes the form
\begin{equation}\label{VJHyart13}
\begin{split}
\Omega^\dagger V_J \Omega &=
2E_{J,0}(\Phi_b)(1-\cos\phi_b)
\\ &~~~+\sum_{k=1}^3 2 E_{J,k}(\Phi_k)(1-\cos\phi_k),
\end{split}
\end{equation}
for magnetic flux $\Phi_k$ and Joesphson energy $E_{J,k}(\Phi_k)$.  The Majorana-Josephson potential is given by
\begin{equation}\label{VMHyart13}
\begin{split}
V_M=E_M\Big[&i\gamma_1'\gamma_2'\cos\Big(\frac{\phi_1-\phi_2}{2}+\alpha_{12}\Big)
\\ &+i\gamma_2'\gamma_3'\cos\Big(\frac{\phi_2-\phi_3}{2}+\alpha_{23}\Big)
\\ &+i\gamma_3'\gamma_1'\cos\Big(\frac{\phi_3-\phi_1}{2}+\alpha_{31}\Big)\Big]
\\ +E_M\Big[&i\gamma_b'\gamma_g'\cos\Big(\frac{\phi_b}{2}+\alpha_{bg}\Big)
\\&+i\gamma_g'\gamma_1\cos\Big(\alpha_{g1}-\frac{\phi_1}{2}\Big)
\\ &+i\gamma_1\gamma_b'\cos\Big(\frac{\phi_1-\phi_b}{2}+\alpha_{1b}\Big)\Big],
\end{split}
\end{equation}
where $E_M$ characterizes the strength of the Majorana-Josephson energy and the $\alpha$s denote the Aharanov-Bohm phase shifts between different islands, for instance
\begin{equation}
\begin{split}
\alpha_{12} &= \frac{\pi}{2\Phi_0}(\Phi_1+\Phi_2)
\\ \alpha_{23} &=\frac{\pi}{2\Phi_0}(\Phi_2+\Phi_3)
\\ \alpha_{31}&= -\frac{\pi}{2\Phi_0}(-\Phi_1+2\Phi_2+\Phi_3).
\end{split}
\end{equation}

The essential assumption is
\begin{equation}\label{Ineq}
E_J(\Phi)\gg E_M, E_C,
\end{equation}
 where $E_C$ is the single-electron charging energy for each junction and $0\leq \Phi\leq \Phi_{\text{max}}<\frac{1}{2}\Phi_0$.  The range of values of $\Phi$ is to keep $E_{J,k}(\Phi_k)$ strictly positive (recall that $E_{J,k}(\Phi_k)=E_{J,k}(0)\cos(\pi\frac{\Phi_k}{\Phi_0})$).  There is a tradeoff in how close  $\Phi_{\text{max}}$ is to $\frac{1}{2}\Phi_0$, which we will address later.  We see that the action is minimized when all superconductors are in phase.  At the minimum $\phi_z=0$, both $T$ and $V_J$ vanish and as is easily seen from Eq.~(\ref{Omega}), $\Omega^\dagger V_M\Omega|_{\phi_z=0}=V_M|_{\phi_z=0}$.  The low energy Hamiltonian will simply contain $V_M$ and terms accounting for phase fluctuations.

The amplitude of a phase slip from $0$ to $2\pi$ at a junction $k$ is $U_k=$
\begin{equation}\label{U}
\begin{split}
	&16 \left( \frac{E_{C,k} E_{J,k}(\Phi_k)^3}{2\pi^2}\right)^{1/4} e^{-\sqrt{\frac{2E_{J,k}( \Phi_k)}{E_{C,k}}}}\cos\Big(\frac{q_k \pi}{e}\Big).
\end{split}
\end{equation}
The WKB-like form can be understood from an analogy between a Cooper-pair box Hamiltonian and a quantum rotor model, discussed in Ref.~\onlinecite{Koch07}.  We can see from the exponential dependence on the ratio $E_{J,k}(\Phi_k)/E_{C,k}$ that we only need to account for phase slips at junctions with the smallest ratio of Josephson energy to charging energy.

 During braiding the Josephson energy between the bus and the ground is maximized, therefore we can set $\phi_b=0$ and ignore fluctuations.  We do need to account for fluctuations about junctions $i=1,2,3$ where the flux is being tuned such that the Josephson energy of a particular junction is being minimized at certain points of the braiding protocol.  The low-energy effective Hamiltonian thus takes the form
\begin{equation}\label{HyartHeff}
H_{\text{eff}}=-\sum_{k=1}^3 i U_k \gamma_k\gamma_k'+\Omega^\dagger V_M \Omega|_{\phi_k=0}.
\end{equation}

\begin{figure}[t!]
	\includegraphics[width=0.7\columnwidth]{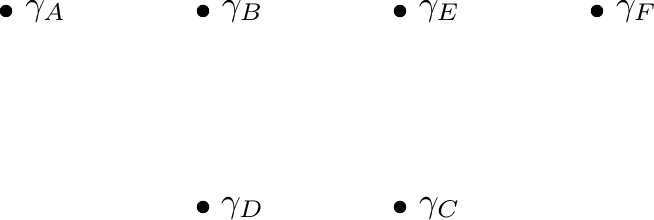}
\caption{Low energy MZM picture.}
\label{PiSimple}
\end{figure}

Making the assumption that tunnel couplings are much stronger than Coulomb couplings, $E_M \gg U_k$, and only keeping terms to first order in $U_k$, yields the low energy Hamiltonian
\begin{equation}\label{HHyart13}
\begin{split}
H&= -i\Delta_1 \gamma_B\gamma_E -i\Delta_2 \gamma_E\gamma_F-i\Delta_3\gamma_E\gamma_C
\end{split}
\end{equation}
where $\gamma_F=\gamma_2, \gamma_C=\gamma_3$ and $\gamma_B$ is a linear combination of the MZMs at the first junction ($\gamma_1, \gamma_b', \gamma_g'$) and $\gamma_E$ is a linear combination of the MZMs at the second junction ($\gamma_1', \gamma_2', \gamma_3'$), see Fig.~\ref{PiSimple}.  The coupling strengths $\Delta_k=U_k f(\alpha)$, where $f(\alpha$) is some function of the single electron Aharanov-Bohm phase shifts which is $\mathcal{O}(1)$ for the allowed range of $\Phi$.  The specific form of $f$ and of $\gamma_B, \gamma_E$ is known \cite{Hyart13}, but is not important for our discussion.  Note that the closer $\Phi_{\text{max}}$ is to $\frac{1}{2}\Phi_0$, the larger the value of $\Delta_{\text{max}}=\Delta_i(\Phi_{\text{max}})$ and the smaller the ratio $\Delta_{\text{min}}/\Delta_{\text{max}}$.  Ref.~\onlinecite{Heck12} finds the unitary evolution operator is equivalent to the Berry matrix for braiding MZMs up to corrections of order $\Delta_{\text{min}}/\Delta_{\text{max}}$.  It is therefore important for this ratio to be small to ensure quantization of the braiding phase.

For readout we set $\Phi_b=\Phi_{\text{max}}$ and $\Phi_k=0, ~k=1,2,3$.  We can ignore phase fluctuations about $\phi_k=0, \dot{\phi}_k=0$ and therefore the Lagrangian becomes
\begin{equation}\label{HyartReadout}
\begin{split}
	\mathcal{L}&= \frac{\hbar^2}{8e^2}C\dot{\phi}_b^2+\frac{\hbar}{2e}\Big[q_b+\frac{e}{2}(1-i\gamma_b\gamma_b')\Big]\dot{\phi}_b
\\ &-E_{J,b}(1-\cos\phi_b)-\Omega^\dagger V_M \Omega|_{\phi_k=0}.
\end{split}
\end{equation}

The Lagrangian is that of a top-transmon~\cite{Hassler11} with the extra term,  $V_M$.  A top-transmon is a hybrid topological-superconducting qubit for which the bus hosts MZMs, the parity of which splits each transmon energy level. The transmon is set inside a transmission line resonator, for instance a coplanar waveguide with interrupted feedline.  The authors of Ref.~\onlinecite{Hyart13} derive the readout Hamiltonian under the assumptions that the transmon remains in its lowest two energy levels and the resonator coupling can be described by the Jaynes-Cummings Hamiltonian:
\begin{equation}\label{HReadout}
\begin{split}
H&= \sigma_z\big[\frac{\hbar}{2} \Omega_0 +i\gamma_b \gamma_b' \delta_+\cos(\pi q_b/e)\big]+i\gamma_b\gamma_b' \delta_- \cos(\pi q_b/e)
\\ &~~~+\Omega^\dagger V_M \Omega|_{\phi_z=0}+\hbar\omega_0 a^\dagger a +\hbar g (\sigma_+ a +\sigma_-a^\dagger).
\end{split}
\end{equation}
In the above, $\sigma_z$ acts on the qubit degree of freedom of the transmon and $\sigma_{\pm}$ are the raising/lowering operators of the transmon state.  $a, a^\dagger$ describe the photons in the resonator, $\omega_0$ is the bare resonator frequency, $\Omega_0$ is the frequency spacing of the two lowest levels of the transmon (with no MZM), and $g$ is the resonator-transmon coupling strength.  The MZM couplings $\delta_{+}, \delta_-\sim e^{-\sqrt{8 E_{J,b}(\Phi_{\text{max}})/E_{C,b}}}$ are the average dispersion of the lowest two transmon states and half the difference in dispersion of the lowest two transmon states, respectively.  That is, if $\delta \varepsilon_0$ is the difference in ground state energy of the transmon when the MZMs are in an even parity state and an odd parity state, and $\delta \varepsilon_1$ is the analogous quantity for the first excited state,
\begin{equation}
\delta_{\pm}= \frac{\delta \varepsilon_1  \pm \delta \varepsilon_0}{2}.
\label{eqn:deltapm}
\end{equation}

\subsection{Energy Subspaces of the Effective Hamiltonian}\label{Subspaces}

The low energy subspace of the Hamiltonian in Eq.~(\ref{HyartHeff}) is given by Eq.~\eqref{HHyart13} neglecting terms $\mathcal{O}\Big(\big( \frac{\Delta}{E_M}\big)^2\Big)$.  At the first turning point, $\Phi_1=\Phi_{\text{max}}, \Phi_2=\Phi_3=0$, and the Hamiltonian is
\begin{equation}
H \approx iU_{\text{max}}\gamma_1\gamma_1'+iE_M\Big[\gamma_2'\gamma_3'+\frac{1}{\sqrt{2}}(\gamma_1'\gamma_2'+\gamma_3'\gamma_1')\Big].
\label{IntermediateHam}
\end{equation}
In the above, we write $U_{\text{max}}=U_i(\Phi_i=\Phi_{\text{max}})$. 

A change of basis allows us to diagonalize Hamiltonian  of Eq.~\eqref{IntermediateHam} as
\begin{equation}
	H= i\varepsilon_1\tilde{\gamma}_0\tilde{\gamma}_1+ i\varepsilon_{2}\gamma_{\text{ex},1}\gamma_{\text{ex},2}.
	\label{eqn:lambda-ham}
\end{equation}
Here
\begin{equation}
\begin{split}
	\varepsilon_1&= \sqrt{2}{U_{\text{max}}} +\mathcal{O}\Big(\frac{U_{\text{max}}}{E_M}\Big)^2
\\ \varepsilon_{2} &= \sqrt{2}E_M +\frac{\sqrt{2}}{8}\frac{U_{\text{max}}^2}{E_M}+\mathcal{O}\Big(\frac{U_{\text{max}}}{E_M}\Big)^2
\end{split}
\end{equation}
 The four (many-body) eigenstates have the following energies to lowest order in $U_{\text{max}}/E_M$:
\begin{equation} \label{T1Energy}
\begin{split}
	\lambda_0&=-\varepsilon_1-\varepsilon_2\\
	\lambda_1&= \varepsilon_1-\varepsilon_2
\\ \lambda_{2} &= \varepsilon_2-\varepsilon_1\\
\lambda_3&=\varepsilon_1+\varepsilon_2
\end{split}
\end{equation}
$\tilde{\gamma}_0,~ \tilde{\gamma}_1,\gamma_{\text{ex},1}$ and $\gamma_{\text{ex},2}$ are linear combinations of $\gamma_1,~\gamma_1',~\gamma_2',$ and $\gamma_3'$, the exact forms of which are unimportant for this discussion.

 $\gamma_2$ and $\gamma_3$ do not appear in the Hamiltonian of Eq.~(\ref{IntermediateHam}), thus occupying their associated fermionic mode has no energy cost. The energy associated with $i\tilde{\gamma}_0\tilde{\gamma}_1$ is on the order of the Coulomb couplings $U_{\text{max}}$, while the energy of the excited modes is on the order of the Majorana couplings, $E_M\gg U_{\text{max}}$.  The gap between the first excited state and ground state is much smaller than the gap between the first and second excited states, thus the most common diabatic errors are when $i\tilde{\gamma}_0\tilde{\gamma}_1$ is flipped.  The system transitions to higher excited state when the parity of $i\gamma_{\text{ex},1}\gamma_{\text{ex},2}$ flips.

At any time during the braiding process, one of the Coulomb couplings $U_i$ is set to its minimum value and the Hamiltonian can be approximated by
\begin{equation}\label{Simple}
\begin{split}
H &= -iU_1\gamma_1\gamma_1'-iU_2\gamma_2\gamma_2'-iE_M(\gamma_1'\gamma_2'+\gamma_2'\gamma_3'+\gamma_3'\gamma_1').
\end{split}
\end{equation}
We have set $U_3=0$ (discarding the exponentially small value of $U_{\text{min}}=U_i(\Phi_i=0)$) and ignored the anisotropy in the Majorana couplings.  A similar change of basis as before allows us to write the Hamiltonian in terms of effective MZM operators.  Writing the Hamiltonian again in the form of Eq.~(\ref{eqn:lambda-ham}) we find
\begin{equation}
\begin{split}
	&\varepsilon_1' =\frac{1}{\sqrt{3}}\sqrt{U_1^2+U_2^2} + \mathcal{O}\Big(\frac{U^2}{E_M}\Big).
\\ & \varepsilon_2'=\sqrt{3}E_M\Big(1+\frac{2}{9}\frac{U_1^2+U_2^2}{E_M^2}\Big).
\end{split}
\end{equation}
Note that during braiding the energy gaps are the same order of magnitude as at the turning points and as $\gamma_3$ does not appear in the Hamiltonian in Eq.~(\ref{Simple}), each energy level is two-fold degenerate .

If we consider coupling the MZM system to a dissipative bath, the bath can relax the system from a highly excited state into a lower state.  Assuming the bath cannot change the total parity of the MZM system and that it cannot affect the qubit, which is stored in the decoupled MZM pair, then we see that the bath can relax the system from the third excited state to the ground state, $\ket{3}\to \ket{0}$, or from the second excited state to the first excited state $\ket{2}\to \ket{1}$.

\subsection{Modified Architecture}\label{Modified}

\begin{figure}
\includegraphics[width=\columnwidth]{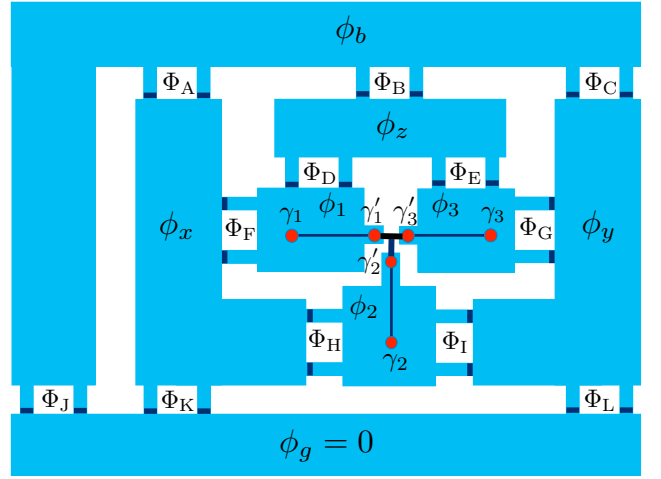}
\caption{$T$-junction architecture with parity check ability. Every MZM island can be connected to the bus (ground) with the other two MZM islands connected to the ground (bus).  No MZM islands share a split Josephson junction, thus the Coulomb couplings for different MZM pairs are independent of each other. }
\label{fig:ModifiedTPCheck}
\end{figure}

We now show that the effective Hamiltonian of Fig.~\ref{HyartMod} is of the form of equation~\ref{HyartHeff}, with the essential feature that the $U_k$ can be tuned between exponentially separated minimum and maximum values.

The microscopic Lagrangian is given by Eq.~(\ref{Lagrangian}), with modified Josephson potential and charging energy:
\begin{equation}\label{ModifiedVJandT}
\begin{split}
\Omega^\dagger V_J \Omega &= E_{J,A}(\Phi_A)(1-\cos(\phi_b-\phi_1))
\\ &+ E_{J,B}(\Phi_B)(1-\cos(\phi_b-\phi_3))
\\ &+E_{J,0}(\Phi_0)(1-\cos\phi_b)
\\ &
+\sum_{k=1}^3 E_{J,k}(\Phi_k) (1-\cos\phi_k)
\\ \Omega^\dagger T\Omega &= \frac{\hbar^2}{2e}(C_{b,g}\dot{\phi}_b^2 +\sum_{k=1}^3 \left( C_{k,g}\dot{\phi}_k^2 +C_{k,b}(\dot{\phi}_b -\dot{\phi}_k)^2\right)
\\ &+\frac{\hbar}{2e} \left( q_b \dot{\phi_b} +\sum_{k=1}^3 (q_k +\frac{e}{2}(1-i\gamma_k\gamma_k'))\dot{\phi}_k \right)
\end{split}
\end{equation}
The Majorana-Josephson potential is unchanged up to appropriate redefinition of the single electron Aharanov-Bohm phase shifts.

The system operates in the regime $E_J(\Phi)\gg E_C, E_M$, thus the action is minimized when all superconducting islands are in phase ($V_J$ and $T$ vanish).  We account for phase fluctuations about the minima $\phi_k=0$ by considering phase slips from $0$ to $2\pi$ at each junction.  The tunneling amplitude associated with a phase slip from $0$ to $2\pi$ at junction $Z$ is given by $U_Z(\Phi_Z) =$
\begin{equation}
16 \left(\frac{E_{C,Z}E_{J,Z}(\Phi_Z)^3}{2\pi^2} \right)^{1/4}
e^{-\sqrt{\frac{8 E_{J,Z}(\Phi_Z)}{E_{C,Z}}}}\cos\Big(\frac{q_Z\pi}{e}\Big)
\end{equation}
where $q_Z$ is the difference in induced charge and $E_{C,Z}$ is the difference in charging energies between the two islands on either side of the junction.  As is easily seen from the above, if certain junctions have a larger ratio of $E_J(\Phi)/E_C$ than others, the amplitude of phase slips at these junctions is exponentially smaller and can thus safely be ignored.

During the braiding process, we maximize $E_{J,0}(\Phi_0)$ and ignore phase fluctuations about the minima $\phi_b=\phi_g=0$.  Note that every other junction (for which we do need to account for phase fluctuations) has a Majorana wire on one of its neighboring islands and not on the other.  The low energy Hamiltonian for the modified architecture is
\begin{equation}
\begin{split}
H_{\text{eff}}&= \Omega^\dagger V_M \Omega|_{\phi_i=0}-i( U_1 +U_A)\gamma_1\gamma_1'
\\ &~~~-iU_2\gamma_2\gamma_2'-i(U_3+U_B)\gamma_3\gamma_3'
\\ &= \Omega^\dagger V_M \Omega|_{\phi_z=0}-i\sum_{k=1}^3 \tilde{U}_k \gamma_k\gamma_k',
\end{split}
\end{equation}
which is of the same form as Eq.~(\ref{HyartHeff}).  Each $\tilde{U}_k$ can be independently tuned between exponentially separated values $\tilde{U}_{\text{min}}$.

To understand the last point more explicitly, note that threading zero flux through any junction maximizes the Josephson energy of that junction while threading flux $\Phi_{\text{max}}$ minimizes the Josephson energy.  Thus if we compare the value $\tilde{U}_1(\Phi_1=0, \Phi_A=0)$ with $\tilde{U}_1(\Phi_1=\Phi_{\text{max}},\Phi_A=\Phi_{\text{max}})$, the former is the sum of two numbers, each of which is exponentially smaller than the corresponding term in the latter as $\sqrt{E_J(\Phi_{\text{max}})/E_C}\ll \sqrt{E_J(0)/E_C}$.

We now show that for each parity measurement in the modified system the Langrangian takes the same form as Eq.~(\ref{HyartReadout}).  First, to measure $i\gamma_1\gamma_0$ we tune $\Phi_A=\Phi_2=\Phi_3=0$ to maximize the Josephson energies of the corresponding junctions, which sets $\phi_b=\phi_1$ and $\phi_2=\phi_3=0$.  We ignore phase fluctuations about these minima.  The Lagrangian becomes
\begin{equation}\label{LagMeasurement1}
\begin{split}
	\mathcal{L}&= \frac{\hbar^2}{8e^2}C_1\dot{\phi}_1^2+\frac{\hbar}{2e}\Big[q'_1+\frac{e}{2}(1-i\gamma_1\gamma_1')\Big]\dot{\phi}_1
\\ &~~~-E_{J,1}(\Phi_{\text{max}})(1-\cos\phi_1) -\Omega^\dagger V_M\Omega|_{\phi_i=0},
\end{split}
\end{equation}
where $C_1=C_{b,g}+C_{1,g}+C_{2,b}+C_{3,b}$, $E_{J,1}(\Phi_{\text{max}})=E_{J,0}(\Phi_{\text{max}})+E_{J,1}(\Phi_{\text{max}})+E_{J,B}(\Phi_{\text{max}})$, and $q_1'=q_1+1_b$.  Eqs.~(\ref{LagMeasurement1}) and (\ref{HyartReadout}) have the same form, indicating the same arguments given in Ref.~\onlinecite{Hyart13} apply to our modified architecture. The same analysis applies for measuring $i\gamma_3\gamma_0$ by interchanging the roles of junctions $A$ and $B$ and of junctions $1$ and $3$.

To measure $i\gamma_2\gamma_0$ tune $\Phi_A=\Phi_B=\Phi_2=0$ to maximize the Josephson energies of the corresponding junctions and tune all remaining fluxes to $\Phi_{\text{max}}$.  Now $\phi_b=\phi_1=\phi_3$ and $\phi_2=0$.  Again, ignoring phase fluctuations about these mininima, the Lagrangian has become
\begin{equation}\label{LagMeasurement2}
\begin{split}
\mathcal{L}&= \frac{\hbar^2}{8 e^2}C_2\dot{\phi}_b^2
-E_{J,2}(\Phi_{\text{max}})(1-\cos\phi_b)-\Omega^\dagger V_M \Omega|_{\phi_i=0}
\\ &+\frac{\hbar}{2e}\Big[q_b+q_1+q_3+\frac{e}{2}(2-i\gamma_1\gamma_1'-i\gamma_3\gamma_3')\Big]\dot{\phi}_b.
\end{split}
\end{equation}
In the above, $C_2=C_{b,g}+C_{1,g}+C_{3,g}+C_{2,b}$.  We assume that the total parity of the MZM system is known, thus if
\begin{equation}
 x=\frac{1}{2}(3-i\gamma_1\gamma_1'-i\gamma_2\gamma_2'-i\gamma_3\gamma_3')
\end{equation}
we know whether $x$ is an even or odd number.   This allows us to rewrite the Lagrangian as
\begin{equation}\label{LagMeasurement2p}
\begin{split}
	\mathcal{L}&= \frac{\hbar^2}{8 e^2}C_2\dot{\phi}_b+\frac{\hbar}{2e}\Big[q'_2-\frac{e}{2}(1-i\gamma_2\gamma_2')\Big]\dot{\phi}_b
\\&-E_{J,2}(\Phi_{\text{max}})(1-\cos\phi_b)-\Omega^\dagger V_M \Omega|_{\phi_i=0},
\end{split}
\end{equation}
where $q_2'=q_b+q_1+q_3+ex$.  Eq.~(\ref{LagMeasurement2p}) takes the same form as Eq.~(\ref{HyartReadout}) up to an unimportant sign difference.

The preceding analysis relies on the assumption that the total parity of the MZM system has not changed during the braiding and measurement process (if it were to change, measuring the parity of $\frac{1}{2}(2-i\gamma_1\gamma_1'-i\gamma_3\gamma_3')$ would not tell us the correct value fo the parity of $\frac{1}{2}(1-i\gamma_2\gamma_2')$).  We can further modify the architecture to that of Fig.~\ref{fig:ModifiedTPCheck} to allow us to check the total parity at each turning point.  The key features are of the more complex architecture are that each MZM island can be coupled to the bus with the other two MZM islands coupled to the ground and each MZM island can be coupled to the ground with the other two MZM islands coupled to the bus.  For instance, we can measure $1/2(1-i\gamma_1\gamma_1')$ and then measure $\frac{1}{2}(2-i\gamma_2\gamma_2'-i\gamma_3\gamma_3')$.  It is also possible to connect all three MZM islands to the bus as a further check of the total parity.

The analysis that the geometry of Fig.~\ref{fig:ModifiedTPCheck} leads to the same effective Hamiltonian for the braiding process and Lagrangian for measurement as Eqs.~(\ref{HyartHeff}) and (\ref{HyartReadout}) respectively is much the same as the preceding analysis for Fig.~\ref{HyartMod}, therefore we will just outline the key points rather than going through the full derivation.  There are no Josephson junctions directly connecting two MZM islands, thus each Coulomb coupling $\Delta_i$ can be independently tuned between exponentially separated minimum and maximum values.  For readout, each island containing a MZM wire can be phase locked to the bus (ground) with the other two MZM islands phase locked to the ground (bus).

Finally, Ref.~\onlinecite{Heck12} notes that the order of flux-tuning matters.  Fixing the ground state degeneracy to two when switching the Coulomb coupling in island $k$ off and the coupling in island $k'$ on requires increasing $|\Phi_k'|$  before decreasing $|\Phi_k|$.

\section{Measurement}\label{Measurement}

\begin{figure}[t]
	\includegraphics[width=0.8\columnwidth]{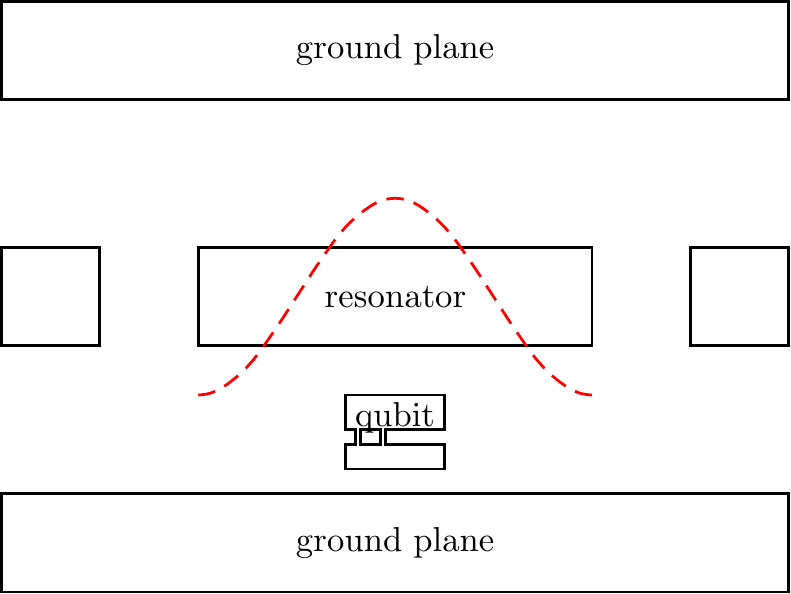}
\caption{Schematic illustration of the transmon qubit\cite{Koch07}.  The qubit (lower-middle) is embedded inside a coplanar waveguide (two ground planes with a feedline running through the center).  The short section of feedline traps a standing wave (shown in red) and can thus be thought of as a resonator.  The qubit (in this case two superconductors connected by a split Josephson junction) is capacitively coupled to the resonator, and slightly shifts the resonantor's frequency. A wave traveling down the feedline will experience an impedance mismatch due to the resonator/qubit.  This will result in a transmission amplitude $S_{21}$ whose value depends on the state of the qubit.  }
\label{transmon}
\end{figure}

We now give a more detailed explanation of the measurement described Section~\ref{sec:SC-measurement} of the main text.  It might appear rather extraordinary that one can measure the parity of $i\gamma_i\gamma_0$, even though $\gamma_0$ is a linear combination of three MZMs located on different superconductors. The key insight is that the Majorana-Josephson coupling $\mathcal{O}(E_M)$ remains the same order of magnitude throughout the braiding and measurement process and is much stronger than the Coulomb couplings.  Thus even when the superconducting islands are disconnected from each other, the inner MZMs are still coupled and affect the parity of the superconducting island being measured.

This measurement is employed in superconducting qubit experiments to readout the qubit state \cite{Sank14}.  A qubit is capacitively coupled to a transmission line resonator, for instance a coplanar waveguide with interrupted feed line.  The setup is schematically shown in Fig.~\ref{transmon}.  A microwave traveling down the transmission line experiences an impedance mismatch at the point where the line couples to the resonator.  The microwave is partially reflected and partially transmitted, with its transmission amplitude, $S_{21}$, dependent on both the probe wave's frequency and the frequency of the resonator.  The resonator's frequency is shifted from its bare value, $\omega_0,$ by the state of the qubit.  Using standard signal processing techniques one can extract $S_{21}$ and from this value infer the state of the qubit.  A note on terminology: in the superconducting qubit literature this measurement is known as a ``dispersive measurement" because it infers the state of the qubit from a shift in the resonator's frequency.  In this paper we call this measurement projective because its utility for our system is that it projects the MZM into a definite energy state.

In preparation of a measurement, our system is tuned such that one MZM island is connected to the bus and the remaining MZM islands are connected to phase ground.  The system will look like a transmon with MZMs, i.e. a top-transmon \cite{Hassler11}.  The MZMs split the transmon's ground state  into four energy levels, each of which is two-fold degenerate (these energies are reported in Appendix \ref{Subspaces}).  Measurement of $S_{21}$ projects the system into one of these eigenstates.  With sufficient resolution, we can readout this state and thus detect an error from transitions out of the ground state.

Let $\omega_{\ket{j}}$ denote the effective frequency of the resonator when the transmon is in its ground state and the MZM system is in state $\ket{j}$.  Tuning the probe frequency to be directly between $\omega_{\ket{0}}$ and $\omega_{\ket{1}}$, the resonator's effective frequencies when qubit is in its lowest two energy states, allows for the greatest separation between the two most probable measurement outcomes~\cite{Jeffrey14}.  That is,
\begin{equation}
\omega_{\text{probe}}=\frac{\omega_{\ket{0}}+\omega_{\ket{1}}}{2}.
\end{equation}
The measurement result is plotted in the IQ plane (I=Re$[S_{21}]$, Q=Im[$S_{21}$]).  In a noiseless system there would be exactly one point for each of the system's energy states, and the measurement result would be one of these points.
In reality, noise from circuit elements and from the finite bandwidth of the incoming microwave smears the possible outcomes into a distribution centered about the noiseless point.  To determine which energy state a given outcome corresponds to, these distributions are projected onto the line connecting the noiseless points for the two lowest-energy states.  For long enough measurement time, the distributions will be normal.  The intersection of two distributions denotes the dividing line between the two measurement results: a result to the left of the dividing line is interpreted to mean the system is in the state whose distribution peaks to the left of the intersection point.  As long as the peaks of the distributions are well separated, only the  tails overlap and measurement errors are exponentially small.  As mentioned in Section~\ref{sec:discussion}, if a measurement returns a result near the dividing line, we can always repeat the measurement to find an unambiguous results.

The separation of the noiseless points is directly proportional to
\begin{equation}
\label{freqsep}
 \frac{\omega_{\text{probe}}-\omega_{\ket{j}}}{\omega_{\ket{j}}},
\end{equation}
and the proportionality depends on the quality factors of the resonator and transmon.  The system is in the dispersive regime, that is for resonator-qubit coupling strength $g$ and detuning $\delta\omega=\Omega_0-\omega_0$ (recall that $\Omega_0$ is the transmon frequency while $\omega_0$ is the bare resonator frequency), $\delta\omega\gg g$.
As will be shown, $\omega_{\ket{j}}=\omega_0$ to lowest order in $g/\delta\omega$, therefore the distance between noiseless points in the IQ plane is approximately proportional to the frequency difference between the probe microwave and the effective resonator frequency (the numerator of Eq.~(\ref{freqsep}).  We can therefore use this frequency separation to estimate the necessary resolution for a measurement to distinguish the different MZM states.  We proceed by finding the effective frequencies of the resonator.

We begin with the full readout Hamiltonian given in Ref.~\onlinecite{Hyart13}
\begin{equation} \label{Hr}
\begin{split}
H_{r}&= \hbar \omega_0 a^\dagger a +\hbar g (\sigma_+a+\sigma_- a^\dagger)
+\sigma_z\Big(\frac{\hbar}{2}\Omega_0+i\gamma_1\gamma_1'\delta_+\Big)
\\ &~~+i\gamma_1\gamma_1'\delta_-+V_M,
\end{split}
\end{equation}
where $a$ is the annihilation operator for a photon in the resonator, $\sigma_z$ describes the qubit degree of freedom, and $g$ is the transmon-resonator coupling strength.

We define
\begin{equation}
\label{HMZMpm}
H_{\text{MZM}\pm} = i\gamma_1\gamma_1'( \delta_- \pm \delta_+)+V_M.
\end{equation}
Note that this Hamiltonian takes the same form as the Hamiltonian in Eq.~(\ref{IntermediateHam}) with $U_{\text{max}}\to \delta_- \pm \delta_+$.

Let $\lambda_{j_\pm}$ be the $j$th eigenvalue of $H_{\text{MZM}\pm}$ corresponding to eigenstate $\ket{j_\pm}$.  The $\lambda_{j_\pm}$s can easily be deduced from the results of Appendix \ref{Subspaces}. We can write our basis as $\ket{n, \pm ,j}\equiv \ket{n}\otimes \ket{\pm}\otimes\ket{j_\pm}$.  Then
\begin{equation}
\begin{split}
H_r \ket{n,+,j} &= \Big[\hbar \omega_0 n +\frac{\hbar}{2} \Omega_0 +\lambda_{j,+}\Big]\ket{n,+,j}
\\ &\quad +\hbar g \sqrt{n+1} \sum_k\langle k_-|j_+\rangle\ket{n+1,-,k}
\\ H_r \ket{n+1,-, j} &= \Big[ \hbar \omega_0 (n+1) -\frac{\hbar}{2}\Omega_0 +\lambda_{j,-} \Big] \ket{n+1,-,j}
\\ &\quad+\hbar g \sqrt{n+1} \sum_k\langle k_+|j_-\rangle\ket{n,+,k}.
\end{split}
\end{equation}
The Hamiltonian can be analyzed using non-degenerate perturbation theory, and we find the associated energies up order $\left( g/\delta \omega \right)^2$ are
\begin{equation}
\begin{gathered}
	\begin{split}
	\epsilon_{n,+,j}=&\Big(n+\frac{1}{2}\Big)\omega_0+\frac{\delta\omega}{2}\\
	&+g^2(n+1)\sum_{k}\frac{|\langle j_+|k_-\rangle|^2}{\delta\omega+\lambda_{j,+}-\lambda_{k,-}}
	\end{split}
	\\
	\begin{split}
	\epsilon_{n+1,-,j}=&\Big(n+\frac{1}{2}\Big)\omega_0-\frac{\delta\omega}{2}\\
	&-g^2(n+1)\sum_{k}\frac{|\langle j_-|k_+\rangle|^2}{\delta\omega+\lambda_{k,+}-\lambda_{j,-}}
	\end{split}
\end{gathered}
\end{equation}
Because the system has positive detuning, $\Omega_0>\omega_0$, we can assume that the transmon remains in its ground state (only consider $\epsilon_{n,-,j}$).  The effective resonator frequency is given by
\begin{equation}
\begin{split}
	\omega_{\ket{j}} &= \epsilon_{n+1,-,{j}}-\epsilon_{n,-,{j}}
\\ &= \omega_0 - g^2\sum_{k}\frac{|\langle j_-|k_+\rangle|^2}{\delta\omega+\lambda_{k,+}-\lambda_{j,-}}
\end{split}
\end{equation}
Notice that although in an eigenstate $\ket{n,-,j}$ are mixed with $\ket{n+1,+, k}$, but the majority of the weight is still in $\ket{n,-,j}$ in the regime $g\ll \delta\omega$ and $\delta_\pm \ll E_M$ (e.g. the overlap $\langle 1_+|2_-\rangle$ are of the order $(\delta_\pm/E_M)^2$), so it still makes sense to label the eigenstates and the resonator frequency by $\ket{j}$.

Therefore, the frequency separation for the three lowest MZM states, keeping only leading order terms in $g/\delta\omega$ and first order terms in $\frac{\delta_-}{E_M}$:
\begin{equation}
\begin{split}
&\omega_{\text{probe}} -\omega_{\ket{0}} =  -\left( \omega_{\text{probe}}-\omega_{\ket{1}}\right) \sim \frac{2\sqrt{2} g^2 \delta_+}{\delta \omega^2-8\delta_+^2}\\
&\omega_{\text{probe}} -\omega_{\ket{2}}
\\ &\sim \frac{2\sqrt{2} g^2 \delta_+}{\delta \omega^2-8\delta_+^2}\left( 1+\frac{8\delta_+^2+2\sqrt{2}\delta_+\delta\omega+\delta\omega^2}{2(8\delta_+^2-\delta\omega^2)}\frac{\delta_-}{E_M}\right),
\end{split}
\end{equation}
 where we have used the results of Appendix \ref{Subspaces}.  Except near the pole $\delta_+=2\sqrt{2}\delta\omega$, the second term in parenthesis of $\omega_{\text{probe}}-\omega_{\ket{2}}$ is $\mathcal{O}(1)$, thus we see the $\ket{0}$ and $\ket{2}$ states are separated in the IQ plane by a factor of $\delta_-/E_M\ll 1$ less than the $\ket{0}$ and $\ket{1}$ states.  This can in part be understood by considering the Hamiltonian in Eq.~(\ref{Hr}).  The resonator couples to the transmon energy levels and the transmon couples to the parity of $i\gamma_1\gamma_1'$.  $\gamma_2'$ and $\gamma_3'$ couple to $\gamma_1'$ through $V_M$.  The parity of $i\gamma_1\gamma_0$, which distinguishes states $\ket{0}$ and $\ket{1}$ but not states $\ket{0}$ and $\ket{2}$, is thus coupled more strongly to the resonator (and has a greater effect on the transmission amplitude) than the parity of $i\gamma_{\text{e}x,1}\gamma_{\text{ex},2}$, which is only coupled to the resonator through higher order processes.  Since $\ket{0}$ and $\ket{2}$ are distinguished by this parity, they have a smaller difference in their effect on the transmission amplitude.

\section{Error Correction}
\label{MeasurementError}

We now derive Eq.~(\ref{eqn:meas-error-ineq}) of Section~\ref{sec:discussion}.  If $0<\varepsilon_{\text{meas}}\ll \varepsilon$, then the probability of having an error after $n$ recovery steps is the probability of the ancillas having odd parity after $n$ recovery steps, $\varepsilon[t_u](2\varepsilon[t_u](1-\varepsilon[t_u]))^n$ to lowest order, plus the probability that after $j<n$ steps the ancillas had odd parity but we interpreted the measurement result incorrectly.  The probability that after the initial unitary evolution the ancillas have odd parity but we measure even parity is $\varepsilon[t_u]\varepsilon_{\text{meas}}$.
The probability that after $j>1$ steps the ancillas have odd parity but we measure even parity is (dropping the explicit dependence on $t_u$ for the moment)
\begin{equation}
\begin{split}
&(\varepsilon_{\text{meas}}(1-\varepsilon)+\varepsilon(1-\varepsilon_{\text{meas}}))
\\ &\times (2\varepsilon(1-\varepsilon)(1-\varepsilon_{\text{meas}})+\varepsilon_{\text{meas}}(1-\varepsilon)^2+\varepsilon_{\text{meas}}\varepsilon^2)^{j-1} \\ &\times 2\varepsilon \varepsilon_{\text{meas}}(1-\varepsilon)
\\ &\approx \varepsilon_{\text{meas}}\varepsilon(2\varepsilon)^j.
\end{split}
\end{equation}
The first line corresponds to the probability that after the initial evolution we measure the ancillas to have odd parity, the second line is the probability that $j-1$ recovery steps return an odd parity measurement, and the third line is the probability that in the final recovery step the ancillas have odd parity but we measure even parity.  The approximation in the last line keeps only the leading order term.

Therefore, the leading order correction from measurement is simply $\varepsilon_{\text{meas}}\varepsilon[t_u]$, coming from misinterpreting the result of the first measurement. We can safely ignore measurement at the $n^{\text{th}}$ recovery step provided that
\begin{equation}
\varepsilon_{\text{meas}}\varepsilon[t_u]\ll \varepsilon[t_u](2\varepsilon[t_u](1-\varepsilon[t_u]))^n,
\end{equation}
which together with our initial assumption tells us it is safe to ignore measurement error when
\begin{equation}
\label{eq:meas-error-bound}
\varepsilon_{\text{meas}}\ll \text{min}_{n\in\mathbb N}\left( \varepsilon[t_u], (2\varepsilon[t_u](1-\varepsilon[t_u]))^n\right).
\end{equation}
For small $\varepsilon[t_u]$, $\langle n [t_u]\rangle\ll 1$, and $\varepsilon_{\text{meas}}\ll \varepsilon[t_u]$ is sufficient.

\section{Reality Check}
\label{Reality}

The following inequalities must be satisfied
\begin{equation}
 \delta\omega\gg g
\end{equation}
and
\begin{equation}
\Delta_{\text{SC}},\Delta_{\text{wire}}>E_J,\hbar\Omega_0,\hbar\omega_0 \gg E_M,\Delta \gg k_B T, \Delta_{\text{min}},
\label{eqn:ineqs}
\end{equation}
 where $\Delta_{\text{SC}}$ is the superconducting gap of the island, $\Delta_{\text{wire}}$ is the superconducting gap in the nanowire, and $\Delta$ ($\Delta_{\text{min}}$) is the maximum (minimum) Coulomb coupling between MZMs on the same wire.  We satisfy these inequalities with the physically reasonable frequency estimates
\begin{equation}
\begin{split}
\Delta_{\text{wire}},\Omega_0, \omega_0 &\sim 100 \,\text{GHz}
\\ E_M &\sim 50\,\text{GHz}
\\ \Delta &\sim 10 \,\text{GHz}
\\ \frac{\delta\omega}{2\pi}&=\frac{\Omega_0-\omega_0}{2\pi}\sim 200\,\text{MHz}
\\ \frac{\delta_+}{2\pi},\frac{\delta_-}{2\pi}&\sim 150\,\text{MHz}
\\ \frac{g}{2\pi}&\sim 40\,\text{MHz}.
\label{FreqEstimates}
\end{split}
\end{equation}
The estimates of $\Delta_{\text{wire}},\Omega_0,\omega_0,\Delta,\delta_{\pm}$ are taken from Appendix F of Ref.~\onlinecite{Hyart13}.  We have chosen $\delta\omega$ and $g$ to be comparable to the values listed in Ref.~\onlinecite{Koch07}.  $E_M$ was chosen to satisfy the inequality, its actual value is exponentially sensitive to the length scales of the physical system.

It is worth noting that some of the constraints are threshold inequalities while others involve orders of magnitude difference.  For example, we need $\Delta_{\text{wire}},\Delta_{\text{SC}}>\omega_0$ so that performing the measurement does not induce bulk quasiparticles, but a factor of two is probably sufficient.  On the other hand, the relative phase has corrections $\mathcal{O}(\Delta_{\text{min}}/\Delta)$, thus we need this ratio to be as small as possible (a reasonable ratio of Josephson energy to charging energy for a transmon is $E_J/E_C\sim 50$~\cite{Koch07}, corresponding to $\Delta_{\text{min}}/\Delta \sim e^{-20}$, so we expect errors from a finite value of $\Delta_{\text{min}}$ to be negligible).  The frequency estimates above do not take into account these subtleties, so while it appears that both $\omega_0$ and $\Delta_{\text{wire}}$ have the same frequency estimate, in an actual experiment one would carefully track $\mathcal{O}(1)$ factors and choose experimental parameters (such as the appropriate superconductor) to ensure this is not the case.  We also note that throughout this paper we assume $E_M\gg \Delta$, but as noted in Ref.~\onlinecite{Hyart13} the results must remain valid when $E_M$ and $\Delta$ are comparable, due to the topological nature of the braiding.

We use the values listed in Eq.~(\ref{FreqEstimates}) to estimate the measurement time needed to resolve the energy levels of the MZM system.  In the previous section, we found the magnitude of the frequency splitting of the two possible measurement outcomes is
\begin{equation}
|\frac{\omega_{\ket{0}}-\omega_{\ket{1}}}{2}| \sim \frac{2\sqrt{2} g^2 \delta_+}{8\delta_+^2-\delta \omega^2}\approx 30\,\text{MHz}.
\end{equation}
(Note that for these values of $g$ and $\delta\omega$ the dispersive shift for a transmon is $\sim \frac{g^2}{\delta\omega}\approx 50$\,MHz.)

We assume that noise in the system results in a normal distribution centered at the noiseless point for each state of the MZM system.  To distinguish the measurement results of the resonator's effective frequency, the width of these peaks, $\sigma$, must be smaller than the separation between the two peaks.   For the ground and first excited states we have the condition
\begin{equation}
\frac{\left( \omega_{\ket{0}}-\omega_{\ket{1}}\right)^2}{8\sigma^2}> 1,
\end{equation}
which implies $\sqrt{2}\sigma < 30\,$MHz.  Too large a value of $\sigma$ could result in measurement error or could require taking multiple measurements. The uncertainty relation sets a lower bound on the measurement time necessary to place a measurement outcome within one of these Gaussian peaks:
\begin{equation}
\label{eq:meas-time-bound}
\sigma t_{\text{meas}} =\frac{1}{2}\Rightarrow t_{\text{meas}}  > 20\,\text{ns}.
\end{equation}

In order for the measurement to resolve all four energy states of the MZM system, $\sigma$ must decrease by a factor of $\delta_-/E_M$.  Thus $t_{\text{meas}}$ must increase by $E_M/\delta_-\sim 50$ (for our frequency estimates), which sets the lower bound on the  measurement time around 1\,$\mu$s.

The bound on the measurement time does not appear to be a fundamentally limiting factor; for comparison, experiments have reported quasiparticle poisoning times ranging from 10ms to 1 minute \cite{Higginbotham15, Woerkom15}.  Note that the usual decoherence times (relaxation time $T_1$ and decoherence time $T_2$) that affect superconducting qubits do not apply here as we simply want the transmon to remain in its ground state.  The relevant source of error is instead the thermal population of the excited state, which for a system at 20 mK is between 5-10\%~\cite{Jeffrey14}.

\end{document}